\documentclass[preprint,5p,times,twocolumn]{elsarticle}

\usepackage{amssymb}
\usepackage{amsmath}

\usepackage{tabularx}
\usepackage[utf8]{inputenc}
\usepackage{longtable}
\usepackage{siunitx}
\usepackage{float}
\usepackage{upgreek}
\usepackage{algorithm}
\usepackage{soul}
\usepackage{array}
\usepackage{algpseudocode}
\usepackage[none]{hyphenat}
\usepackage{microtype}
\usepackage{stfloats} 
\usepackage{bm}
\usepackage{bbm}
\usepackage[table]{xcolor}
\usepackage{booktabs}
\usepackage{pdflscape}
\usepackage{enumitem}
\usepackage{svg}
\usepackage{caption}
\usepackage{subcaption}
\PassOptionsToPackage{hyphens}{url}\usepackage{hyperref}
\usepackage{xurl}

\hypersetup{colorlinks}

\definecolor{specialred}{rgb}{0.69, 0.17, 0.15} 
\definecolor{specialblue}{rgb}{0.16, 0.26, 0.40} 
\definecolor{specialcyan}{rgb}{0.65, 0.87, 0.95} 

\journal{Applied Energy}

\begin{document}

\begin{frontmatter}

\titlebeforetext{Offshore Wind Turbine Tower Design and Optimization: A Review and AI-Driven Future Directions}

\title{Offshore Wind Turbine Tower Design and Optimization: \\A Review and AI-Driven Future Directions}

\affiliation[inst1]{organization={Department of Mechanical Engineering, Massachusetts Institute of Technology},
            city={Cambridge},
            state={MA},
            country={USA}}

\affiliation[inst2]{organization={LAETA-INEGI, Faculty of Engineering, University of Porto},
            city={Porto},
            country={Portugal}}
            
\affiliation[inst3]{organization={Faculty of Mechanical Engineering, Delft University of Technology},
            city={Delft},
            country={Netherlands}}

\affiliation[inst6]{organization={School of Engineering, Brown University},
            city={Providence},
            state={RI},
            country={USA}}
            
\affiliation[inst7]{organization={CONSTRUCT, Faculty of Engineering, University of Porto},
            city={Porto},
            country={Portugal}}
            
\affiliation[inst4]{organization={TEMA, University of Aveiro},
            city={Aveiro},
            country={Portugal}}

\affiliation[inst5]{organization={Department of Mechanical Engineering, and Department of Electrical and Computer Engineering, University of Texas at Dallas},
           city={Richardson},
            state={TX},
            country={USA}}

\cortext[cor1]{Corresponding author.}
 
\author[inst1,inst2]{João Alves Ribeiro\corref{cor1}}\ead{jpar@mit.edu}
\author[inst3,inst6]{Bruno Alves Ribeiro}
\author[inst7]{Francisco Pimenta}
\author[inst4]{Sérgio M. O. Tavares}
\author[inst5]{Jie Zhang}
\author[inst1]{Faez Ahmed}

\begin{abstract}
Offshore wind energy leverages the high intensity and consistency of oceanic winds, playing a key role in the transition to renewable energy. As energy demands grow, larger turbines are required to optimize power generation and reduce the Levelized Cost of Energy (LCoE), which represents the average cost of electricity over a project's lifetime. However, upscaling turbines introduces engineering challenges, particularly in the design of supporting structures, especially towers. These towers must support increased loads while maintaining structural integrity, cost-efficiency, and transportability, making them essential to offshore wind projects' success. This paper presents a comprehensive review of the latest advancements, challenges, and future directions driven by Artificial Intelligence (AI) in the design optimization of Offshore Wind Turbine (OWT) structures, with a focus on towers. It provides an in-depth background on key areas such as design types, load types, analysis methods, design processes, monitoring systems, Digital Twin (DT), software, standards, reference turbines, economic factors, and optimization techniques. Additionally, it includes a state-of-the-art review of optimization studies related to tower design optimization, presenting a detailed examination of turbine, software, loads, optimization method, design variables and constraints, analysis, and findings, motivating future research to refine design approaches for effective turbine upscaling and improved efficiency. Lastly, the paper explores future directions where AI can revolutionize tower design optimization, enabling the development of efficient, scalable, and sustainable structures. By addressing the upscaling challenges and supporting the growth of renewable energy, this work contributes to shaping the future of offshore wind turbine towers and others supporting structures.

\end{abstract}

\begin{graphicalabstract}
\includegraphics[width=1\textwidth]{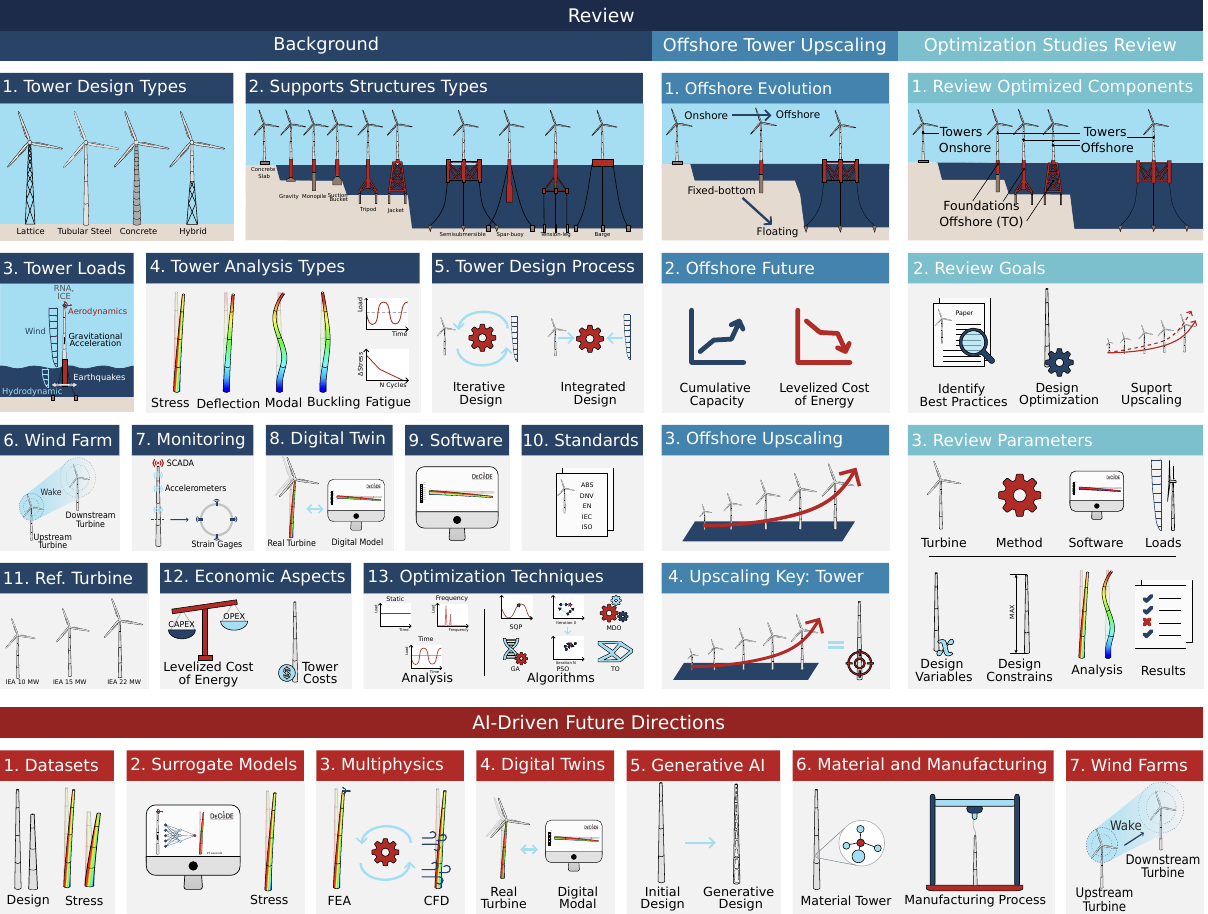}
\end{graphicalabstract}

\onecolumn
\begin{highlights}
\item Summarize foundational background on offshore wind turbine tower design and optimization.
\item Highlight the tower’s role as a crucial element in wind turbine upscaling.
\item Review key studies on optimization design for offshore wind turbine towers.
\item Present future AI-driven research directions in offshore wind turbine design and optimization.
\end{highlights}
\twocolumn


\end{frontmatter}

\section{Introduction}

\textbf{\textit{The Role of Wind Energy in Sustainable Development:}} 
Wind energy is emerging as a cornerstone in the global effort to combat climate change, offering unique opportunities to drastically reduce carbon emissions, improve public health, and preserve vital resources. By 2050, wind power is expected to account for over 30\% of global electricity production, with the potential to reduce CO\(_2\) emissions by up to 14,871 million tonnes annually, equivalent to the total emissions of China in 2021 \cite{Long_Chen_Xu_Li_Liu_Wang_2023}. This remarkable impact on emissions is not just a crucial step in mitigating climate change but also brings profound health benefits. For example, in the United States, wind power in 2014 alone resulted in an estimated 2.0 billion USD in health savings due to improved air quality, a figure that could rise to 8.4 billion USD if the displacement of fossil fuel generators prioritized those with higher health impacts \cite{Qiu_Zigler_Selin_2022}. Furthermore, wind energy's resource efficiency is unparalleled, requiring only 0.004 liters of water per kilowatt-hour, in stark contrast to the 1.9 liters for oil and 2.3 liters for nuclear power \cite{Saidur_Rahim_Islam_Solangi_2011}. This significant reduction in water consumption is especially critical in the face of growing global water scarcity. 

\textbf{\textit{The Role of Wind Turbines in the Energy Revolution:}}
Central to this energy revolution is the wind turbine, a technology that converts the kinetic energy of wind into electrical power and has significantly evolved since its origins in the 18th century \cite{Gipe_Möllerström_2022}. This early prototype laid the foundation for modern wind energy systems, which are now deployed at large scales for commercial electricity generation, both onshore and, more recently, offshore. While onshore wind technology is well-developed and widely implemented, offshore wind energy is rapidly emerging as a major growth sector.

\textbf{\textit{The Transition from Onshore to Offshore Wind Turbines:}}
The transition from onshore to Offshore Wind Turbines (OWTs) was driven by the need to access stronger and more consistent winds, which are typically available at sea, presenting benefits such as higher and constant energy production and reduced visual and noise impacts. Initially, offshore wind farms utilized fixed foundations like monopiles and jackets in shallow waters, but as the industry sought to exploit deeper waters with stronger wind resources, the limitations of fixed structures became evident. This led to the development of floating OWT, suitable for deployment in deeper waters beyond 50 meters, expanding the geographic scope of offshore wind projects and accessing superior wind conditions. 

\textbf{\textit{The Advancements in Floating Wind Turbine Technology:}}
As the floating wind turbine technology has advanced and become increasingly viable, the focus shifted to optimizing performance, with a key goal being the reduction of the Levelized Cost of Energy (LCoE), representing the average cost of electricity over a project's lifetime. To achieve this, the industry has increased turbine power ratings, hub heights, and rotor diameters, adopted larger turbines, capturing more wind and generating more energy \cite{doi:10.1126/science.aau2027}. The increased hub height, along with longer blade lengths, requires turbines to be mounted on taller towers to access stronger and more stable winds at higher altitudes, while ensuring adequate clearance for a larger swept area. This configuration maximizes energy capture, reduces the total number of turbines required, and decreases capital and operational costs, collectively lowering the LCoE and enhancing the economic feasibility of floating wind projects.

\textbf{\textit{The Role of the Tower for Offshore Wind Turbine Upscaling:}}
As wind turbines increase in size, the tower becomes a critical element, required to support the added weight of larger blades and withstand the higher forces generated by stronger wind speeds. From the early days of the wind industry, towers have played a fundamental role for two key reasons: they raise the rotor to capture optimal wind resources and provide a reliable load path from the turbine to the foundation \cite{Ng_Ran_2016}. While the objective of achieving strength with minimal mass remains central to reducing material and labor costs, the push for taller turbines and greater hub heights has intensified the demands on tower design. Towers must balance reduced weight with the ability to handle diverse operating conditions and withstand extreme events throughout their lifetime, while also remaining practical for manufacturing and transportation. Therefore, optimizing tower design is crucial to fully realize the benefits of larger turbines, such as reduced LCoE and improved efficiency, without compromising structural integrity or economic viability.

\begin{figure*}[b!]
    	\centering
        \includegraphics[width=1\textwidth]{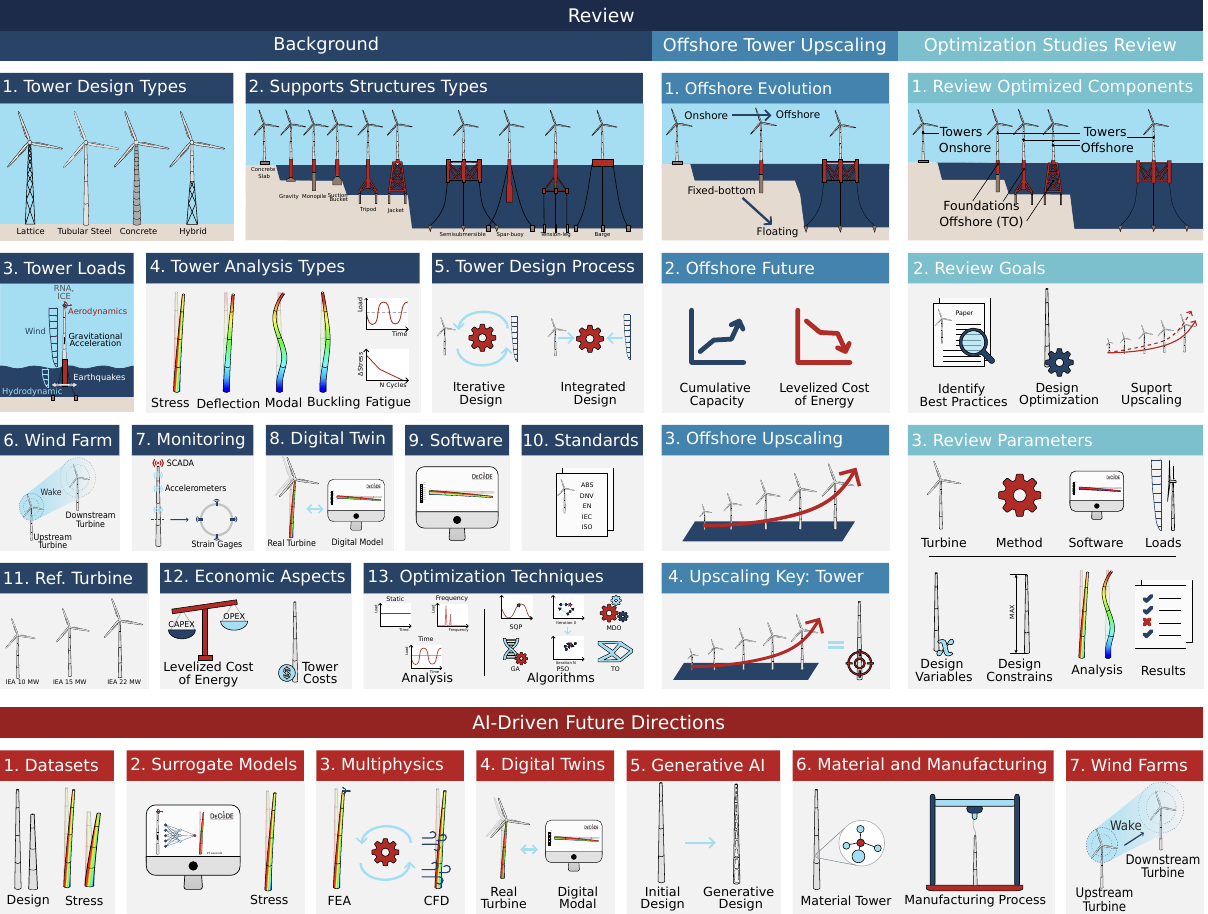}
        \caption{OWT Design and Optimization: Review and AI-Driven Future Directions. Review: (1) Background on tower and support structure design,  load types, analysis methods, isolated and wind-farm tower design processes, monitoring systems, digital twin, software, standards, reference turbines, economic factors, and optimization techniques; (2) Offshore Tower Upscaling, exploring its evolution, future prospects, and its critical role in turbine scaling; (3) Optimization Studies Review, highlighting significant contributions to the design optimization of towers and foundations. AI-Driven Future Directions: dataset generation, surrogate models, multiphysics simulations, digital twins, generative AI, advanced materials and manufacturing, and wind farm design.
}
        \label{fig:flowchart}
\end{figure*}

\textbf{\textit{The Optimization Techniques in Wind Turbine Design:}}
Various optimization algorithms have significantly advanced wind turbine design by improving performance and efficiency. Techniques such as Sequential Quadratic Programming (SQP) \cite{Häfele_Gebhardt_Rolfes_2019}, Genetic Algorithms (GA) \cite{Chen_Li_He_2020}, Particle Swarm Optimization (PSO) \cite{Li_Xu_Yuan_2024}, Multidisciplinary Design Optimization (MDO) \cite{Cui_Allison_Wang_2021}, and Topology Optimization (TO) \cite{Marjan_Huang_2023} have played a crucial role in refining turbine structures by balancing key parameters such as weight, cost, and structural integrity. These foundational algorithms in design optimization have achieved greater precision with advancements in computational power, though this has led to rising computational costs. Emerging Artificial Intelligence (AI) techniques offer promising pathways to enhance efficiency and reduce the computational demands of design optimization.

\textbf{\textit{The Role of Artificial Intelligence in Engineering Design:}}
AI, particularly through Machine Learning (ML) and Deep Learning (DL), are transforming engineering design processes. Techniques such as Deep Neural Networks (DNNs) \cite{10.1063/1.1144830}, Diffusion Models (DFs) \cite{10.1145/3626235}, and Large Language Models (LLMs) \cite{10.1145/3641289} are driving significant innovation across various industries. Additionally, Digital Twin (DT)  technologies \cite{9899718}, which simulate real-world systems in a virtual environment, are increasingly being integrated with AI to enhance predictive capabilities and optimize designs \cite{LENG2021119, LO2021101297}. These advanced methods have shown remarkable success in fields such as aerospace (e.g., aircraft design \cite{designs8020029, 10.1115/1.4064673}), automotive (e.g., vehicle design \cite{10.1115/1.4065063, elrefaie2024drivaernetlargescalemultimodalcar}), energy (e.g., power transformer and battery design \cite{doi:10.1177/1464420721992445, sung2024coolingguidediffusionmodelbattery}), shipbuilding (e.g., ship design \cite{jmse11122215, bagazinski2023shipdshiphulldataset}), and personal mobility (e.g., bicycle design \cite{10.1115/1.4052585, regenwetter2024bikedmultimodaldataset14}). 

\textbf{\textit{The Role of AI in Design Optimization of the OWT:}} 
AI techniques have demonstrated transformative impacts across multiple fields and are now positioned to significantly enhance OWT design optimization by enabling faster and more efficient design processes. AI models offer valuable insights into complex interactions among turbine structures within wind farms, dynamic environmental loads, and varying operational conditions. This capability facilitates structural performance optimization, improves fatigue prediction accuracy, and strengthens overall reliability. When integrated within DT frameworks, AI further enhances predictive capabilities, enabling real-time monitoring and predictive maintenance, which collectively reduce downtime and extend the lifespan of critical components. Additionally, AI-driven approaches can streamline computational resources by approximating complex simulations, accelerating design iterations, and supporting more efficient wind energy systems. Consequently, these advancements contribute to the development of more resilient, scalable, and economically viable OWT designs.

\subsection{Objectives and Contributions} 

The primary objective of this paper is to provide a comprehensive review of current advancements and challenges and AI-driven future directions in OWT design optimization, with a particular focus on tower. 

To support this objective, the paper offers a detailed background on OWT towers to establish a foundational understanding. Building on recent trends in the field, it further highlights the tower’s critical role as a central factor in upscaling turbines, demonstrating its importance in the development of larger and more efficient OWTs.

The key contributions of this paper are as follows:
\begin{itemize} 
    \item \textbf{\textit{Comprehensive Background on OWT Tower Design:}} Provides an in-depth overview of tower and support structure designs, load types, analysis methods, isolated and wind-farm tower design processes, monitoring systems, DT technologies, design software, standards, reference wind turbines, economic factors, and optimization techniques, serving as a foundational resource for researchers.
    
        \pagebreak

    \item \textbf{\textit{The Tower’s Role in OWT Upscaling:}} Contributes to OWT upscaling by identifying the tower as a critical element for scaling efficiency. This analysis emphasizes tower optimization as a key strategy to support the development of larger turbines, grounded in a review of historical developments and future trends in wind turbine evolution.

    \item \textbf{\textit{Tower Optimization Studies Review:}} Offers a comprehensive review of optimization techniques for OWT towers, integrating insights from onshore turbine research to address the limited offshore-specific studies. The review also covers TO for foundations, given the structural similarities between towers and foundations. It highlights best practices and key challenges in tower design optimization, establishing a foundation for future research. Key analysis areas include turbine types, software, load conditions, optimization methods, design variables and constraints, analysis techniques, and main findings.

    \pagebreak
    \item \textbf{\textit{AI-Driven Future Directions in Tower Design:}} Proposes seven AI-driven research directions to advance OWT tower design, including dataset generation for AI-driven optimization, development of surrogate models, integration of multiphysics simulations, DT technologies, generative AI applications, material and manufacturing innovations, and optimization of tower design within wind farms.

\end{itemize}

\begin{table*}[b!]
\scriptsize	
\centering
\caption{Overview of types of designs for wind turbine towers (adapted from \cite{Ng_Ran_2016, Hernandez-Estrada_Lastres-Danguillecourt_Robles-Ocampo_Lopez-Lopez_Sevilla-Camacho_Perez-Sariñana_Dorrego-Portela_2021}).}
\begin{tabularx}{\textwidth}{@{}p{1.25cm}p{0.95cm}p{1.75cm}p{5.5cm}p{5cm}p{0.5cm}p{0.9cm}@{}}
\toprule
\textbf{Design} & \textbf{Material} & \textbf{Application [m]} & \textbf{Advantages} & \textbf{Disadvantages} & \textbf{Fig.} & \textbf{Refs.} \\ \midrule
 Lattice   & Steel & 20-160 & 
\begin{minipage}[t]{5.5cm}
    \begin{itemize}[label={-}, leftmargin=*, topsep=0pt, partopsep=0pt, parsep=0pt, itemsep=0pt]
        \item Lower construction and transport costs.
        \item Superior stiffness.
        \item Reduced wind load.
    \end{itemize}
\end{minipage}
&
\begin{minipage}[t]{5cm}
    \begin{itemize}[label={-}, leftmargin=*, topsep=0pt, partopsep=0pt, parsep=0pt, itemsep=0pt]
        \item Susceptible to fatigue.
        \item Aesthetically less appealing.
        \item Causes turbulence to blades.
        \item Ice load in cold places.
        \item Attracts bird nests, increasing bird strike risk.
        \item Difficult and costly to clear debris or ice.
    \end{itemize}
\end{minipage}
& \ref{fig:design_features_wtts}-a & \cite{Zwick_Muskulus_Moe_2012, Long2012LatticeTF} 
\\
\addlinespace[0.2cm]
Tubular  & Steel   & 20-120 & 
\begin{minipage}[t]{5.5cm}
    \begin{itemize}[label={-}, leftmargin=*, topsep=0pt, partopsep=0pt, parsep=0pt, itemsep=0pt]
        \item Aesthetically pleasing.
        \item Uniform rigidity in all directions.
        \item Predictable dynamic and fatigue properties.
        \item Safe and convenient maintenance.
        \item Encloses power electronics and switchboards.
    \end{itemize}
\end{minipage}
&
\begin{minipage}[t]{5cm}
    \begin{itemize}[label={-}, leftmargin=*, topsep=0pt, partopsep=0pt, parsep=0pt, itemsep=0pt]
        \item High transport and assembly costs.
        \item Expensive for very tall towers.
    \end{itemize}
\end{minipage}
& \ref{fig:design_features_wtts}-b & \cite{Uys_Farkas_Jármai_van_Tonder_2007, Hu_Baniotopoulos_Yang_2014}
\\
\addlinespace[0.2cm]
On-site/ \newline Precast  & Concrete  & 60-120 & 
\begin{minipage}[t]{5.5cm}
    \begin{itemize}[label={-}, leftmargin=*, topsep=0pt, partopsep=0pt, parsep=0pt, itemsep=0pt]
        \item High durability.
        \item Superior stiffness.
        \item Easy installation.
    \end{itemize}
\end{minipage}
&
\begin{minipage}[t]{5cm}
    \begin{itemize}[label={-}, leftmargin=*, topsep=0pt, partopsep=0pt, parsep=0pt, itemsep=0pt]
        \item Vulnerable to weather conditions.
        \item Vulnerable joints.
        \item High transport and assembly costs.
    \end{itemize}
\end{minipage}
& \ref{fig:design_features_wtts}-c &\cite{Ma_Meng_2014, Paredes_Barbat_Oller_2011}
\\
\addlinespace[0.2cm]
Hybrids  & Hybrids  & 80-150 & 
\begin{minipage}[t]{5.5cm}
    \begin{itemize}[label={-}, leftmargin=*, topsep=0pt, partopsep=0pt, parsep=0pt, itemsep=0pt]
        \item Low construction and transport costs.
        \item High resistance.
        \item Potential to overcome previous design weaknesses.
    \end{itemize}
\end{minipage}
&
\begin{minipage}[t]{5cm}
    \begin{itemize}[label={-}, leftmargin=*, topsep=0pt, partopsep=0pt, parsep=0pt, itemsep=0pt]
        \item Experimental stage.
    \end{itemize}
\end{minipage}
& \ref{fig:design_features_wtts}-d & \cite{Chen_Li_He_2020, Alvarez-Anton_Koob_Diaz_Minnert_2016}
\\ \bottomrule
\end{tabularx}
\label{table:design_features_wtts}
\end{table*}

\subsection{Work Outline}
The remainder of the paper is organized as follows. \autoref{sec:background} presents a detailed background on OWT structures design, focusing on towers. \autoref{sec:upscaling} discusses the evolution and future prospects of offshore wind energy, emphasizing the tower's critical role in supporting upscaling efforts. \autoref{sec:review} provides a comprehensive review of optimization techniques for OWT towers. Lastly, \autoref{sec:futureprespectives} explores future trends in tower design and optimization, highlighting AI's role in driving innovation, design efficiency, and accuracy.

\section{Background} \label{sec:background}

\subsection{Tower Design Types} \label{sec:tower_design_types}

\begin{figure}[htt!]
	\centering
    \includegraphics[width=0.485\textwidth]{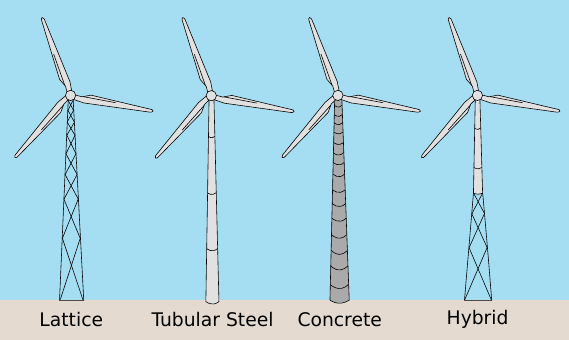}
    \caption{Types of design for wind turbine towers: (a) Lattice, (b) Tubular Steel, (c) Concrete, and (d) Hybrid.}
    \label{fig:design_features_wtts}
\end{figure}

Wind turbines are predominantly built from steel or concrete due to their exceptional strength and durability. Steel towers are generally designed as either lattice or tubular structures, while concrete towers are reinforced with steel and can be either built on-site or delivered as prefabricated units. Hybrid towers, which combine steel and reinforced concrete, are also available \cite{Hernandez-Estrada_Lastres-Danguillecourt_Robles-Ocampo_Lopez-Lopez_Sevilla-Camacho_Perez-Sariñana_Dorrego-Portela_2021}. Each type of tower design is suited to different height requirements and applications, as shown in \autoref{table:design_features_wtts} and \autoref{fig:design_features_wtts}.

In remote or off-grid areas of developing countries, the high cost of transporting steel or concrete structures can be prohibitive due to accessibility issues. For these regions, alternative materials such as timber and bamboo are being explored to reduce costs and increase sustainability \cite{Pourrajabian_Dehghan_Javed_Wood_2019}. Research has shown that natural materials like bamboo can be effective for constructing lattice towers, providing a cost-efficient and environmentally friendly solution \cite{Adhikari_Wood_Sudak_2015}. This approach helps to minimize manufacturing and transportation expenses while maintaining structural integrity.

\subsection{Support Structures Types}
Wind turbine support structures are broadly classified into onshore and offshore foundations, each optimized to address specific environmental and technical demands. Onshore structures are primarily determined by soil properties, whereas offshore foundations are selected based on water depth and seabed characteristics.

\begin{figure}[htt!]
    \centering
    \includegraphics[width=0.485\textwidth]{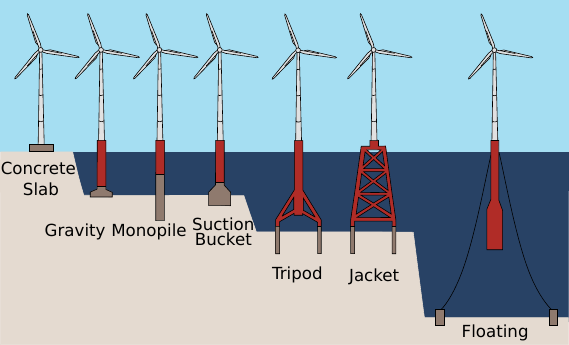}
    \caption{Types of support structures for wind turbines, Onshore: (a) Concrete Slab; and Offshore: (b) Gravity, (c) Monopile, (d) Suction Bucket, (e) Tripod, (f) Jacket, and (g) Floating.
    }
    \label{fig:types_support_wtt}
\end{figure}

\begin{figure}[htt!]
	\centering
\includegraphics[width=0.485\textwidth]{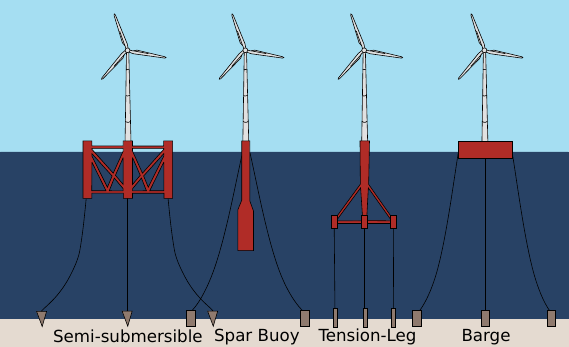}
\caption{Types of support structures for floating offshore 
wind turbines: (a) Semi-submersible, (b) Spar Buoy, (c) Tension-Leg Platform, and (d) Barge.}
    \label{fig:types_support_wtt_floating}
\end{figure}

\autoref{table:types_support_structures} and \autoref{fig:types_support_wtt} present a detailed overview of these structures, including concrete slabs, gravity, monopiles, suction buckets, tripods, jackets, and floating foundations. For floating OWTs, \autoref{table:floating_substructures} and \autoref{fig:types_support_wtt_floating} provide an overview of floating foundation types, such as semi-submersibles, spar buoys, tension leg platforms, and barges. 

Additionally, \citet{Faraggiana_Giorgi_Sirigu_Ghigo_Bracco_Mattiazzo_2022} offer an in-depth comparison of design methodologies, numerical modeling tools, and optimization techniques for floating OWT support structures, highlighting current limitations in achieving optimal designs.

\begin{table*}[htt!]
\scriptsize	
\centering
\caption{Overview of types of support structures for wind turbines (adapted from \cite{Hernandez-Estrada_Lastres-Danguillecourt_Robles-Ocampo_Lopez-Lopez_Sevilla-Camacho_Perez-Sariñana_Dorrego-Portela_2021, Chen_Kim_2022, Henderson_Collu_Masciola_2016}).}

\begin{tabularx}{\textwidth}{@{}p{0.9cm}p{2.5cm}p{1.5cm}p{4.9cm}p{4.9cm}p{0.25cm}p{0.8cm}@{}}
\toprule
\textbf{Support} & \textbf{Description} & \textbf{Use} & \textbf{Advantages} & \textbf{Disadvantages} & \textbf{Fig.} & \textbf{Refs.} \\ \midrule

Concrete Slab & Concrete slab anchored by gravity, reinforced with piles in soft soil. 
& Onshore & 
\begin{minipage}[t]{4.9cm}
\begin{itemize}[label={-}, leftmargin=*, topsep=0pt, partopsep=0pt, parsep=0pt, itemsep=0pt]
        \item Stable and secure foundation.
    \end{itemize}
\end{minipage} & 
\begin{minipage}[t]{4.9cm}
\begin{itemize}[label={-}, leftmargin=*, topsep=0pt, partopsep=0pt, parsep=0pt, itemsep=0pt]
        \item Requires suitable soil.
    \end{itemize}
\end{minipage} & \ref{fig:types_support_wtt}-a & \cite{Currie_Saafi_Tachtatzis_Quail_2015, Mohamed_Austrell_2018}
\\
\addlinespace[0.2cm]
Gravity & Circular pile and concrete slab, deeply buried and weight-anchored. 
& Offshore: shallow waters & 
\begin{minipage}[t]{4.9cm}
\begin{itemize}[label={-}, leftmargin=*, topsep=0pt, partopsep=0pt, parsep=0pt, itemsep=0pt]
        \item Structurally safe.
    \end{itemize} 
\end{minipage} & 
\begin{minipage}[t]{4.9cm}
\begin{itemize}[label={-}, leftmargin=*, topsep=0pt, partopsep=0pt, parsep=0pt, itemsep=0pt]
        \item Expensive for deep waters.
    \end{itemize}
\end{minipage} & \ref{fig:types_support_wtt}-b & \cite{Esteban_Couñago_López-Gutiérrez_Negro_Vellisco_2015, Esteban_López-Gutiérrez_Negro_2019} 
\\
\addlinespace[0.2cm]

Monopile & Circular pile buried in the seabed at water depth. & Offshore: depths $\leq$ 30m & 
\begin{minipage}[t]{4.9cm}
    \begin{itemize}[label={-},leftmargin=*, topsep=0pt, partopsep=0pt, parsep=0pt, itemsep=0pt]
        \item Effective in sand/gravel, no seabed preparation needed.
        \item Simple, economical and quick installation.
        \item Adaptable to various depths.
    \end{itemize}
\end{minipage} & 
\begin{minipage}[t]{4.9cm}
    \begin{itemize}[label={-}, leftmargin=*, topsep=0pt, partopsep=0pt, parsep=0pt, itemsep=0pt]
        \item Noisy installation, impacts fishing/environment.
        \item Susceptible to wind, wave, and seismic loading.
        \item Needs erosion protection.
        \item High costs and risks for deeper installations.
    \end{itemize}
\end{minipage} & \ref{fig:types_support_wtt}-c & \cite{Gupta_Basu_2020, Arshad_O’Kelly_2016} 
\\
\addlinespace[0.2cm]

Suction Bucket & Inverted bucket structure buried in the seabed. & Offshore: 5–60m depths& 
\begin{minipage}[t]{4.9cm}
    \begin{itemize}[label={-}, leftmargin=*, topsep=0pt, partopsep=0pt, parsep=0pt, itemsep=0pt]
        \item High resistance, low cost and less material.
        \item Quiet installation, less material.
    \end{itemize}
\end{minipage} & 
\begin{minipage}[t]{4.9cm}
    \begin{itemize}[label={-}, leftmargin=*, topsep=0pt, partopsep=0pt, parsep=0pt, itemsep=0pt]
        \item Uncertain in deep waters, needs more research.
    \end{itemize}
\end{minipage} & \ref{fig:types_support_wtt}-d & \cite{Wang_Yang_Zeng_2017, Wang_Yang_Zeng_2017_2} 
\\
\addlinespace[0.2cm]

Tripod & Three-legged structure distributing the load uniformly. & Offshore: deep waters & 
\begin{minipage}[t]{4.9cm}
    \begin{itemize}[label={-}, leftmargin=*, topsep=0pt, partopsep=0pt, parsep=0pt, itemsep=0pt]
        \item Low weight.
        \item Suitable for various soils.
        \item Extra stability.
        \item Potential for reef creation.
    \end{itemize}
\end{minipage} & 
\begin{minipage}[t]{4.9cm}
    \begin{itemize}[label={-}, leftmargin=*, topsep=0pt, partopsep=0pt, parsep=0pt, itemsep=0pt]
        \item Noisy installation, impacts fishing/environment.
        \item Higher construction/maintenance costs.
        \item Needs erosion protection.
        \item Potential invasive species issues.
    \end{itemize}
\end{minipage} & \ref{fig:types_support_wtt}-e & \cite{Yeter_Garbatov_Guedes_Soares_2015, Ma_Yang_Chen_2018}
\\
\addlinespace[0.2cm]

Jacket & More complex than a lattice tower, requiring advanced structural analysis. & Offshore: deep waters & 
\begin{minipage}[t]{4.9cm}
    \begin{itemize}[label={-}, leftmargin=*, topsep=0pt, partopsep=0pt, parsep=0pt, itemsep=0pt]
        \item Low weight.
        \item Ideal for deeper waters.
        \item Extra stability.
        \item Potential for reef creation.
        \item Suitable for various soils.
    \end{itemize}
\end{minipage} & 
\begin{minipage}[t]{4.9cm}
    
    \begin{itemize}[label={-}, leftmargin=*, topsep=0pt, partopsep=0pt, parsep=0pt, itemsep=0pt]
        \item Advanced structural analysis.
        \item Higher construction/maintenance costs.
        \item Noisy installation, impacts fishing/environment.
        \item Maintenance problems like grout joint issues.
    \end{itemize}
\end{minipage} & \ref{fig:types_support_wtt}-f & \cite{Saha_Gao_Moan_Naess_2014, Shi_Park_Chung_Shin_Kim_Lee_Kim_2015} 
\\
\addlinespace[0.2cm]

Floating & Floating base anchored to the seabed with mooring. & Offshore: very deep waters (100–200m) 
& 
\begin{minipage}[t]{4.9cm}
\begin{itemize}[label={-}, leftmargin=*, topsep=0pt, partopsep=0pt, parsep=0pt, itemsep=0pt]
        \item Ideal for deep waters.
    \end{itemize}
\end{minipage} & 
\begin{minipage}[t]{4.9cm}
\begin{itemize}[label={-}, leftmargin=*, topsep=0pt, partopsep=0pt, parsep=0pt, itemsep=0pt]
        \item High costs.
    \end{itemize}
\end{minipage} & \ref{fig:types_support_wtt}-g & \cite{article_offshore, Salic_Charpentier_Benbouzid_Le_Boulluec_2019} 
\\ \bottomrule
\end{tabularx}
\label{table:types_support_structures}
\end{table*}

\begin{table*}[htt!]
\scriptsize	
\centering
\caption{Overview of types of support structures for floating OWTs (adapted from \cite{Hernandez-Estrada_Lastres-Danguillecourt_Robles-Ocampo_Lopez-Lopez_Sevilla-Camacho_Perez-Sariñana_Dorrego-Portela_2021, Chen_Kim_2022, Henderson_Collu_Masciola_2016}).}

\begin{tabularx}{\textwidth}{@{}p{0.9cm}p{2.5cm}p{1.5cm}p{4.9cm}p{4.9cm}p{0.25cm}p{0.8cm}@{}}
\toprule
\textbf{Support} & \textbf{Description} & \textbf{Use} & \textbf{Advantages} & \textbf{Disadvantages} & \textbf{Fig.} & \textbf{Refs.} \\ \midrule

Semi- \newline submersible & 
Three-legged floating foundation. & 
Offshore: depth waters (50–300m) & 
\begin{minipage}[t]{4.9cm}
    \begin{itemize}[label={-}, leftmargin=*, topsep=0pt, partopsep=0pt, parsep=0pt, itemsep=0pt]
        \item Small motion and good stability.
        \item Simple quayside assembly with wet towing.
        \item Effective yaw motion and torque.
    \end{itemize}
\end{minipage} & 
\begin{minipage}[t]{4.9cm}
    \begin{itemize}[label={-}, leftmargin=*, topsep=0pt, partopsep=0pt, parsep=0pt, itemsep=0pt]
        \item High manufacturing costs.
        \item Challenging natural frequency.
    \end{itemize}
\end{minipage} & 
\ref{fig:types_support_wtt_floating}-a & 
\cite{Liu_Li_Yi_Chen_2016, Zhang_Shi_Karimirad_Michailides_Jiang_2020} 
\\
\addlinespace[0.2cm]

Spar Buoy & 
Deep-draft vertical cylinder offering buoyancy. & 
Offshore: depths $>$ 150m & 
\begin{minipage}[t]{4.9cm}
    \begin{itemize}[label={-}, leftmargin=*, topsep=0pt, partopsep=0pt, parsep=0pt, itemsep=0pt]
        \item Low cost.
        \item Small wave forces due to low surface volume.
        \item Easy installation with catenary mooring.
        \item Favorable natural period.
    \end{itemize}
\end{minipage} & 
\begin{minipage}[t]{4.9cm}
    \begin{itemize}[label={-}, leftmargin=*, topsep=0pt, partopsep=0pt, parsep=0pt, itemsep=0pt]
        \item Large motion.
        \item Stability relies on buoyancy/weight distribution.
        \item Large yaw motion and torque.
    \end{itemize}
\end{minipage} & 
\ref{fig:types_support_wtt_floating}-b & 
\cite{Namik_Stol_2014, Tomasicchio_D’Alessandro_Avossa_Riefolo_Musci_Ricciardelli_Vicinanza_2018} 
\\
\addlinespace[0.2cm]

Tension-Leg Platform (TLP) & 
Tethered platform with minimal heave, pitch, and roll movements. & 
Offshore: depths $>$ 50m & 
\begin{minipage}[t]{4.9cm}
    \begin{itemize}[label={-}, leftmargin=*, topsep=0pt, partopsep=0pt, parsep=0pt, itemsep=0pt]
        \item Small motion and high stability.
        \item Small wave forces due to low surface volume.
        \item Favorable natural period.
        \item Effective yaw motion and torque.
    \end{itemize}
\end{minipage} & 
\begin{minipage}[t]{4.9cm}
    \begin{itemize}[label={-}, leftmargin=*, topsep=0pt, partopsep=0pt, parsep=0pt, itemsep=0pt]
        \item High manufacturing costs.
        \item Depends on positive mooring tension.
\item Requires tensioned tethers and costly anchors.
    \end{itemize}
\end{minipage} & 
\ref{fig:types_support_wtt_floating}-c & 
\cite{osti_973961, Bachynski_Moan_2012} 
\\
\addlinespace[0.2cm]

Barge & 
Extensive shallow-draft barge structure. & 
Offshore: calm seas/harbors depths $>$ 50m & 
\begin{minipage}[t]{4.9cm}
    \begin{itemize}[label={-}, leftmargin=*, topsep=0pt, partopsep=0pt, parsep=0pt, itemsep=0pt]
        \item High buoyancy and stability.
        \item Effective yaw motion and torque.
        \item Simple installation with standard mooring lines.
    \end{itemize}
\end{minipage} & 
\begin{minipage}[t]{4.9cm}
    \begin{itemize}[label={-}, leftmargin=*, topsep=0pt, partopsep=0pt, parsep=0pt, itemsep=0pt]
        \item Large motion.
        \item Large wave forces due to large surface volume.
        \item High manufacturing costs.
        \item Difficult to manage natural frequency.
    \end{itemize}
\end{minipage} & 
\ref{fig:types_support_wtt_floating}-d & 
\cite{Hu_Wang_Chen_Li_Sun_2018, Kosasih_Suzuki_Niizato_Okubo_2020} 
\\ \bottomrule
\end{tabularx}
\label{table:floating_substructures}
\end{table*}

\subsection{Tower Load Types} \label{sec:towerloads}

The structural analysis of wind turbine towers involves multiple load types. Key loads include wind loads acting directly on the tower, aerodynamic loads transferred from the rotor blades, hydrodynamic loads caused by waves and ocean currents in offshore turbines, and gravitational loads due to the mass of the Rotor-Nacelle Assembly (RNA) and the tower, resulting from gravitational acceleration.

\begin{figure}[htt!]
	\centering
    \includegraphics[width=0.325\textwidth]{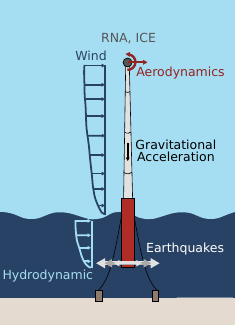}
    \caption{Types of loads on a wind turbine tower: wind loads, aerodynamics loads, hydrodynamic loads, gravitational loads, and in some cases, ice loads and earthquake effects.}
    \label{fig:types_loads_wtt}
\end{figure}

Additional loads, such as ice accumulation in cold climates and seismic effects in high-risk earthquake zones, can also be significant. \autoref{fig:types_loads_wtt}  illustrates these different types of loads. For further details, see the reviews by 
\citet{Ng_Ran_2016} and \citet{Hernandez-Estrada_Lastres-Danguillecourt_Robles-Ocampo_Lopez-Lopez_Sevilla-Camacho_Perez-Sariñana_Dorrego-Portela_2021}.

\subsection{Tower Analysis Types} \label{sec:analysis}

Different types of analyses are performed on wind turbine towers to ensure their structural integrity and performance. These analyses include stress analysis, deflection, modal, buckling, and fatigue. \autoref{table:analysis_comparison} provides a comprehensive overview of these techniques. The relevant loads combination and partial safety coefficients for each analysis are different and should be defined in accordance with the design code used (see \autoref{sec:standards}).

\begin{table*}[b!]
\scriptsize	
\centering
\caption{Overview of typical analysis techniques applied to wind turbine towers (adapted from \cite{Hernandez-Estrada_Lastres-Danguillecourt_Robles-Ocampo_Lopez-Lopez_Sevilla-Camacho_Perez-Sariñana_Dorrego-Portela_2021, Petrini_Manenti_Gkoumas_Bontempi_2010}).}

\begin{tabularx}{\textwidth}
{@{}p{1.5cm}p{6cm}p{8.7cm}p{0.9cm}@{}}
\toprule
\textbf{Analysis Type} & \textbf{Description} & \textbf{Simplified Equations} & \textbf{Refs.}  \\ \midrule
Stress  \newline       Analysis & 
Calculates the tower's maximum resistance under static and dynamic loads, including the weight of the tower, rotor, and nacelle.
& 
Maximum stress: $\sigma_{\text{max}} = \dfrac{F_a}{A_b} \pm \dfrac{My}{I}$ 
\newline 
$F_a$: axial load; $A_b$: tower cross-sectional area; $M$: bending moment; $I$: moment of inertia; $y$: distance to the neutral axis.&  \cite{Wiegard_König_Lund_Radtke_Netzband_Abdel-Maksoud_Düster_2021, Bazeos_Hatzigeorgiou_Hondros_Karamaneas_Karabalis_Beskos_2002}
\\
\addlinespace[0.2cm]
Deflection Analysis &
Assesses tower top deflection due to wind and gravitational loads, with the base fixed. Ensures curvature stays within material limits. Total displacement must be under 2\% of tower height to avoid foundation failure and tipping.
& 
Displacement from rotor load: $\delta_r = \dfrac{F_RL^3}{3EI}$ 
\newline 
Displacement from wind load: $\delta_t = \dfrac{5F_T L^3}{48EI}$ 
\newline
$F_R$: thrust load; $L$: tower height; $E$: elasticity modulus; $I$: moment of inertia; $F_T$: aerodynamic wind load on the tower.
&  \cite{Dagli_Tuskan_2018, Fan_Li_Zhang_2019}
\\
\addlinespace[0.2cm]
Modal Analysis & Ensures the tower's natural frequency doesn't match operational frequencies, avoiding resonance and excessive vibrations from wind or rotor spin.
&
First mode natural frequency: $\omega_1 = \sqrt{\dfrac{3EI}{(0.2235 m_T + m_{RNA})L^3}}
$ 
\newline 
$E$: elasticity modulus; $I$: moment of inertia; $m_T$: tower mass; $m_{RNA}$: RNA mass; $L$: tower height. &  \cite{Kandil_Saudi_Eltaly_El-khier_2016, Bajrić_Høgsberg_Rüdinger_2018}
\\
\addlinespace[0.2cm]
Buckling Analysis & 
Evaluates the risk of tower buckling caused by the nacelle and rotor mass, which increases when the load is misaligned with the center of gravity.

& 
Load capacity of a thin cylindrical tower: $P_{\text{cr}} = \dfrac{\pi^2 EI}{4L^2}$ 
\newline 
$E$: elasticity modulus; $I$: moment of inertia; $L$: tower height.&  \cite{Ma_Martinez-Vazquez_Baniotopoulos_2020, Lee_Bang_2012} 
\\
\addlinespace[0.2cm]
Fatigue Analysis & Examines the tower's long-term structural behavior under cyclical loads, turbulence, and wind direction variations. 
& 
$\log _{10}^{N}=\log _{10}^a-m \log _{10}\left(\Delta \sigma\left(\frac{t}{t_{r e f}}\right)^{k}\right)$
\newline 
Cumulative damage Palmgren-Miner rule:
\newline 
$D=\sum \frac{n_i}{N_i}=\frac{1}{\bar{a}} \sum n_i \Delta \sigma_i^m\left(\frac{t}{t_{r e f}}\right)^{k \cdot m}$
\newline 
$N$: fatigue cycles; $m$: SN curve slope; $t_{\text{ref}}$: reference thickness; $k$: thickness exponent; $t$: fracture thickness; $\Delta \sigma$: equivalent forces from wind and wave action; $D$: fatigue damage.
&  \cite{Li_Hu_Wang_Meng_2018, 10.1115/1.4028340}
\\
\bottomrule
\end{tabularx}
\label{table:analysis_comparison}
\end{table*}

\subsection{Tower Design Process} \label{sec:tower_desing_process}
\subsubsection{Tower Design Process Types}
The design of wind turbine towers requires meticulous consideration of structural integrity, material efficiency, and adaptability to diverse environmental conditions. Two primary methodologies are prevalent in the design process: iterative and integrated design approaches \cite{Barter_Robertson_Musial_2020}.

\begin{figure*}[htt!]
    \centering
    \begin{minipage}{0.95\textwidth}
        \centering
        \includegraphics[width=0.435\textwidth]{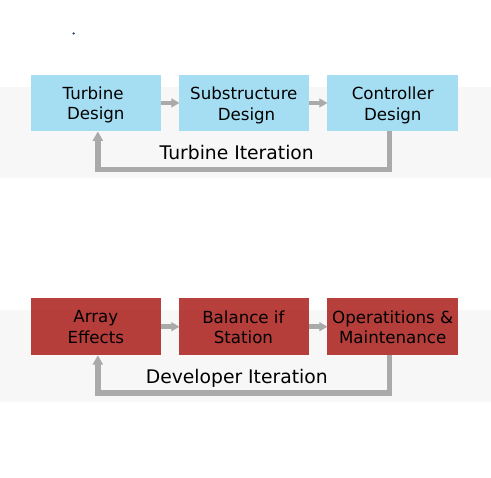}
        \hspace{0.05\textwidth}
        \includegraphics[width=0.435\textwidth]{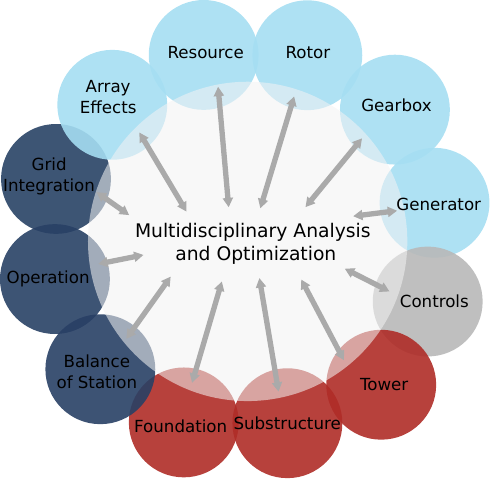}
    \end{minipage}
    \caption{Wind plant design: (a) Iterative Design; (b) Integrated Design (adapted from \cite{Barter_Robertson_Musial_2020}).}
    \label{fig:iterative_vs_integrated}
\end{figure*}

\autoref{table:design_comparison} and \autoref{fig:iterative_vs_integrated} compare iterative and integrated design approaches, highlighting their usage, advantages, disadvantages, and examples from relevant works.

\begin{table*}[htt!]
\scriptsize	
\centering
\caption{Overview of wind turbine design process types.}
\label{table:design_comparison}
\begin{tabularx}{\textwidth}{@{}p{0.9cm}p{3cm}p{3.55cm}p{3.5cm}p{3.5cm}p{0.5cm}p{0.9cm}@{}}
\toprule
\textbf{Design} & \textbf{Description} & \textbf{Use} & \textbf{Advantages} & \textbf{Disadvantages} & \textbf{Fig.} & \textbf{Refs.} \\
\midrule
Iterative & 
\begin{minipage}[t]{3cm}
    A collaborative process where stakeholders exchange ideas and feedback, refining the design through repeated iterations, guided by expertise and business objectives.
\end{minipage} &
\begin{minipage}[t]{3.55cm}
    \begin{itemize}[label={-}, leftmargin=*, topsep=0pt, itemsep=0pt, partopsep=0pt, parsep=0pt]
        \item Projects with evolving requirements.
        \item Situations demanding high adaptability.
        \item Environments benefiting from collaborative innovation.
    \end{itemize}
\end{minipage} &
\begin{minipage}[t]{3.5cm}
    \begin{itemize}[label={-}, leftmargin=*, topsep=0pt, itemsep=0pt, partopsep=0pt, parsep=0pt]
        \item High adaptability to changes.
        \item Promotes innovation through stakeholder collaboration.
        \item Allows incremental improvements.
    \end{itemize}
\end{minipage} &
\begin{minipage}[t]{3.5cm}
    \begin{itemize}[label={-}, leftmargin=*, topsep=0pt, itemsep=0pt, partopsep=0pt, parsep=0pt]
        \item Time-consuming due to continuous iterations.
        \item Potentially costly due to iterative nature.
        \item Possible conflicts of interest among stakeholders.
    \end{itemize}
\end{minipage} &
\ref{fig:iterative_vs_integrated}-a & \cite{Fuglsang_Bak_Schepers_Bulder_Cockerill_Claiden_Olesen_VanRossen_2002, Huang_Li_Zhou_Wang_Zhu_2022} 
\\
\addlinespace[0.2cm]

Integrated & 
\begin{minipage}[t]{3cm}
    Employs MDAO to optimize all aspects of the design simultaneously, integrating multiple disciplinary models.
\end{minipage} &
\begin{minipage}[t]{3.55cm}
    \begin{itemize}[label={-}, leftmargin=*, topsep=0pt, itemsep=0pt, partopsep=0pt, parsep=0pt]
        \item Projects with stable and well-defined parameters.
        \item Systems requiring comprehensive, system-level optimizations.
        \item Complex problems with significant interdisciplinary coupling.
    \end{itemize}
\end{minipage} &
\begin{minipage}[t]{3.5cm}
    \begin{itemize}[label={-}, leftmargin=*, topsep=0pt, itemsep=0pt, partopsep=0pt, parsep=0pt]
        \item Streamlines the design process.
        \item Reduces time and overall costs.
        \item Enhances overall system efficiency and effectiveness.
    \end{itemize}
\end{minipage} &
\begin{minipage}[t]{3.5cm}
    \begin{itemize}[label={-}, leftmargin=*, topsep=0pt, itemsep=0pt, partopsep=0pt, parsep=0pt]
        \item High initial complexity.
        \item Requires substantial initial investment in computational resources.
        \item Demands robust integration of diverse models.
    \end{itemize}
\end{minipage} &
\ref{fig:iterative_vs_integrated}-b & \cite{10.1115/OMAE2014-24244, Bortolotti_Bottasso_Croce_2016} 
\\
\bottomrule
\end{tabularx}
\end{table*}

\subsubsection{Iterative Tower Design}
The offshore wind industry has seen significant growth through iterative design methods, which refine designs based on market feedback. However, this approach often leads to sub-optimal solutions for floating offshore wind plants due to the complexity of the physical environment and engineering challenges. Typically, different entities design the turbine, tower, and substructure, requiring multiple iterations to optimize controllers for load minimization and power maximization. A fully integrated systems-engineering approach is necessary to achieve significant cost reductions and performance improvements \cite{Barter_Robertson_Musial_2020}.

\subsubsection{Integrated Tower Design}
A multidisciplinary approach to floating OWT optimization is essential and challenging, requiring advanced coupled models of economics, aerodynamics, hydrodynamics, structures, and control to capture complex interactions while maintaining reasonable computational costs. Design constraints for cost-competitive floating turbine systems include maximizing net energy production, regulatory compliance, and site adaptability. Designs must be economically manufacturable, transportable, and stable for assembly and tow-out, with efficient maintenance strategies minimizing labor and heavy lift vessel use. Scalability and weight minimization are crucial, with effective corrosion control around the waterline and consideration for decommissioning or repowering to reduce costs \cite{Barter_Robertson_Musial_2020, Patryniak_Collu_Coraddu_2022}.

The integrated approach is especially effective in the conceptual design stage, where critical decisions are made. Early insights are essential for later phases, where modifications become costly. This process narrows the trade-space to achieve cost-optimized designs by initially exploring all options within safety standards, then focusing on cost-competitive choices that incorporate practical insights \cite{Patryniak_Collu_Coraddu_2022}.

\subsubsection{Multidisciplinary Design Analysis and Optimization}
Multidisciplinary Design Analysis and Optimization (MDAO) combines and analyzes the interactions of various disciplinary models (such as structural, aerodynamic, and acoustic models) into a cohesive framework, ensuring that all aspects are considered and optimized simultaneously \cite{Ojo_Collu_Coraddu_2022}. Several libraries facilitate MDAO, with the most widely used being OpenMDAO (Python) \cite{Gray_Hwang_Martins_Moore_Naylor_2019}, DAKOTA (C++) \cite{osti_991840}, and DAFoam (Python, C++) \cite{doi:10.2514/1.J058853}. Following MDAO, the design process incorporates MDO, which focuses on optimizing design variables across multiple disciplines to achieve superior operational performance \cite{Ashuri_Zaaijer_Martins_Zhang_2016}. Additionally, Multidisciplinary System Design Optimization (MSDO) extends this optimization to the system level, often encompassing economic, environmental, and lifecycle considerations \cite{Maki_Sbragio_Vlahopoulos_2012}.

\subsubsection{Tower Design Cycle}
\begin{figure}[htt!]
	\centering
    \includegraphics[width=0.435\textwidth]{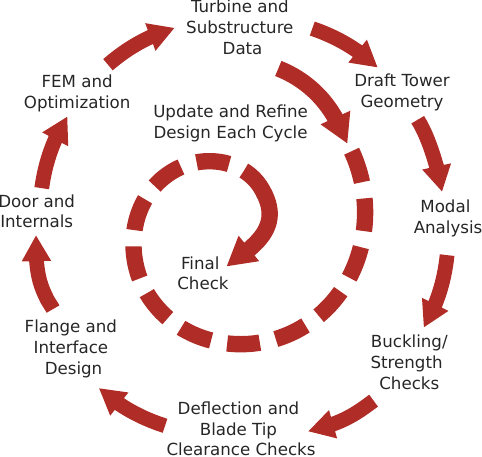}
    \caption{Typical design cycle for the tower (adapted from \cite{Ng_Ran_2016}).}
    \label{fig:design_process}    
\end{figure}

\citet{Ng_Ran_2016} present a design cycle for wind turbine towers, beginning with preliminary designs based on turbine specifications, environmental conditions, standards (see \autoref{sec:standards}), and initial concepts. These preliminary designs are subject to structural checks, modal assessments, and Ultimate Limit State (ULS) evaluations, guided by engineering expertise. Iterative adjustments in geometry and load analyses refine the design toward an optimized layout, as shown in \autoref{fig:design_process}.

Each iteration applies the Load and Resistance Factor Design (LRFD) framework to verify compliance with ULS, Fatigue Limit State (FLS), and Service Limit State (SLS) criteria. Fatigue loads are validated through type certification, while ULS loads are confirmed by numerical analysis. A Certified Verification Agent (CVA) provides final certification, ensuring the design meets all standards and enhancing the OWT tower's reliability and safety.

In the early stages of OWT design, simpler configurations may meet initial requirements, though increasing complexity calls for integrated approaches. Extensive simulations and Design Load Cases (DLCs) optimize the design for scenarios like parked, operational, start-up, shut-down, and fault states. Thousands of simulations assess diverse wind and wave conditions to refine final configurations through detailed load analyses.

\subsection{Wind Farm Tower Design} \label{sec:wind_farm}

\subsubsection{Aerodynamic Wake Effects in Wind Farms
}
\begin{figure}[htt!]
	\centering
    \includegraphics[width=0.475\textwidth]{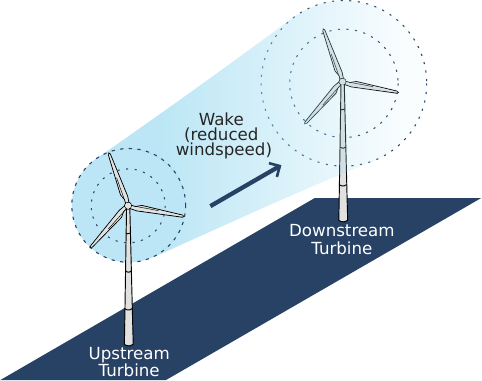}
    \caption{Wake effect from the upstream turbine reducing wind speed and impacting downstream turbine performance.}
    \label{fig:wake}    
\end{figure}

Wind farms encounter significant aerodynamic challenges, particularly the formation of wakes, which are areas of reduced wind speed behind upstream turbines, as shown in \autoref{fig:wake} \cite{Porte-Agel2020, wes-7-2271-2022}. Downstream turbines operating within these wakes experience lower wind speeds, reducing power output, and face increased structural stress due to turbulent forces on the towers \cite{https://doi.org/10.1002/we.2310}. In many wind farms, limited space leads to consistent wake effects on some turbines, resulting in a 10\% to 15\% reduction in total energy production \cite{9374111}. Optimizing the layout to minimize wake effects can improve power generation, reduce mechanical stress, and extend turbine lifespans \cite{chowdhury2012unrestricted, chowdhury2013optimizing, wes-4-99-2019, Swamy_2020}.

\subsubsection{Conventional vs. Integrated Tower Design Processes}
The conventional wind farm tower design process typically follows a sequential method, optimizing the layout first, then designing the towers. However, this method often leads to suboptimal results by neglecting interdependencies between turbine placement and tower design. An integrated approach that optimizes both layout and tower design simultaneously can potentially reduce costs by determining the optimal number, size, and height of wind turbines based on wind conditions and overall farm configuration \cite{Swamy_2020, Ashuri_Ponnurangam_Zhang_Rotea_2016}.

\pagebreak

\subsubsection{Tower Design and Layout Optimization in Wind Farms} \label{sec:wf_opt}
Wind farms often use uniform turbine towers for simplicity, which explains why approximately 68\% of published studies focus on uniform wind farm optimization \cite{TAO2020118339}. However, wind conditions can vary considerably across a farm, making uniform tower designs less efficient. Employing different tower types and hub heights can more effectively capture wind resources at varying elevations and locations, leading to reduced energy costs and enhanced efficiency \cite{https://doi.org/10.1002/we.2310, wes-4-99-2019, Swamy_2020, TAO2020118339}. For example, \citet{https://doi.org/10.1002/we.2310} reported a 4.9\% reduction in energy costs by implementing two different tower sizes.

Various optimization techniques enhance both turbine layout and tower design. Genetic algorithms \cite{WANG2018136, CHEN201356, ABDULRAHMAN2017267} and greedy algorithms \cite{CHEN2016676} are used to optimize layout and hub height, while MDO frameworks address the integration of layout with support structure height, diameter, and thickness \cite{Ashuri_Ponnurangam_Zhang_Rotea_2016}. Gradient-based methods further refine layout, tower diameter at key sections, shell thickness, and hub height \cite{https://doi.org/10.1002/we.2310}. The WindFLO framework \cite{REDDY2020115090}, which employs parallel computing to optimize both layout and turbine characteristics like rotor diameter and tower height, has led to improved energy production, cost savings, and increased efficiency. For a comprehensive review of non-uniform wind farm layout optimization, see \cite{TAO2020118339}.

\subsubsection{Simulation Challenges in Large Wind Farms}
Although customizing wind turbine towers alongside layout optimization offers significant benefits, it is computationally demanding, particularly with Computational Fluid Dynamics (CFD) simulations. CFD provides precise wind flow modeling but becomes costly for large-scale optimization, with challenges growing in larger wind farms as wake effect simulations increase quadratically with the number of turbines \cite{wes-4-99-2019}. Additionally, variable and uncertain wind conditions mean a layout optimized for current conditions may not perform optimally in future years \cite{Swamy_2020}. These factors complicate the simultaneous optimization of tower design and layout, despite potential efficiency gains.

\subsubsection{AI for Wind Farm Tower Design} \label{sec:wf_ai}
AI has begun to play a significant role in various aspects of wind farm design, particularly in accelerating wind flow simulations and optimizing turbine layouts. AI-driven models, such as Artificial Neural Networks (ANNs) \cite{TI2020114025}, Physics-Informed Neural Networks (PINNs) \cite{ZHANG2021116641}, Generative Adversarial Networks (GANs) \cite{ZHANG2022121747}, and Graph Neural Networks (GNNs) \cite{LI2023120928}, have been employed to improve the speed and accuracy of wake effect simulations. These models can learn complex relationships between input parameters and system performance, significantly reducing the time required for traditional CFD simulations. A comprehensive review of these techniques can be found in \cite{10.1063/5.0091980}. Additionally, these ML-based wake models have been utilized to optimize wind farm layouts \cite{YANG2023105280, YANG2023119240, YANG2024120161}.

\subsection{Monitoring Systems for Offshore Wind Turbine Towers} \label{sec:monitoring}

\subsubsection{Failure Mechanisms in Offshore Wind Turbines}
OWTs operate in demanding environments where mechanical, electrical, and structural components encounter multiple failure challenges, including long-term fatigue, electrical issues, installation errors, and severe weather \cite{XIA2023113322}. They are subject to continuous cyclic loads from wind, waves, and extreme events, such as storms, earthquakes, and heavy snowfall. Over time, these dynamic loads lead to cumulative fatigue that weakens critical components, including the tower \cite{https://doi.org/10.1155/2021/5548727, LI2020107291}. Corrosion reduces material thickness, while marine collisions further increase structural vulnerability. In floating solutions, the mooring system is also affected by marine growth and accumulated fatigue in hard-to-inspect elements. Without timely detection and intervention, these issues can lead to catastrophic failures, such as structural collapse \cite{LI2020107291}. Between 2014 and 2019, over 700 OWT accidents were reported, underscoring the urgent need for advanced monitoring and intervention \cite{su11020494}.

\subsubsection{Impact of Failures on Operation and Maintenance Costs
}
Failures in OWTs drive operations and maintenance (O\&M) costs from 19\% to 30\% of total project expenses, rising to 35\% as turbines age due to increased maintenance demands \cite{MAIENZA2020114716, 5976683, doi:10.1080/15567249.2018.1461150, articleoffom, en7042595, 6104127}. Tower failures incur high costs due to complex logistics and repair demands, with offshore repairs often doubling expenses compared to onshore projects \cite{5976683, doi:10.1080/15567249.2018.1461150}. Additionally, the trend toward larger OWTs, while increasing energy capacity, makes these structures more vulnerable to wind and wave forces, further raising failure risks and O\&M expenses \cite{LI2020107291, su11020494, doi:10.1177/1475921714556568}. Reducing these costs is crucial for wind power competitiveness, achievable through technological advances, early failure detection, and efficient maintenance strategies \cite{XIA2023113322, LIU2023115009, en13123132}. Remote monitoring systems enable early structural change detection, optimized maintenance planning, and extended OWT lifespan, ultimately reducing O\&M expenses \cite{https://doi.org/10.1155/2021/5548727, doi:10.1177/1475921714556568, MARTINEZLUENGO201691}.

\subsubsection{Structural Health Monitoring for Offshore Wind Turbines}
Among monitoring systems for OWTs, Structural Health Monitoring (SHM) is crucial for early damage detection, preventing catastrophic failures. A particularly effective approach for continuously monitoring the dynamic properties of structures is Operational Modal Analysis (OMA) \cite{MAGALHAES20111431}. OMA can act as a warning system to detect structural misbehavior, using methods such as Frequency Domain Decomposition (FDD) \cite{Brincker_2000}, Enhanced Frequency Domain Decomposition (EFDD) \cite{brincker2001damping}, and the poly-reference Least Squares Complex Frequency Domain method (p-LSCF), also known commercially as PolyMAX \cite{guillaume2003poly,peeters2004polymax,peeters2005polymax}, for frequency-domain applications. For time-domain analyses, Covariance-driven Stochastic Subspace Identification (SSI-COV) \cite{peeters2000system} and data-driven Stochastic Subspace Identification (SSI) \cite{van2012subspace} are widely recognized.

OMA methods estimate natural frequencies, damping ratios, and mode shapes by leveraging environmental excitation, in contrast to Experimental Modal Analysis, which requires controlled and measurable dynamic excitations and thus relies on heavy and expensive equipment \cite{cunha_2006}. While OMA enables continuous and efficient dynamic monitoring of structures, it assumes that excitation is temporally and spatially random, stationary, and that the structural response is linear. Although these assumptions are rarely fully satisfied, OMA methods have demonstrated their ability to provide meaningful results for onshore wind turbines \cite{PEREIRA2023115120}, fixed offshore wind turbines \cite{doi:10.1177/1475921714556568}, and, more recently, floating wind turbine applications \cite{KIM2019106226, PIMENTA2024117367}.

SHM supports real-time data collection, enabling condition-based maintenance, reducing inspection times, avoiding unnecessary replacements, and allowing remote supervision. It optimizes repairs during low-wind periods and informs next-generation turbine designs with lighter towers and reduced safety margins \cite{MARTINEZLUENGO201691, Ciang_2008, Wondra2019WirelessHA, WANG2021109168}.

SHM also facilitates the estimation of the remaining useful life of turbines, extending maintenance intervals and supporting permit extensions \cite{Wondra2019WirelessHA}. This enhances turbine reliability, cost-efficiency, and sustainability, highlighting SHM's role in optimizing OWT operations and extending service life. Effective SHM relies on optimal sensor placement, robust data acquisition, and advanced analytics \cite{LI2020107291, LIU2023115009, Ciang_2008}. Techniques like the inverse Finite Element Method (iFEM) further improve real-time monitoring, providing precise measurements of displacements, strains, and stresses in OWT towers \cite{LI2020107291}.

\subsubsection{SCADA and Sensor Integration in Wind Turbine Towers}
Modern wind turbines are equipped with Supervisory Control and Data Acquisition (SCADA) systems to monitor environmental and operational parameters, such as wind field properties, rotor speed, blade angle, reactive power, and generator temperature evolution. SCADA systems typically record each turbine's history by capturing key values (averages, extremes, standard deviations) at 10-minute intervals, though higher-frequency sampling (15 s or 1 s) is also used. However, challenges persist in data standardization and access to high-frequency data \cite{su11020494, wes-5-983-2020}. Although SCADA data provides valuable insights, it may not fully capture structural damage or other critical data for such issues.

To address SCADA’s limitations, complementary sensors like strain gauges and accelerometers are integrated into SHM to improve damage detection and structural analysis. Strain gauges measure deformation, enabling early damage detection and extending turbine tower lifespan \cite{Wondra2019WirelessHA}. Accelerometers, measuring acceleration relative to gravity, monitor vibrations and modal parameters, providing insights into the tower’s dynamic behavior \cite{MARTINEZLUENGO201691}. Installed across tower sections, these sensors enhance understanding of turbine dynamics and support structural and fatigue assessments \cite{wes-5-983-2020, dt_portugal, Simoncelli2024}. Integrating sensors with SCADA enables comprehensive monitoring of parameters like vibration and acceleration, supporting real-time fault detection and diagnostics. This approach was demonstrated in \cite{PACHECO2023115913}, which used SCADA and bending moment data to assess fatigue in onshore wind farms. For floating OWT, semi-analytical model based on tower-top accelerations was proposed and compared to SCADA-based methods \cite{PIMENTA2024119981}.

\subsubsection{AI for Monitoring Offshore Wind Turbines}
Integrating AI with SCADA enhances fault detection, sensor efficiency, and system performance \cite{en13123132}. The extensive data provided by SCADA, along with systematic advancements in AI and DL techniques, has been utilized for wind turbine condition monitoring \cite{tautz_2017}, particularly focusing on the rotor’s mechanical and electrical components. Neural Networks (NNs) applied to SCADA data also enable power curve analysis, identifying unhealthy turbines \cite{gottschall_2008, papatheou_2014}. Additionally, SCADA data supports data-driven approaches for fatigue evaluation and internal force estimation \cite{https://doi.org/10.1002/we.2888}. Various NNs have been utilized to monitor wind turbine generators through the analysis of electrical signals \cite{yan2024bi}.

In floating OWTs, harsh marine conditions challenge monitoring components, making AI-based techniques essential for tracking key parameters. ANN models can predict tower top acceleration and root forces by analyzing interactions between tower forces and foundation movements \cite{WANG2022112105} and detect structural damage using FEM results and experimental data from offshore jacket structures \cite{mousavi2021deep}. The complex dynamics of these structures require advanced AI models for accurate real-time monitoring \cite{LIU2023115009}.
\citet{STETCO2019620} provides an extensive survey of ML methods for wind turbine condition monitoring.

\subsection{Digital Twin for Offshore Wind Turbine Towers} \label{sec:digitaltwin}

\subsubsection{Digital Twin in Offshore Wind Turbines} \label{sec:dt_of}
A DT is a digital representation of a physical asset, built from data and simulation models to enable real-time monitoring, predictive analysis, control, and decision-making throughout the asset's lifecycle \cite{10268901}. In wind energy, DTs digitally represent turbines and farms, supporting continuous data exchange and analysis, typically integrated with SCADA systems \cite{XIA2023113322, 9696318}. For example, \citet{wes-5-1155-2020} estimated tower loads on onshore turbines using SCADA data, while \citet{Branlard_2020} combined SCADA with OpenFAST models to estimate wind speed, thrust, torque, and loads with 10\% accuracy, relying only on existing measurements instead of additional sensors. 

The development of a DT encompasses several steps: creating a virtual model that accurately represents the physical asset, integrating sensor data and historical records, optimizing data flow, facilitating communication between modules, and applying intelligent control strategies to improve asset performance \cite{WANG2021109168}. 

For a comprehensive review of DT technology in wind energy, consult \cite{10268901}, which details capability levels, challenges, solutions, and a roadmap for future research.

\subsubsection{Motivation for Digital Twins in Offshore Wind Turbines}
Monitoring systems in OWTs often lack the ability to integrate real-time data and conduct comprehensive analyses, resulting in increased O\&M costs and greater safety risks for both operators and marine ecosystems \cite{XIA2023113322}. Additionally, the FEM, widely used to assess the dynamic behavior of OWTs, faces limitations due to its high computational demands and inability to predict future system states, making it less effective for proactive decision-making \cite{WANG2021109168, ZHAO2023531}. These challenges underscore the need for an approach that integrates real-time monitoring with predictive capabilities. DTs address this gap by combining data acquisition, diagnostics, and predictive modeling into a unified framework. For instance, \citet{ZHAO2023531} introduced a DT based on a Reduced Order Model (ROM) that is 650 times faster than FEM, with an error margin below 0.2\%, enhancing monitoring precision and accelerating decision-making for efficient OWT management.

\begin{table*}[b!]
\scriptsize
\centering
\caption{Overview of the most commonly used software for the structural analysis of wind turbine towers (adapted from \cite{Hernandez-Estrada_Lastres-Danguillecourt_Robles-Ocampo_Lopez-Lopez_Sevilla-Camacho_Perez-Sariñana_Dorrego-Portela_2021}).}
\begin{tabularx}{\textwidth}{@{}p{2.2cm}p{10.5cm}p{1.5cm}p{0.8cm}p{1.8cm}@{}}
\toprule
\textbf{Software} & \textbf{Analysis Type} & \textbf{Category} & \textbf{Refs.} & \textbf{Applications}\\ \midrule
Abaqus & Finite element analysis for dynamic, static, and thermal calculations. & Commercial & \cite{0b112d0e5eba4b7f9768cfe1d818872e} & \cite{Qiu_Lu_Sun_Qu_Xue_Tong_2020, Gücüyen_2017}
\\
\addlinespace[0.2cm]
Ansys & Finite element simulation for structures and fluids. & Commercial & \cite{ansys} & \cite{Huo_Tong_2020, Wang_Zhou_Wan_2012}
\\
\addlinespace[0.2cm]
HAWC2 & Time-domain simulations with external effects, load applications, structural dynamics, and control systems. & Commercial & \cite{18aac95355e641309b54a6830618c5ca} & \cite{Karimirad_Moan_2012, Smilden_Sørensen_Eliassen_2016}
\\
\addlinespace[0.2cm]
MATLAB-Simulink & Dynamic system modeling. & Commercial & \cite{matlab} & \cite{Mohammadi_Fadaeinedjad_Moschopoulos_2018, Tong_Zhao_Zhao_2017}
\\
\addlinespace[0.2cm]
OpenFAST &  Multiphysics, multi-fidelity tool for simulating the coupled dynamic response of wind turbines. & Open source & \cite{Jonkman_2013} & \cite{PIMENTA2024119981, Jonkman_Branlard_Jasa_2022}
\\
\addlinespace[0.2cm]
OpenSees & Structural analysis for earthquake resistance. & Open source & \cite{McKenna_Scott_Fenves_2010} & \cite{Mo_Kang_Li_Zhao_2017, MohammadiDarestani_Shafieezadeh_Cha_2020}
\\ \bottomrule
\end{tabularx}
\label{table:structural_analysis_software}
\end{table*}

\subsubsection{Benefits of Digital Twins in Offshore Wind Turbines} \label{sec:dt_of_benefits}

DTs offer substantial benefits for OWTs, facilitating continuous monitoring, predictive maintenance, and enhanced performance across their lifecycle. For example, \citet{reviewdt} created a DT model for predicting the lifespan of fixed and floating OWTs, while \citet{wes-9-1-2024} implemented a DT to estimate fatigue loads on floating OWTs, achieving an accuracy range of 5\%-15\%. Additionally, \citet{doi:10.1177/0309524X211060550} introduced a pioneering DT focused on monitoring and maintaining wind turbine tower surface protection, integrating virtual twins, image processing, corrosion resistance, and sensor data. Furthermore, \citet{JORGENSEN2023103806} proposed a DT for fatigue assessment of bolted ring-flanges, using Gaussian Process models to handle uncertainties and enhance real-time monitoring.

DTs not only improve reliability but also optimize turbine performance by increasing energy capture and minimizing structural loads through precise wind field mapping \cite{ZHANG2023117507}. In terms of support structure reliability, DTs demonstrate strong potential. \citet{WANG2021109168} reviewed recent advancements, highlighting DT frameworks that enable real-time monitoring, fault diagnosis, and operation optimization for OWT support structures, offering promising applications in reliability analysis.

Moreover, DTs align design expectations with operational performance, optimizing for example turbine height, enhancing visual integration based on public preferences, and ensuring efficient allocation of O\&M resources \cite{XIA2023113322}. They also contribute to reduced O\&M costs, as demonstrated by \citet{wes-9-1-2024}, who proposed a DT-based framework integrating failure analysis and value-driven model development to refine maintenance strategies \cite{XIA2023113322}.

\subsubsection{Challenges of Digital Twins in Offshore Wind Turbines} \label{sec:dt_of_challenges}
Despite recent advancements, DTs for OWTs still require significant development to realize their full potential \cite{XIA2023113322}. A key challenge lies in the lack of high-resolution, multiphysics simulations that can capture the complex interactions between fluid dynamics, structural behavior, control systems, and environmental variability \cite{jmse10101357}.

Additionally, the limitations in data fusion restrict the integration of SCADA and sensor data, while issues in signal processing and noise management reduce the accuracy of real-time insights necessary for effective operational decision-making. These issues impair the ability of DTs to deliver actionable information in the dynamic offshore environment \cite{MARTINEZLUENGO2019867}.

Furthermore, the limited availability of open and accessible datasets restricts the development and validation of DT models, limiting their effectiveness for predictive maintenance and operational analysis \cite{HAMEED20112154}. Additionally, the lack of standardization in SCADA data reduces the effectiveness of DT frameworks by compromising data quality and consistency, which are essential for reliable and efficient use \cite{en13123132}.

\subsubsection{AI for Digital Twin Offshore Wind Turbines} \label{sec:dt_of_ai}
AI plays a vital role in advancing DT models, driven by developments in Industry 4.0, big data, and ML technologies \cite{WANG2021109168}. In wind turbine DT applications, AI facilitates SCADA system monitoring by leveraging regression models for accurate predictions. For instance, Temporal Convolution Networks (TCN) are employed by \citet{9696318} to forecast wind speed and power output, while \citet{JIANG2019207} uses ANN to predict the real-time thrust on the rotor of floating OWTs based on aero-hydro-servo-elastic coupled simulations. Moreover, \citet{ZHANG2023117507} introduces the first DT for onshore/offshore wind farm flow systems using PINN, enabling predictions of the spatiotemporal wind field across the entire wind farm. AI models in DTs are also effective for detecting anomalies. For example, Deep Convolution Long Short-Term Memory Neural Networks (DCLSTMNN) are applied by \citet{MOUSAVI2024116563} to detect damage in complex floating wind turbine structures under varying uncertainties. These implementations emphasize AI's crucial role in enhancing the reliability, efficiency, and predictive accuracy of DTs in wind energy systems.

\subsection{Design and Simulation Software for Wind Turbines} \label{sec
}\label{sec:software}
\subsubsection{Limitations of Full-Scale and Scaled Experimental Test} 
Full-scale experiments and real-world data offer valuable insights into wind turbine tower performance, such as in fatigue assessment \cite{Pacheco_Pimenta_Pereira_Cunha_Magalhaes_2022} and static tests for evaluating flexural strength and failure modes \cite{Sim_Prowell_Elgamal_Uang_2014}. However, these methods are costly and typically require access to wind farms, which is often restricted. Scaled experimental tests, such as wind tunnel experiments \cite{Dou_Guala_Lei_Zeng_2019}, dynamic response tests of floating OWTs \cite{Cao_Xiao_Cheng_Liu_Wen_2020}, and buckling tests of wind turbine towers \cite{Dimopoulos_Gantes_2012}, provide practical approximations. Despite their value, simulations remain essential for extensive testing, design optimization, and early issue identification, without the significant costs and time demands of physical tests.

\subsubsection{Software for Structural Analysis of Wind Turbines}
Essential software tools such as Ansys, Abaqus, and OpenFAST, as shown in \autoref{table:structural_analysis_software}, provide comprehensive capabilities for analyzing structural dynamics and performance, supporting the development of efficient and reliable wind turbine designs. OpenFAST, in particular, provides a robust simulation framework to advance research and innovation in wind turbine design and performance. Additionally, tools like Amr-Wind \cite{AMR-Wind}, a parallel, adaptive-mesh incompressible flow solver for wind turbine simulations, enable high-fidelity flow simulations in wind farms using high-performance computing. Furthermore, the Inductiva API \cite{Sarmento2024} and Amazon Web Services (AWS) \cite{aws} facilitate the execution of open-source physical simulators like OpenFAST on the cloud, allowing for parallel simulations across hundreds of CPU cores \cite{InductivaOpenFAST, AWSOpenFAST}.

\subsubsection{Computer-Aided Design and Aero-Hydro-Servo-Elastic} 
Computer-aided design and manufacturing (CAD/CAM) models are essential for parametric storage, updating, and efficient exchange of geometric, material, and infrastructure data, facilitating collaborative design of OWT components.

The tower must be designed to withstand lifetime loads, meeting safety margins and reliability targets, and evaluated under real-life scenarios and environmental exposures through Aero-Hydro-Servo-Elastic (AHSE) simulations. Specialized CAD tools for AHSE modeling cover aerodynamics, hydrodynamics, structural dynamics, and control system dynamics for each relevant DLC. For an overview of AHSE capabilities and commonly used software, refer to \cite{https://doi.org/10.1002/wene.52}.

\begin{table*}[b!]
\scriptsize	
\centering
\caption{Overview for OWT tower design standards (adapted from \cite{Ng_Ran_2016}).}
\begin{tabularx}{\textwidth}{@{}p{6.2cm}p{10.7cm}p{0.8cm}@{}}
\toprule
\textbf{Standard} & \textbf{Description} & \textbf{Refs.} \\ \midrule

ABS Bottom-founded offshore wind turbines. & Provides criteria for the design, construction, installation, and survey of bottom-founded offshore wind turbines. & \cite{ABS-Bottom-Founded-Offshore-Wind-Turbines}
\\
\addlinespace[0.2cm]
ABS Floating offshore wind turbines. & Provides criteria for the design, construction, installation, and survey of floating offshore wind turbines. & \cite{ABS-Floating-Offshore-Wind-Turbines}
\\
\addlinespace[0.2cm]
ANSI/AISC 360-16. Specification for structural steel buildings. & Provides specifications for the design, fabrication, and erection of structural steel buildings. & \cite{ANSI-AISC-360-16}
\\
\addlinespace[0.2cm]
ANSI/ACP OCRP-1-2022. Offshore compliance recommended practices. & Guide to offshore wind farm assets that extract wind energy, transmit electricity to shore-based grids, and/or store energy using offshore or onshore facilities or devices. & \cite{ANSI-ACP-OCRP-1-2022}
\\
\addlinespace[0.2cm]
API Recommended Practice 2A-WSD. Planning, designing, and constructing fixed offshore platforms-working stress design. & Guide for the design and construction of new platforms and for the relocation of existing platforms used for the drilling, development, and storage of hydrocarbons in offshore areas. & \cite{API-RP-2A-WSD}
\\
\addlinespace[0.2cm]
DNV-RP-C203. Fatigue design of offshore steel structures. & Presents recommendations in relation to fatigue analyses based on fatigue tests (SN data) and fracture mechanics. & \cite{DNV-RP-C203}
\\
\addlinespace[0.2cm]
DNV-ST-0119. Floating wind turbine structure. & Specifies general principles and requirements for the structural design of floating wind turbine structures. & \cite{DNV-ST-0119}
\\
\addlinespace[0.2cm]
DNV-ST-0126. Support structures for wind turbines. & Specifies general principles and guidelines for the structural design of wind turbine support structures. & \cite{DNV-ST-0126}
\\
\addlinespace[0.2cm]
DNV-ST-0437. Loads and site conditions for wind turbines. & Provides design requirements and guidelines to be used for the determination of loads and site conditions for onshore and offshore wind turbines. & \cite{DNV-ST-0437}
\\
\addlinespace[0.2cm]
EN 1993-1-6. Eurocode 3: Design of steel structures - Part 1-6: Strength and stability of shell structures. & Provides guidelines for the design, strength, and stability of steel shell structures. & \cite{Eurocode3-Part1-6}
\\
\addlinespace[0.2cm]
EN 1993-1-9. Eurocode 3: Design of steel structures - Part 1-9: Fatigue. & Provides guidelines for the design of steel structures with considerations for fatigue. & \cite{Eurocode3-Part1-9}
\\
\addlinespace[0.2cm]
IEC 61400-1:2019. Wind energy generation systems - Part 1: Design requirements. & Specifies essential design requirements for the structural integrity of wind turbines, covering all subsystems to protect against damage throughout their lifetime. & \cite{IEC61400-1}
\\
\addlinespace[0.2cm]
IEC TS 61400-3-2:2019. Wind energy generation systems - Part 3-2: Design requirements for floating offshore wind turbines. & Specifies requirements for assessing external conditions and ensuring the engineering integrity of floating offshore wind turbines, providing protection against all hazards during their lifetime. & \cite{IEC61400-3-2}
\\
\addlinespace[0.2cm]
ISO 19901-3:2024. Petroleum and natural gas industries - Specific requirements for offshore structures - Part 3: Topsides structure. & Provides requirements, guidance and information for the design and fabrication of topsides structure for offshore structures, including in-service, pre-service and post-service conditions. & \cite{ISO19901-3-2024} 
\\
\addlinespace[0.2cm]
ISO 19902:2020. Petroleum and natural gas industries - Fixed steel offshore structures. & Specifies requirements and provides recommendations applicable to the following types of fixed steel offshore structures for the petroleum and natural gas industries. & \cite{ISO19902-2020}
\\ \bottomrule
\end{tabularx}
\label{table:design_guidance_references}
\end{table*}

\begin{table*}[b!]
\scriptsize
\centering
\caption{Overview of OWT reference models.}
\begin{tabularx}{\textwidth}{@{}p{2.5cm}p{4cm}p{2.3cm}p{2.3cm}p{1.7cm}p{2cm}p{1cm}@{}}
\toprule
\textbf{Wind Turbine Model} & \textbf{Entity} & \textbf{Rated Capacity (MW)} & \textbf{Rotor Diameter (m)} & \textbf{Hub Height (m)} & \textbf{Tower Mass (ton)} & \textbf{Refs.} \\ \midrule
NREL 5 MW & NREL  & 5 & 126 & 90  & 347.46 & \cite{osti_947422} 
\\
\addlinespace[0.2cm]
LEANWIND 8 MW & University College Cork and DNV-GL  & 8 & 164 & 110  & 558 & \cite{Desmond_Murphy_Blonk_Haans_2016} \\
\addlinespace[0.2cm]
DTU 10 MW & DTU &10  & 178.3 & 119  & 605 & \cite{bak2013dtu} \\
\addlinespace[0.2cm]
IEA 10 MW & IEA & 10 & 190 & 119 &  628.44 & \cite{osti_1529216} \\
\addlinespace[0.2cm]
IEA 15 MW & IEA & 15 & 240 & 150  & 860 & \cite{osti_1603478} \\
\addlinespace[0.2cm]
IEA 22 MW & IEA and DTU & 22 & 284  & 170  & 1574 & \cite{1cd3e417f8854b808c5372670588d3d0}\\ \bottomrule
\end{tabularx}
\label{table:offshore_wind_turbine_ref}
\end{table*}

\subsection{Offshore Wind Turbine Standards} \label{sec:standards}
In the design and certification of OWT towers, compliance with established codes and standards is crucial for ensuring safety and regulatory adherence. These standards are developed by various institutions, including the American Bureau of Shipping (ABS) and the American National Standards Institute (ANSI) in the USA, the International Electrotechnical Commission (IEC) and Det Norske Veritas (DNV) in Europe, and the globally recognized International Organization for Standardization (ISO), among others.

These standards categorize scenarios into DLCs, addressing specific operational and environmental conditions. Each DLC defines the required analysis type, such as fatigue or ultimate limit state, and specifies partial safety factors. \autoref{table:design_guidance_references} summarizes the most relevant standards for OWT tower design, based on \cite{Ng_Ran_2016}, with updates to older standards and the inclusion of additional ones. Other codes may apply to specific structural components, and local regulations might prioritize different standards. For more details, refer to \cite{Ng_Ran_2016}.

\subsection{Reference Wind Turbine Models} \label{sec:wt_reference}
Reference wind turbine models, developed by institutions such as the National Renewable Energy Laboratory (NREL), the International Energy Agency (IEA), and the Danish Technical University (DTU), are essential in advancing wind energy technology. These models serve as benchmarks for simulation, analysis, and comparison, supporting the consistent evaluation of emerging technologies prior to commercial application. For example, some models have been extensively used for code-to-code and code to reduced scale models experimental data comparisons during different working tasks organized by the IEA Wind Technology Collaboration Programme (IEA Wind TCP). \autoref{table:offshore_wind_turbine_ref} highlights the most commonly used reference models for offshore applications.

The adoption of reference models ensures methodological consistency across studies, allowing for direct comparisons of technologies and optimization strategies under controlled, standardized conditions. They validate advanced simulation tools and analytical techniques, establishing reliable benchmarks that enhance accuracy and confidence in research outcomes. Moreover, reference models facilitate collaborative research by providing a shared framework, reducing intellectual property barriers, and improving the safety, reliability, and performance of wind energy systems \cite{IEA_Grand_Challenges_2023}.

\subsection{Economic Aspects of Wind Turbines} \label{sec:economics}

\subsubsection{Levelized Cost of Energy for Wind Turbines}
The economic viability of a wind turbine design can be evaluated through the LCoE, defined as the ratio between the total life cycle cost and the Annual Energy Production (AEP). The total life cycle cost includes the initial Capital Expenditure (CapEx), covering construction and installation costs. It also comprises the Operational Expenditure (OpEx), which encompasses the ongoing costs required to operate and maintain the turbine throughout its life, as well as the Decommissioning Expenditure (DecEx), associated with dismantling and safely disposing of the turbine at the end of its operational life. These costs are discounted over time at a discount rate, $r$, and accumulated over the turbine's operational lifespan, $n$, with $t$ representing each year in that lifespan. This relationship is expressed in \autoref{eq:lcoe} \cite{Patryniak_Collu_Coraddu_2022, Castro-Santos_Diaz-Casas_2015}.

\begin{equation}
LCoE = \dfrac{\sum_{t=1}^{n} \dfrac{CapEx_t + OpEx_t + DecEx_t}{(1+r)^t}}{\sum_{t=1}^{n} \dfrac{AEP_t}{(1+r)^t}}
\label{eq:lcoe}
\end{equation}

LCoE measures the cost per unit of energy produced, making it a critical metric for electricity consumers and influencing the competitiveness of wind energy in the market. Reducing the LCoE can be achieved by either increasing energy production or lowering costs. A study by NREL \cite{osti_2278805} identifies CapEx as the most significant factor impacting the LCoE of wind farms. For land-based wind farms, total CapEx is approximately 1750 USD/kW. For fixed-bottom offshore wind projects, CapEx rises significantly to around 4640 USD/kW, with floating offshore wind projects having the highest CapEx at approximately 6169 USD/kW.

\subsubsection{Costs Overview of Wind Turbine Towers}
The tower is one of the most critical components of a wind turbine, significantly impacting overall costs. According to \cite{osti_2278805}, the average cost is approximately 238 USD/kW. For example, a 12 MW turbine's tower would cost around 2.856 million USD, while a 22 MW turbine's tower would cost about 5.236 million USD, assuming a consistent average price per kW.

For a steel tower, it is possible to estimate the material cost based on the price of steel and the tower's mass, as steel is the primary construction material. As of the first quarter of 2024, the average price of steel was 941 USD/ton \cite{Base_Metals_Archives}. However, forecasts predict a 511\% increase in steel demand in the USA by 2033 \cite{Farina_Anctil_2022}, which could lead to significant price escalations.

\begin{table*}[t!]
\scriptsize	
\centering
\caption{Overview of optimization approaches for wind turbines.}

\begin{tabularx}{\textwidth}{@{}p{1.4cm}p{4.5cm}p{4.85cm}p{4.85cm}p{1.1cm}@{}}
\toprule
\textbf{Optimization Approach} & \textbf{Description} & \textbf{Advantages} & \textbf{Disadvantages} & \textbf{Refs.} \\ \midrule

Static \newline Analysis & 
\begin{minipage}[t]{4.5cm}
    Uses static structural models, typically finite-element analysis.
\end{minipage} &
\begin{minipage}[t]{4.85cm}
    \begin{itemize}[label={-}, leftmargin=*, topsep=0pt, itemsep=0pt, partopsep=0pt, parsep=0pt]
        \item Simple and computationally efficient.
        \item Reduces weight via geometry variation.
        \item Supports stiffness and buckling assessment.
    \end{itemize}
\end{minipage} &
\begin{minipage}[t]{4.85cm}
    \begin{itemize}[label={-}, leftmargin=*, topsep=0pt, itemsep=0pt, partopsep=0pt, parsep=0pt]
        \item Lacks real-world accuracy.
        \item Ignores dynamic effects.
        \item Unsuitable for fatigue evaluation.
        \item Limited to static loads.
    \end{itemize}
\end{minipage} & 
\cite{Bazeos_Hatzigeorgiou_Hondros_Karamaneas_Karabalis_Beskos_2002, Li_Lu_2014} 
\\
\addlinespace[0.2cm]

Frequency \newline Domain \newline Analysis & 
\begin{minipage}[t]{4.5cm}
    Assesses structural performance by frequency, allowing rapid, low-cost approximation of key dynamics, ideal for early design stages.
\end{minipage} &
\begin{minipage}[t]{4.85cm}
    \begin{itemize}[label={-}, leftmargin=*, topsep=0pt, itemsep=0pt, partopsep=0pt, parsep=0pt]
        \item Low computational cost.
        \item Fast calculations.
        \item Effective for fatigue and frequency analysis.
    \end{itemize}
\end{minipage} &
\begin{minipage}[t]{4.85cm}
    \begin{itemize}[label={-}, leftmargin=*, topsep=0pt, itemsep=0pt, partopsep=0pt, parsep=0pt]
        \item Limited accuracy.
        \item Cannot capture complex or nonlinear effects.
        \item Limited to frequency-domain constraints.
    \end{itemize}
\end{minipage} & 
\cite{Negm_Maalawi_2000, Dou_Pegalajar-Jurado_Wang_Bredmose_Stolpe_2020}
\\
\addlinespace[0.2cm]

Time \newline Domain \newline Analysis & 
\begin{minipage}[t]{4.5cm}
    Conducts comprehensive design assessments using time-domain simulations, suitable for final design stages.
\end{minipage} &
\begin{minipage}[t]{4.85cm}
    \begin{itemize}[label={-}, leftmargin=*, topsep=0pt, itemsep=0pt, partopsep=0pt, parsep=0pt]
        \item Offers precise and detailed evaluations.
        \item Adheres closely to design standards.
        \item Captures complex and nonlinear behaviors.
    \end{itemize}
\end{minipage} &
\begin{minipage}[t]{4.85cm}
    \begin{itemize}[label={-}, leftmargin=*, topsep=0pt, itemsep=0pt, partopsep=0pt, parsep=0pt]
        \item High computational cost.
        \item Slower optimization process.
        \item Requires significant computational resources.
    \end{itemize}
\end{minipage} & 
\cite{Kvittem_Moan_2015, Chen_Jin_Liu_Li_Luo_2021} 
\\ 
\bottomrule
\end{tabularx}
\label{table:optimization_approaches}
\end{table*}

\begin{table*}[t!]
\scriptsize	
\centering
\caption{Overview of optimization algorithms applied in wind turbine design.}
\begin{tabularx}{\textwidth}{@{}p{1.4cm}p{4.5cm}p{5cm}p{5cm}p{1.1cm}@{}}
\toprule
\textbf{Optimization Algorithm} & \textbf{Description} & \textbf{Advantages} & \textbf{Disadvantages} & \textbf{Refs.} \\ \midrule

Sequential Quadratic Programming (SQP) & 
\begin{minipage}[t]{4.5cm}
    Reformulates an optimization problem as a series of quadratic programming subproblems, focusing on approximating the Hessian matrix for efficiently solving constrained nonlinear optimization challenges.
\end{minipage} & 
\begin{minipage}[t]{5cm}
    \begin{itemize}[label={-}, leftmargin=*, topsep=0pt, partopsep=0pt, parsep=0pt, itemsep=0pt]
        \item Exhibits robust convergence properties.
        \item Maintains high computational efficiency.
        \item Effectively manages boundary constraints.
    \end{itemize}
\end{minipage} & 
\begin{minipage}[t]{5cm}
    \begin{itemize}[label={-}, leftmargin=*, topsep=0pt, partopsep=0pt, parsep=0pt, itemsep=0pt]
        \item Necessitates accurate gradient information.
        \item Requires high memory for large-scale use.
        \item Complexity increases with problem scale.
    \end{itemize}
\end{minipage} & 
\cite{Häfele_Gebhardt_Rolfes_2019, Chew_Tai_Ng_Muskulus_2015} 
\\
\addlinespace[0.2cm]

Genetic Algorithm (GA) & 
\begin{minipage}[t]{4.5cm}
    Mimics evolutionary processes like selection, crossover, and mutation to evolve solutions towards optimality, particularly effective in complex, multimodal landscapes.
\end{minipage} & 
\begin{minipage}[t]{5cm}
    \begin{itemize}[label={-}, leftmargin=*, topsep=0pt, partopsep=0pt, parsep=0pt, itemsep=0pt]
        \item Capable of multi-objective optimization.
        \item Excels in navigating complex search landscapes.
        \item Naturally amenable to parallel computation.
    \end{itemize}
\end{minipage} & 
\begin{minipage}[t]{5cm}
    \begin{itemize}[label={-}, leftmargin=*, topsep=0pt, partopsep=0pt, parsep=0pt, itemsep=0pt]
        \item Computationally expensive.
        \item Designing effective fitness functions and genetic operators is intricate.
        \item May converge prematurely.
    \end{itemize}
\end{minipage} & 
\cite{Chen_Li_He_2020, Gentils_Wang_Kolios_2017} 
\\
\addlinespace[0.2cm]

Particle Swarm Optimization (PSO) & 
\begin{minipage}[t]{4.5cm}
    Inspired by the social behaviors of birds and fish, PSO optimizes by moving particles towards the best solutions within a search space, based on their own and their neighbors' experiences.
\end{minipage} & 
\begin{minipage}[t]{5cm}
    \begin{itemize}[label={-}, leftmargin=*, topsep=0pt, partopsep=0pt, parsep=0pt, itemsep=0pt]
        \item Straightforward to implement.
        \item Minimal parameter tuning required.
        \item Typically achieves rapid convergence.
    \end{itemize}
\end{minipage} & 
\begin{minipage}[t]{5cm}
    \begin{itemize}[label={-}, leftmargin=*, topsep=0pt, partopsep=0pt, parsep=0pt, itemsep=0pt]
        \item Risk of premature local optima convergence.
        \item Struggles with complex problem spaces.
        \item Diversity in the swarm is hard to maintain.
    \end{itemize}
\end{minipage} & 
\cite{Li_Xu_Yuan_2024, Häafele_Rolfes_2016} 
\\
\addlinespace[0.2cm]

Multidisciplinary Design Optimization (MDO) & 
\begin{minipage}[t]{4.5cm}
    Integrates multiple disciplinary models to optimize them simultaneously, addressing complex engineering problems efficiently by considering whole system interactions.
\end{minipage} & 
\begin{minipage}[t]{5cm}
    \begin{itemize}[label={-}, leftmargin=*, topsep=0pt, partopsep=0pt, parsep=0pt, itemsep=0pt]
        \item Facilitates comprehensive system interaction consideration.
        \item Potentially reduces overall design cycles.
        \item Optimizes considering cross-disciplinary factors.
    \end{itemize}
\end{minipage} & 
\begin{minipage}[t]{5cm}
    \begin{itemize}[label={-}, leftmargin=*, topsep=0pt, partopsep=0pt, parsep=0pt, itemsep=0pt]
        \item High computational demand.
        \item Requires cross-discipline coordination.
        \item Complex and demanding setup/management.
    \end{itemize}
\end{minipage} & 
\cite{Cui_Allison_Wang_2021, Ashuri_Zaaijer_Martins_van_Bussel_van_Kuik_2014} 
\\
\addlinespace[0.2cm]

Topology Optimization (TO) & 
\begin{minipage}[t]{4.5cm}
    Employs mathematical models to optimize material layout within a predetermined design space, under specific loads and constraints, to maximize system performance.
\end{minipage} & 
\begin{minipage}[t]{5cm}
    \begin{itemize}[label={-}, leftmargin=*, topsep=0pt, partopsep=0pt, parsep=0pt, itemsep=0pt]
        \item Maximizes material efficiency.
        \item Fosters innovative design solutions.
        \item Versatile across various engineering applications.
    \end{itemize}
\end{minipage} & 
\begin{minipage}[t]{5cm}
    \begin{itemize}[label={-}, leftmargin=*, topsep=0pt, partopsep=0pt, parsep=0pt, itemsep=0pt]
        \item High demand for computational resources.
        \item Sensitive to changes in loading conditions.
        \item Resultant shapes can be complex to fabricate.
    \end{itemize}
\end{minipage} & 
\cite{Marjan_Huang_2023, Zakhama_Abdalla_Gürdal_Smaoui_2010} 
\\  
\bottomrule
\end{tabularx}
\label{table:optimization_algorithms}
\end{table*}

\subsection{Optimization of Wind Turbines} \label{sec:opt_wt}

\subsubsection{Optimization of Wind Turbine Components}
Optimization of wind turbine components, including blades, rotors, support structures, and towers, has led to notable improvements in performance and efficiency. 

\textbf{Blade Optimization}: Techniques like TO \cite{Buckney_Green_Pirrera_Weaver_2013, Song_Chen_Wu_Li_2022}, GA \cite{Lee_Shin_2022}, and combined TO-GA methods \cite{Albanesi_Peralta_Bre_Storti_Fachinotti_2020} have resulted in lighter and more efficient blade designs. Additionally, the simultaneous optimization of topology and laminate properties in composite beams has enhanced structural performance \cite{Blasques_Stolpe_2012}.

\textbf{Rotor Optimization}: Optimization has improved mass and power density. For example, \citet{Hayes_Träff_Sørensen_Willems_Aage_Sigmund_Whiting_2023} reported mass reductions of 54-67\% and power density increases of 13-25\% in direct drive electric machines.

\textbf{Support Structure Optimization}: For onshore foundations, \citet{Shen_Vahdatikhaki_Voordijk_vanMeer_2022} developed a metamodel-based generative design framework, achieving a 99.93\% time reduction with minimal error. In offshore structures, design and material optimization have led to significant benefits \cite{Lemmer2017OptimizationOF}, including steel tonnage savings of 10-25\% in monopiles \cite{Kallehave_Byrne_LeBlancThilsted_Mikkelsen_2015} and weight reductions in semi-submersible platforms through TO \cite{Saeed_Gong_Wan_Long_Saeed_Mei_Xiong_Long_Zhou_Li_2024}. Research on spar platforms has demonstrated cost reductions \cite{Pollini_Pegalajar-Jurado_Dou_Bredmose_Stolpe_2021, Hegseth_Bachynski_Martins_2020, Faraggiana_Sirigu_Ghigo_Bracco_Mattiazzo_2022}. Multi-objective design strategies further improved both stability and economic viability in semi-submersible and spar platforms, balancing structural integrity with cost efficiency \cite{Zhang_Wang_Cai_Xie_Wang_Zhang_2022, Karimi_Hall_Buckham_Crawford_2017}.

\textbf{Tower Optimization}: Towers represent a significant portion of OWT costs, making them ideal for optimization. \citet{Negm_Maalawi_2000} proposed five optimization strategies for tower structures. \citet{Gencturk_Attar_2012} examined cost savings in lattice tower designs through optimized foundations and connections, achieving material savings of up to 50\% in tubular towers. Similarly, \citet{Chen_Yang_Ma_Li_2016} proposed a simplified lattice structure with fewer joints, resulting in a lighter configuration. As tower heights increase, associated costs in transportation, assembly, construction, and maintenance also rise. Structural and economic optimization of towers is essential for advancing wind turbine technology and efficiently managing costs \cite{Hernandez-Estrada_Lastres-Danguillecourt_Robles-Ocampo_Lopez-Lopez_Sevilla-Camacho_Perez-Sariñana_Dorrego-Portela_2021}.

\subsubsection{Optimization Approaches}
Various optimization approaches have been introduced to address the complexities of OWT design: Static Analysis, Frequency Domain Analysis, and Time Domain Analysis \cite{Chen_Kim_2022, Patryniak_Collu_Coraddu_2022, Ojo_Collu_Coraddu_2022}. \autoref{table:optimization_approaches} provides a comparative overview of these approaches, detailing their descriptions, advantages, limitations, and key references in OWT design.

\subsubsection{Optimization Algorithms} \label{sec:opt_alg_wt}
Structural optimization in OWT design involves formulating mathematical models to identify optimal solutions via computational algorithms. This approach addresses non-linear programming challenges, employing iterative methods and approximations for feasible solutions \cite{Ng_Ran_2016}. \autoref{table:optimization_algorithms} summarizes key algorithms in OWT design, including SQP \cite{boggs1995sequential}, GA \cite{Katoch_Chauhan_Kumar_2021}, PSO \cite{Wang_Tan_Liu_2018}, MDO \cite{Martins_Lambe_2013}, and TO \cite{Deaton_Grandhi_2014}, with their advantages, limitations, and primary references.

Advancements in optimization algorithms and computational power are advancing OWT design. Emerging ML techniques now improve efficiency in finding optimal global solutions. For instance, \citet{Qian_Bartels_Marx_2024} applied k-nearest neighbors, Random Forest, and XGBoost to enhance offshore substructure design, while \citet{De_Anda_Ruiz_Bojórquez_Inzunza-Aragon_2023} employed Multi-Objective Particle Swarm Optimization (MOPSO) and ANNs to optimize onshore wind turbine towers. However, ML applications in OWT and wind turbine optimization, though promising, remain in early stages and offer scope for further exploration.

\section{Offshore Wind Energy: Evolution, Future Prospects, and Upscaling} \label{sec:upscaling}

\subsection{Evolution of Wind Energy: From Onshore to Offshore} \label{subsec:ontooff}

\subsubsection{Motivation and Development of Offshore Wind Energy}
Offshore wind energy has emerged as a critical component of global renewable energy strategies, driven by increasing energy demands and the limitations of onshore wind systems. Onshore wind farms face significant challenges, including restricted land availability, logistical difficulties in deploying large turbines, and societal concerns related to noise and visual impacts. These limitations have necessitated the development of offshore wind energy as a viable alternative to meet growing energy demands, marking a clear transition from onshore to offshore systems.

\subsubsection{Onshore to Offshore Transition}
The global wind energy capacity reached 900 GW in 2022, with 93\% installed onshore and only 7\% offshore, according to the IEA \cite{iea_site_0}. Onshore wind technology is currently deployed in 115 countries, whereas offshore capacity exists in just 20 countries. Despite this disparity, offshore wind capacity has been steadily increasing, signaling a shift from onshore to offshore wind energy that is expected to accelerate.

In 2022, a U.S. Department of Energy study reported in \cite{DOE2023} a global deployment of 8385 MW of offshore wind energy, marking a 16.6\% increase from the previous year and bringing total offshore capacity to 59009 MW across 292 projects. China led new offshore installations with 5719.6 MW, followed by the UK (1386 MW), France (480 MW), Germany (342 MW), and Vietnam (331 MW). By the end of the year, China held 46.2\% of global offshore capacity, surpassing the UK’s 23.1\%, with Germany, the Netherlands, and Denmark following at 13.5\%, 5.1\%, and 3.9\%, respectively. Most of the offshore capacity is concentrated in Europe (51\%) and Asia (48.9\%), with North America contributing only 0.1\%.

\subsubsection{Offshore Advantages}
Offshore wind farms offer extensive installation areas, typically located far from populated regions, which minimizes noise, environmental, and visual impacts. These areas also enable the deployment of larger turbines that are logistically impractical for onshore sites due to transportation constraints. Moreover, offshore wind resources exhibit superior quality, characterized by higher and more consistent wind speeds with reduced turbulence, resulting in prolonged turbine lifespans. This enhances the reliability and efficiency of power generation, making offshore wind a cost-effective resource when operated with minimal disruptions throughout its lifecycle. Additionally, the resource availability and technological maturity of offshore wind are comparable to onshore wind, frequently exceeding other renewable energy technologies \cite{Petrini_Manenti_Gkoumas_Bontempi_2010, Esteban_Diez_López_Negro_2011}.

\subsubsection{Offshore Challenges}
However, offshore wind development faces significant challenges, including elevated costs associated with permitting, engineering, construction, operation, and maintenance, as well as the need for extensive marine electrical infrastructure. Technological obstacles involve adapting turbines to operate under harsh marine conditions, requiring corrosion-resistant materials and advanced foundation designs. Offshore installations must also contend with heightened wake effects and the high costs of wind resource assessments. Despite these challenges, ongoing advancements in technology and knowledge remain critical to the viability of offshore wind energy \cite{Petrini_Manenti_Gkoumas_Bontempi_2010, Esteban_Diez_López_Negro_2011}.

\subsection{Evolution of Offshore Substructures: From Fixed-bottom to Floating}

\subsubsection{Fixed-bottom to Floating Transitions}
\begin{figure}[htt!]
    \centering
    \subfloat[\label{fig:types_tech_offshore_0}]
    {\includegraphics[width=0.485\textwidth]{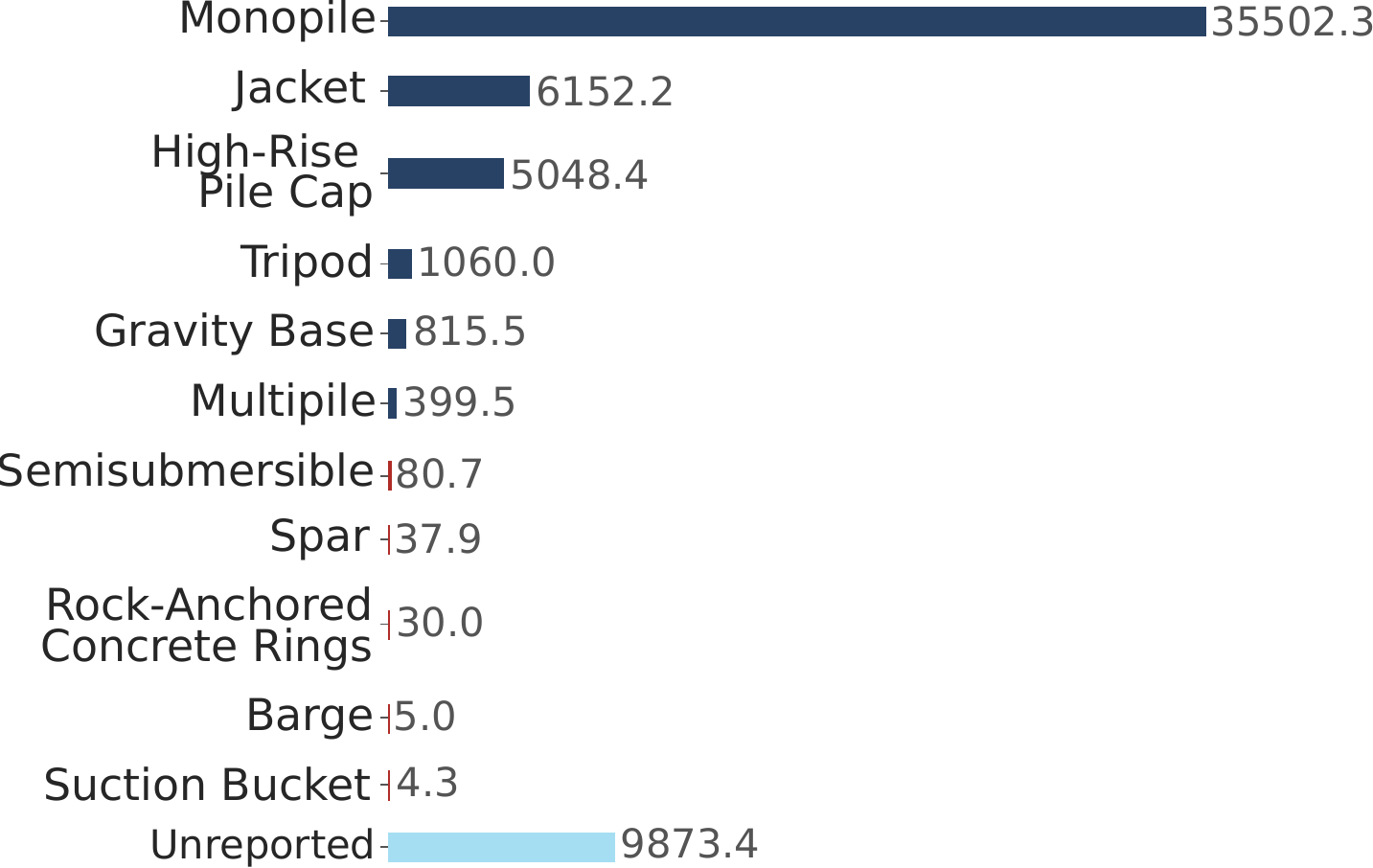}}\hfill
    \vspace{1em}
    \subfloat[\label{fig:types_tech_offshore_1}]
    {\includegraphics[width=0.485\textwidth]{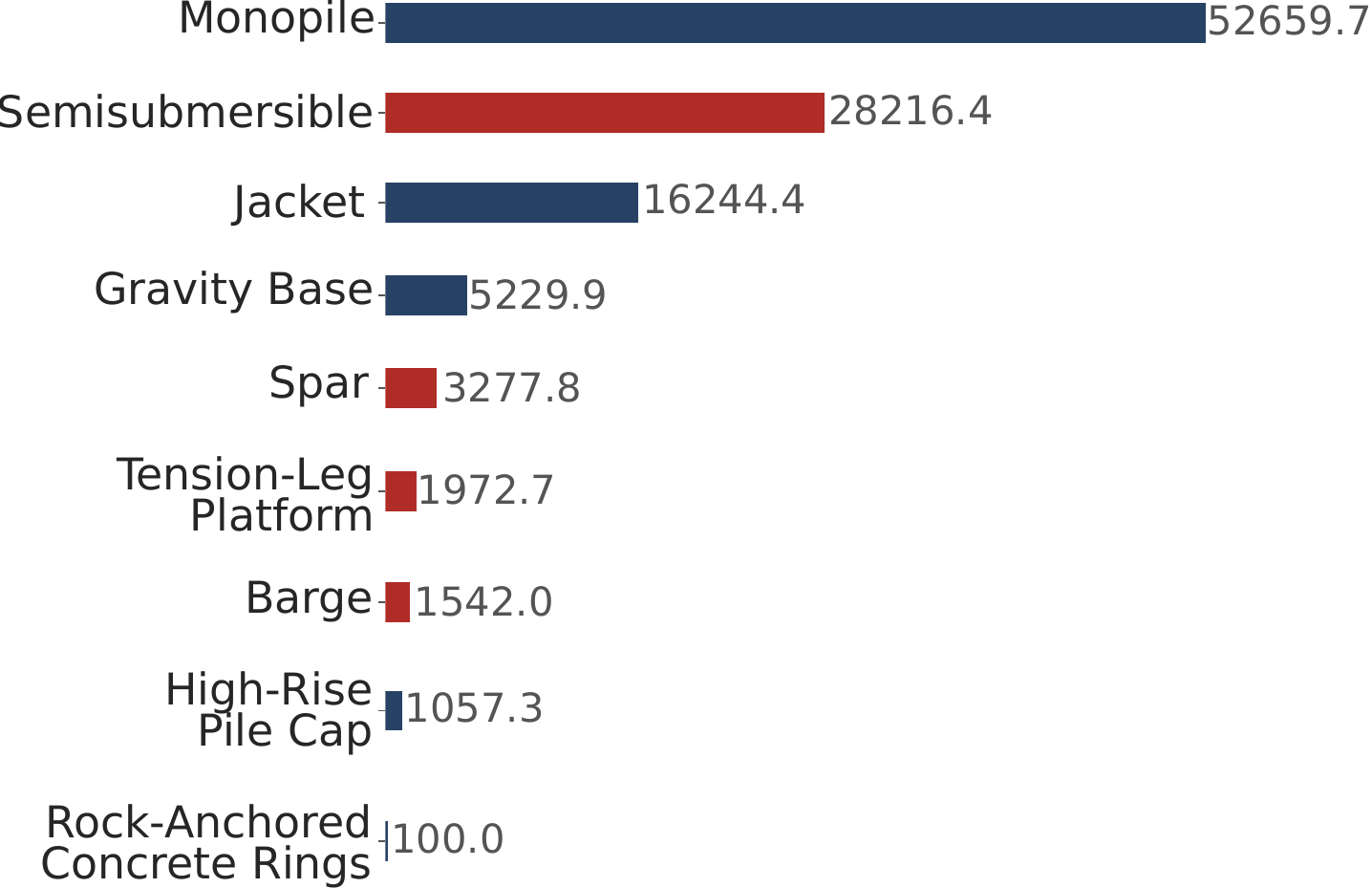}}\hfill
 \includegraphics[width=0.485\textwidth]{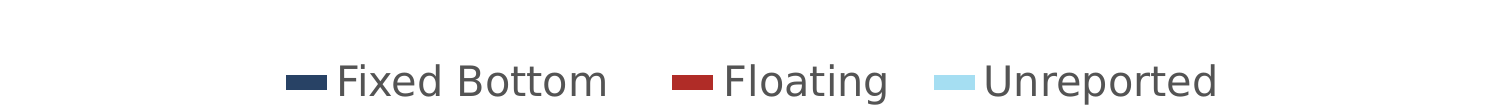}
    \caption{Global offshore wind substructure capacity (MW) for various substructure types in 2022: (a) Operating Projects; (b) Announced Future Projects (adapted from \cite{DOE2023}).}
    \label{fig:types_tech_offshore}
\end{figure}

The U.S. Department of Energy's study \cite{DOE2023} offers an in-depth analysis of offshore wind technologies, revealing that fixed-bottom structures dominated projects in 2022. This predominance is attributed to their straightforward design, cost-effective manufacturing processes, and the support of established, globally standardized supply chains. Monopiles accounted for 60.2\% of substructures, followed by jackets (10.4\%), pile caps (8.6\%), tripods (1.8\%), and gravity-based foundations (1.4\%). Additionally, 16.7\% of structures were unreported, with the remaining percentage including other fixed-bottom technologies and a small portion of floating structures, as illustrated in \autoref{fig:types_tech_offshore_0}.

\begin{figure}[htt!]
	\centering
        \includegraphics[width=0.485\textwidth]{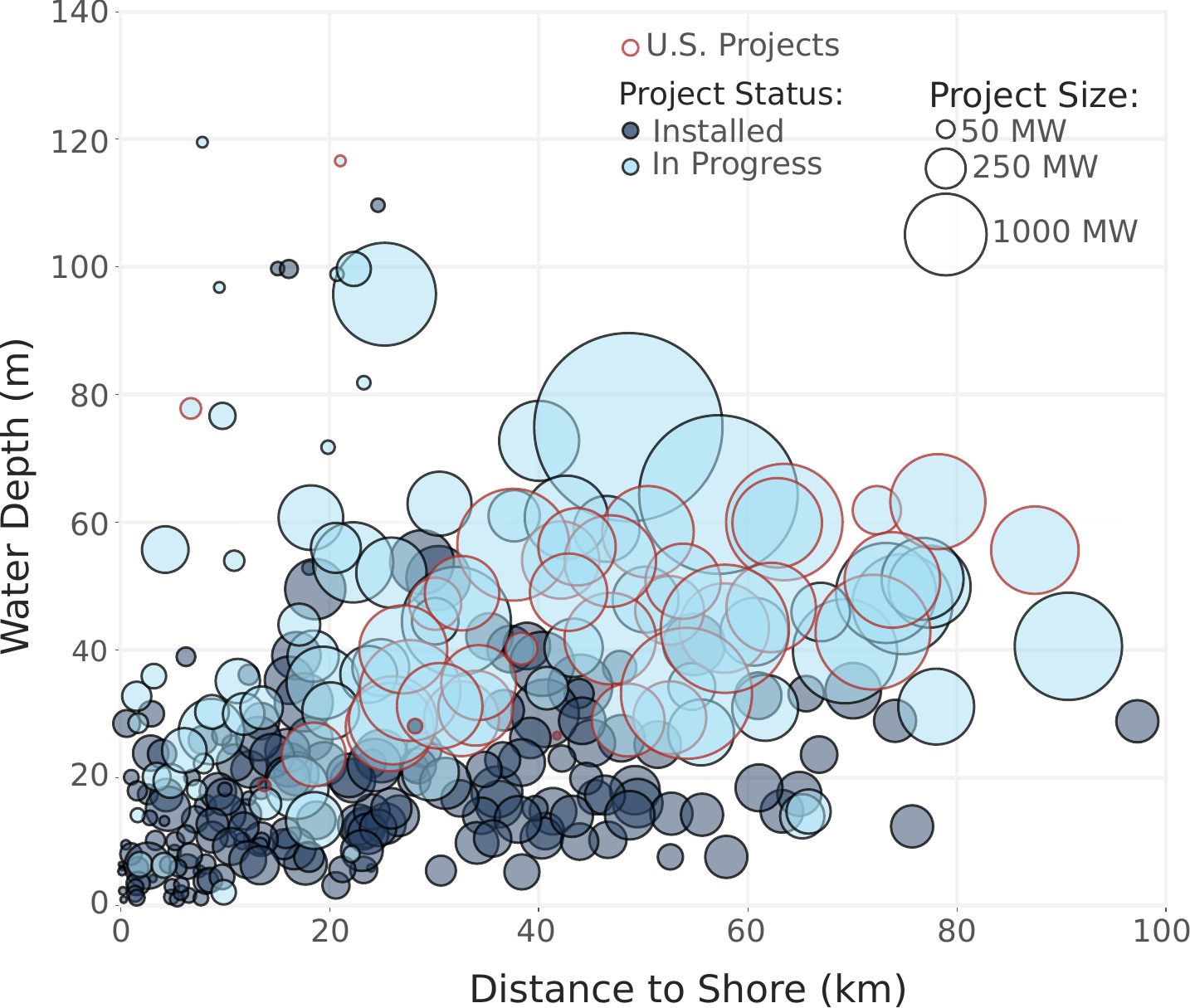}
        \caption{Global offshore wind energy projects by distances to shore (km) and water depths (m) in 2022 (adapted from \cite{DOE2023}).}
    \label{fig:deepvsfar}    
\end{figure}

However, the future dominance of monopiles could be limited by several factors such as the upsizing of wind turbines, unsuitable soil conditions, and construction noise restrictions. Projects are increasingly located in deeper waters and farther from shore due to the scarcity of shallow-water sites, and consequently, to meet decarbonization targets, more offshore wind resource areas are required, leading to larger project sizes, as shown in \autoref{fig:deepvsfar} with the new projects in gray. This drives the need for floating offshore technology, and evidence of this trend is clear when comparing current operational projects with future plans, where the use of floating technology is projected to increase from 0.2\% to 31.7\%, as illustrated in \autoref{fig:types_tech_offshore}.

\subsubsection{Floating Offshore Substructures}
Floating substructures for OWTs include semi-submersibles, spars, tension-leg platforms, and barges. Semi-submersibles, comprising 80\% of projects, are preferred due to their shallow drafts and stability, achieved through large steel columns (typically three) interconnected by tubular members. These substructures can be assembled in port and towed to the site without heavy-lift vessels. Spars, utilized in 9\% of projects, the stability is guaranteed by the inertia for horizontal axis rotations of a massive cylinder whose draft is at least equal to the wind turbine hub height. This design limits their applicability to deeper waters and ports with sufficient depth. Tension-leg platforms, representing 6\% of projects, are more complex to install but feature minimal anchor footprints, making them suitable for deep water and large-scale installations. In this case, the buoyancy exceeds the structure self weight and the remaining stability is guaranteed by tensions cables. Barges, which account for only 4\% of projects, rely on distributed buoyancy to stabilize a massive platform that can support one or more wind turbines. Their application is limited due to technical challenges associated with their large surface volume. These trends are illustrated in \autoref{fig:types_tech_offshore}.

\subsubsection{Floating Offshore Programs}
To advance floating offshore wind technology, several countries have launched pioneering programs. In 2020, Portugal commissioned the WindFloat Atlantic project, the world's first semi-submersible floating offshore wind farm \cite{pt_project}. Similarly, in September 2022, the Biden administration announced the Floating Offshore Wind Shot initiative, aimed at reducing floating offshore wind costs by 70\% to 45 USD/MWh and positioning the U.S. as a leader in turbine design, development, and manufacturing. The initiative aims to deploy 15 GW of floating offshore wind capacity by 2035 \cite{us_project}.

Achieving this level of cost reduction will require simultaneous development and implementation of multiple innovations within the floating system design, focusing on volume production, standardization, industrialization of the supply chain, and leveraging global industry experience. Techno-economic cost modeling suggests that floating offshore wind technology can achieve costs comparable to or lower than fixed-bottom offshore wind \cite{DOE2023}.

\subsection{Future Prospects: Offshore and Floating Offshore} \label{sec:future}

\subsubsection{Cumulative Capacity Prediction}
\begin{figure}[htt!]
    \centering
    \subfloat[\label{fig:estimative_cumulative_capacity_1}]
     {\includegraphics[width=0.485\textwidth]{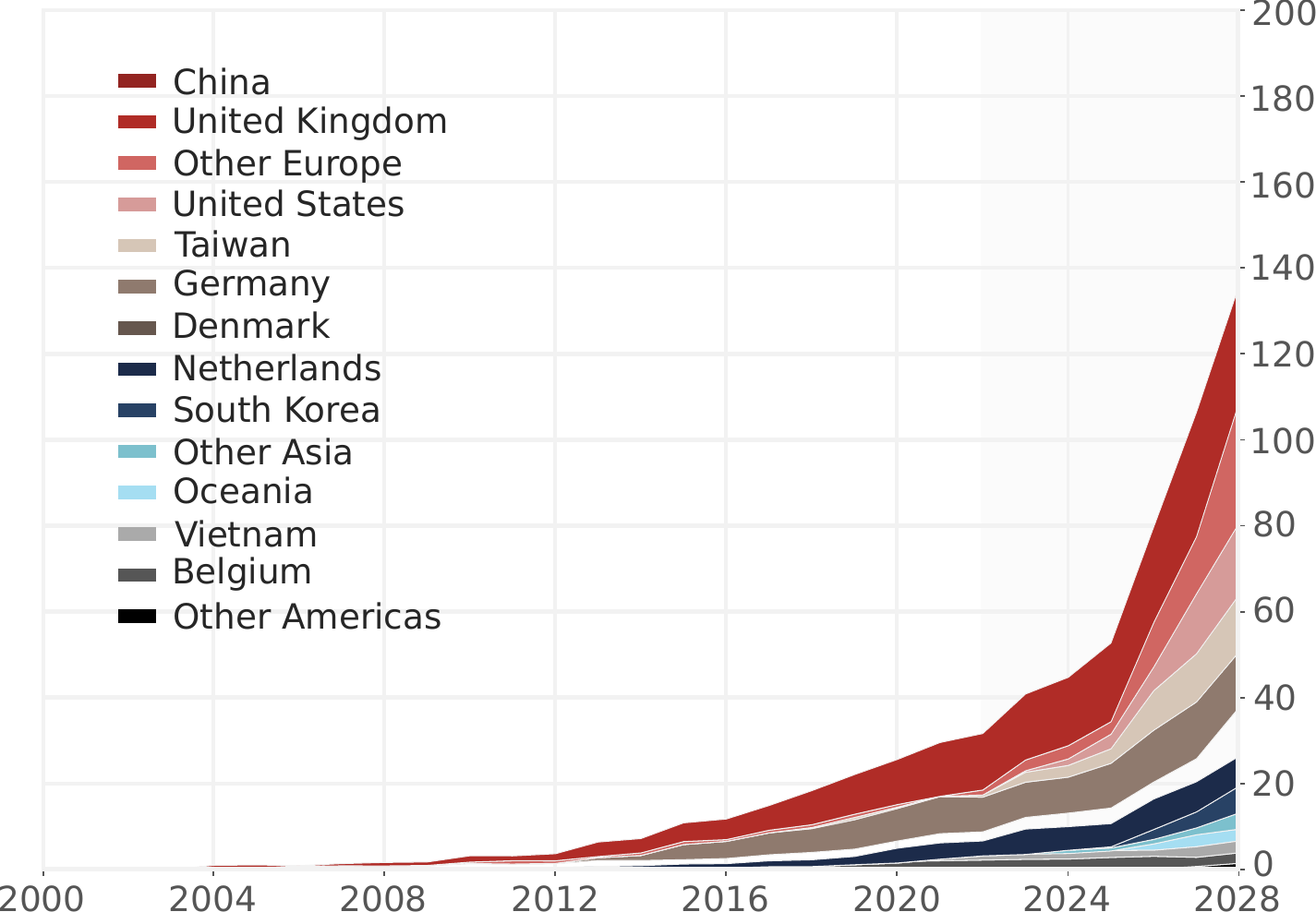}}\hfill
    \vspace{1em}
    \subfloat[\label{fig:estimative_cumulative_capacity_2}]
    {\includegraphics[width=0.485\textwidth]{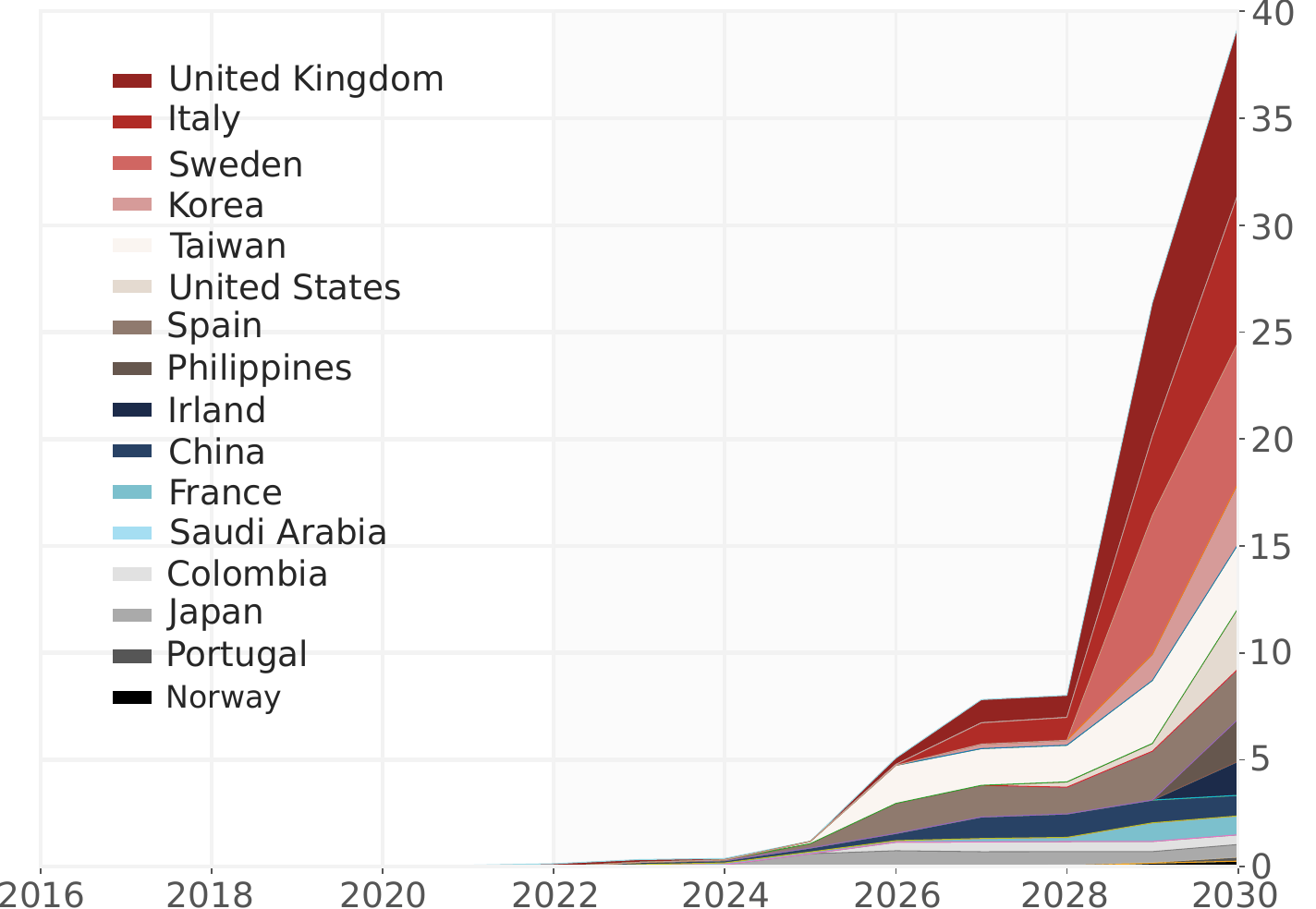}}
    \caption{Estimated cumulative installed wind energy capacity (GW) over the years by country: (a) Offshore, based on developer-announced commercial operation dates; (b) Floating Offshore, based on announced commercial operation dates (adapted from \cite{DOE2023}).}
    \label{fig:estimative_cumulative_capacity}
\end{figure}

Looking to the future, the U.S. Department of Energy study \cite{DOE2023} predicts a significant increase in cumulative wind installations for both offshore and floating offshore wind. Offshore wind capacity is projected to exceed 182 GW by 2028, while floating offshore wind is expected to reach 7957 MW by 2028 and 39385 MW by 2030, as illustrated in \autoref{fig:estimative_cumulative_capacity}.

Additionally, forecasts from BloombergNEF (BNEF) and 4C Offshore predict substantial growth in global offshore wind capacity by 2032, with BNEF estimating 379.5 GW and 4C Offshore projecting 394.4 GW, representing a more than fivefold increase. A key trend is the projected expansion in the Chinese market, expected to install 60.5 to 113 GW. Europe is expected to hold 34–39\% of global capacity, China 19–36\%, other Asian countries (Taiwan, Korea, Japan, and Vietnam) 13–14\%, and the U.S. 11–19\%.

\begin{figure}[htt!]
	\centering
    \includegraphics[width=0.485\textwidth]{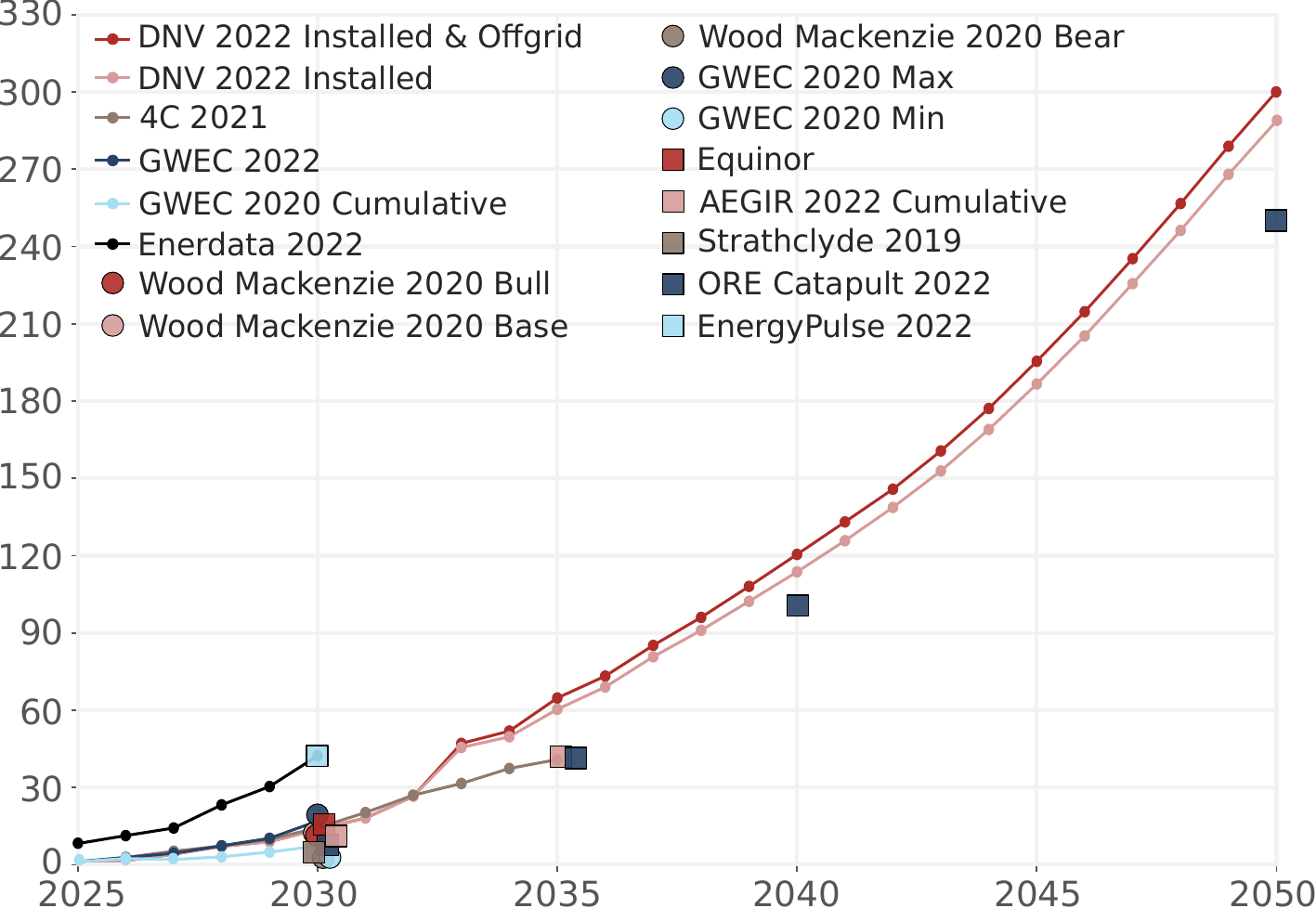}
    \caption{Projections of long-term global cumulative floating offshore wind energy deployment (GW) over the years by various independent groups (adapted from \cite{DOE2023}).}
    \label{fig:estimative_cumulative_capacity_2050}    
\end{figure}

\autoref{fig:estimative_cumulative_capacity_2050}  presents projections for floating offshore wind capacity from 2025 to 2050, estimating growth from 10 GW by 2030 to 300 GW by 2050. This expansion is driven by cost reductions, supply chain maturity, limited shallow-water sites, technological advances, and growing interest from deep-water markets like the U.S. Pacific. These forecasts point to a promising future for offshore wind, supported by substantial investments and innovations, despite typical challenges in scalability and technological progress.

\subsubsection{Levelized Cost of Energy Prediction}
Various research organizations estimate that the LCoE for U.S. floating offshore wind projects will decrease from 82–255 USD/MWh in 2022 to 66–128 USD/MWh by 2030, with variations attributed to differences in pre-commercial and commercial-scale projects. Significant cost reductions are anticipated as floating wind markets expand, following trends typical of early-stage technologies. Moreover, advancements in technology and commercialization from fixed-bottom offshore wind may provide additional benefits to floating systems \cite{DOE2023}.

Key opportunities to reduce LCoE include designing larger rotors to increase energy capture, managing wake effects to improve energy flow across turbines, enhancing component reliability, and optimizing turbine design through refined standards and methods. Additional cost reductions are anticipated from leveraging fixed-bottom system innovations, using established supply chains, minimizing construction steps at sea, automating substructure production, capturing high wind speeds in remote areas, and incorporating lighter, modular components \cite{DOE2023, report_grandevision}. Furthermore, advancements in advanced materials (e.g., low-cost carbon fiber for blades, 3D-printed materials), additive manufacturing, automation, standardization, quality control, on-site manufacturing or assembly, and ML for design optimization support the production of very large components \cite{report_grandevision}.

\subsection{Upscaling Offshore Wind Turbines}
\subsubsection{Upscaling Advantages}
\begin{figure}[htt!]
	\centering
        \includegraphics[width=0.485\textwidth]{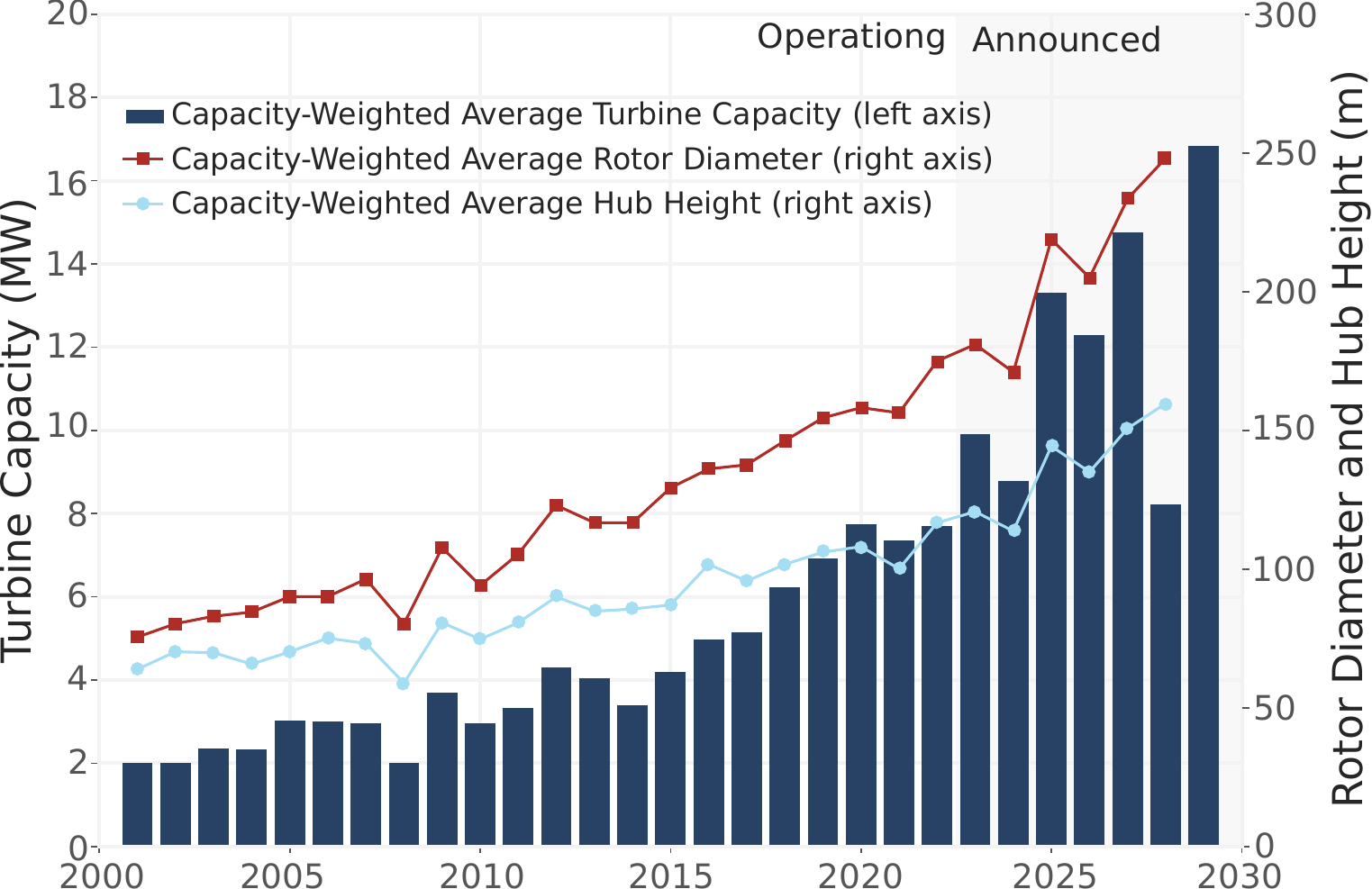}
        \caption{Global average OWT capacities, hub heights, and rotor diameters (adapted from \cite{DOE2023}).}
        \label{fig:offshore_capacities}   
\end{figure}

The development of larger wind turbines is highly valued for its potential to reduce LCoE by decreasing both CapEx and OpEx per unit of energy (\autoref{eq:lcoe}). These cost reductions are achieved by minimizing the number of required installation sites, subsequently decreasing expenses related to installation, maintenance, and components such as array cables and support structures \cite{DOE2023, report_grandevision}. \autoref{fig:offshore_capacities} illustrates the trend in offshore wind energy, highlighting consistent increases in turbine size, hub heights, rotor diameters, and capacities.

\begin{figure}[htt!]
	\centering
    \includegraphics[width=0.485\textwidth]{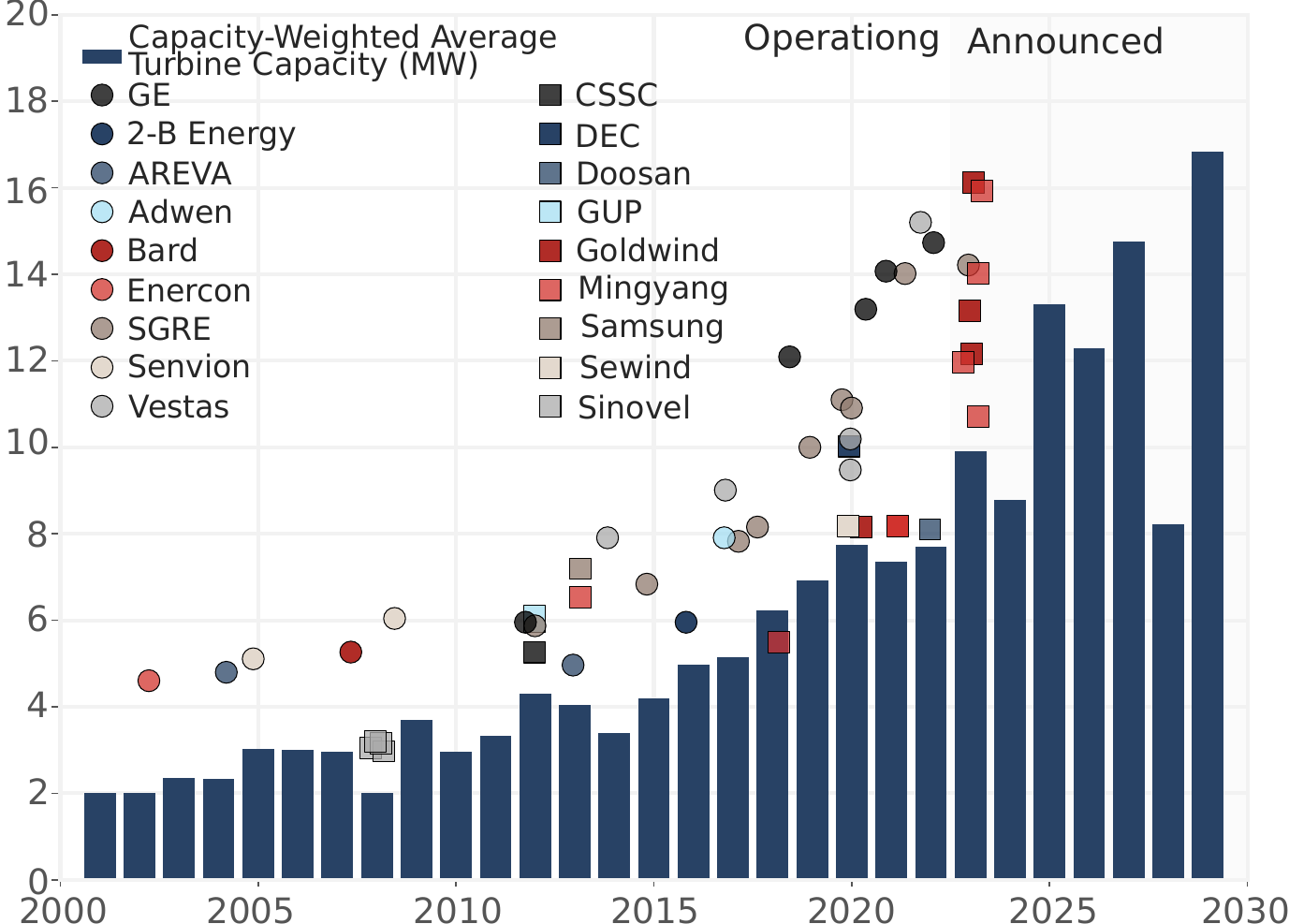}
    \caption{Comparison of OWT prototypes with commercial OWT growth. Note that GE is General Electric, SGRE is Siemens Gamesa Renewable Energy, CSSC is China State Shipbuilding Corporation, DEC is Dongfang Electric Corp., and GUP is Guodian United Power Technology Co., Ltd (adapted from \cite{DOE2023}).}
    \label{fig:offshore_commercial}
\end{figure}

Interest in upscaling is reflected in operational prototypes that frequently exceed the capacity of commercial turbines from their installation year,  as shown in \autoref{fig:offshore_commercial}. While this trend is expected to continue, commercialization from prototype to large scale typically requires years. During this period, turbine reliability must be demonstrated, and infrastructure and supply chains for serial production established. Increasing turbine capacity also demands substantial infrastructure investment, potentially rendering recent investments in smaller turbines obsolete.

\subsubsection{Upscaling Challenges}
Despite the potential of larger wind turbines to reduce costs, there is ongoing debate about whether to continue increasing turbine size or maintain the current 15 MW scale to streamline production and infrastructure. The shift to larger turbines has strained supply chains and extended development timelines as the industry adapts to the latest 15 MW models. While upscaling from 2 MW in the early 2000s to 15 MW by 2023 has reduced turbine numbers and associated costs, further increases in size may yield reduced returns. Additional challenges include heightened technological and market risks, project delays, immature production lines, limited field learning, difficulties with standardization, and higher insurance costs, which could offset potential near-term savings \cite{DOE2023}.

\subsubsection{Manufacturer Competition}
Upscaling is intensifying competition among major wind turbine manufacturers. In June 2024, Dongfang Electric installed the world's first 18 MW turbine \cite{Dongfang}, followed by Mingyang Smart Energy with another 18 MW turbine in China \cite{Mingyang}. Siemens Gamesa has also announced plans for a 21 MW model \cite{Siemens}. Currently, Siemens Gamesa leads the market with 46.2\% share, followed by Vestas (14.0\%), Mingyang (10.1\%), Goldwind (6.6\%), and Envision Energy (5.6\%). Projections suggest Siemens Gamesa will maintain its lead, with a 48.0\% share of future capacity, followed by General Electric (18.1\%), Vestas (17.8\%), Doosan Heavy Industries (4.6\%), and Mingyang (2.4\%).

This competition is driving advancements in turbine size and capacity. Chinese manufacturers, once focused primarily on the domestic market, are now expanding internationally, with companies like Mingyang, Envision Energy, Dongfang, Goldwind, Windey, and Harbin Electric entering markets in Vietnam, Japan, Italy, the United Kingdom, Norway, and France. This expansion represents a significant potential for increased competition in the global OWT industry \cite{DOE2023}.

\subsection{Tower: A Crucial Element in Upscaling} \label{sec:upscaling_tower}

\subsubsection{Upscaling Impact on Tower Mass}
\begin{figure}[htt!]
	\centering
 \includegraphics[width=0.485\textwidth]{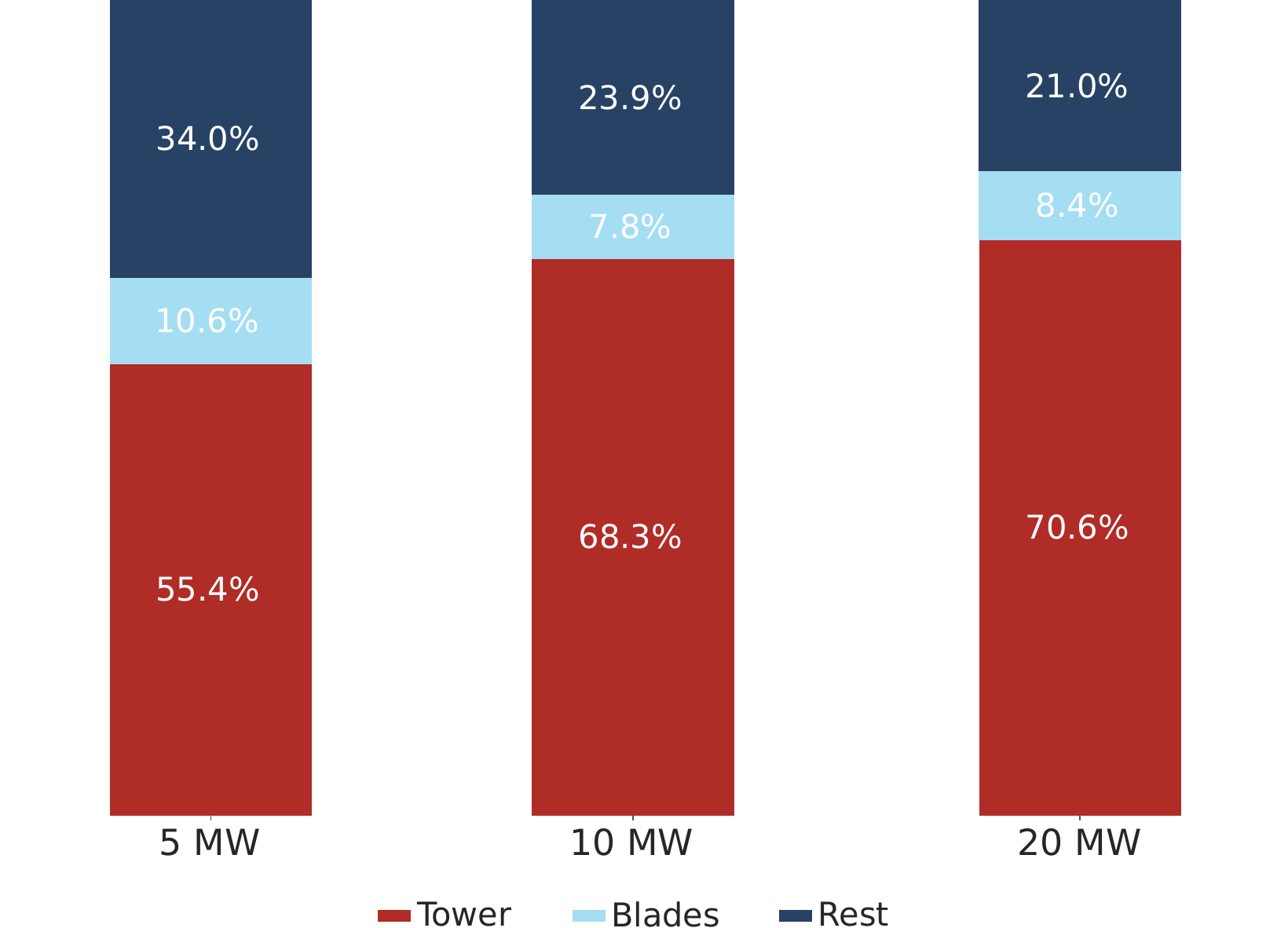}
    \caption{Component mass distribution (\%) for turbines from 5 to 20 MW, highlighting the impact of upscaling on blade and tower mass (adapted from \cite{Ashuri_Zaaijer_Martins_Zhang_2016}).}
    \label{fig:mass_5_to_20}     
\end{figure}

\citet{Ashuri_Zaaijer_Martins_Zhang_2016} uses MDO to analyze 5, 10, and 20 MW wind turbines as data points for studying upscaling trends. By examining the design data and characteristics of these turbines, they derive trends related to loading, mass, and cost. These trends help assess the technical and economic implications of upscaling, particularly its impact on design and cost. The study's findings, shown in \autoref{fig:mass_5_to_20}, highlight the effect of upscaling on component mass. The tower's mass contribution increases from 55.4\% at 5 MW to 70.6\% at 20 MW, reflecting an approximate 27.4\% increase. Conversely, in general, the mass contributions of the blades and other components decrease as turbine size increases.

These findings underscore the significant impact of upscaling on mass distribution within wind turbines, particularly the increasing prominence of the tower's mass. As turbine size grows, the relative mass of the tower becomes more substantial, highlighting the importance of optimizing tower design. Additionally, reducing tower top mass and alleviating loads are emerging needs for large-scale wind turbines. The increase in the tower’s mass percentage suggests that future design and cost considerations will need to prioritize the tower. Understanding these scaling trends is crucial for developing efficient and cost-effective strategies in the design of larger wind turbines.

\subsubsection{Upscaling Impact on Tower Cost} 
The study by \citet{Ashuri_Zaaijer_Martins_Zhang_2016} also analyzes the impact of upscaling on costs. They found that the cost share for blades increases from 7.7\% at 5 MW to 10.7\% at 20 MW, representing a 39\% increase. Similarly, the cost share for towers rises from 7.0\% at 5 MW to 15.6\% at 20 MW, which represents a 123\% increase. The exponent of the power curve fit to these cost elements is provided in \autoref{tab:exponent_curve_fit}. Notably, the tower experiences the highest cost impact, followed by the yaw system, low-speed shaft, and blade, indicating that upscaling significantly affects tower costs more than other components.

\begin{table}[htt!] 
\footnotesize
\centering
\caption{Exponent of the curve fit for 5 to 20 MW wind turbines. The higher the exponent, the higher the negative impact of upscaling (from \cite{Ashuri_Zaaijer_Martins_Zhang_2016}).}
\begin{tabular}{@{}p{3.5cm}p{2.3cm}@{}}
\toprule
\textbf{Cost Component} & \textbf{Trend Exponent} \\ \midrule
Tower & 3.22 \\
\addlinespace[0.1cm]
Yaw system & 2.97 \\
\addlinespace[0.1cm]
Main shaft & 2.89 \\
\addlinespace[0.1cm]
Blades & 2.66 \\
\addlinespace[0.1cm]
Pitch system & 2.66 \\
\addlinespace[0.1cm]
Warranty & 2.55 \\
\addlinespace[0.1cm]
Marinization & 2.55 \\
\addlinespace[0.1cm]
Main bearing & 2.44 \\
\addlinespace[0.1cm]
Hub & 2.39 \\
\addlinespace[0.1cm]
Decommissioning & 2.22 \\
\addlinespace[0.1cm]
Gearbox & 2.19 \\
\addlinespace[0.1cm]
Bedplate & 1.95 \\
\addlinespace[0.1cm]
Railing and platform & 1.95 \\
\addlinespace[0.1cm]
Operation and maintenance & 1.85 \\
\addlinespace[0.1cm]
Foundation system & 1.75 \\
\addlinespace[0.1cm]
Electrical interface & 1.75 \\
\addlinespace[0.1cm]
Installation & 1.75 \\
\addlinespace[0.1cm]
Scour protection & 1.75 \\
\addlinespace[0.1cm]
Generator & 1.75 \\
\addlinespace[0.1cm]
Power electronics & 1.75 \\
\addlinespace[0.1cm]
Electrical connection & 1.75\\
\addlinespace[0.1cm]
Engineering & 1.75 \\
\addlinespace[0.1cm]
Port and staging & 1.75 \\
\addlinespace[0.1cm]
Levelized replacement & 1.75 \\
\addlinespace[0.1cm]
Hydraulic and cooling & 1.75 \\
\addlinespace[0.1cm]
HSS and brake & 1.75 \\
\addlinespace[0.1cm]
Nacelle cover & 1.69 \\
\addlinespace[0.1cm]
Hub cone & 1.18 \\
\addlinespace[0.1cm]
Access equipment & Fixed \\
\addlinespace[0.1cm]
Safety and control & Fixed \\ \bottomrule
\end{tabular}
\label{tab:exponent_curve_fit}
\end{table}

The study also compares Turbine Capital Cost (TCC) with Balance of Station (BOS) costs. TCC includes a variety of components such as blades, hub, pitch mechanism and bearings, nose cone, low-speed shaft and bearings, gearbox, mechanical brake, high-speed shaft and coupling, generator, power electronics, yaw drive and bearing, main frame, electrical connections, hydraulic system, cooling system, nacelle cover, control equipment, safety system, condition monitoring, tower, and marinization. BOS cover the monopile, port and staging equipment, turbine installation, electrical interface and connections, permits, engineering, site assessment, personnel access equipment, scour protection, transportation, offshore warranty premium, and decommissioning. For the 5 MW design,  TCC and BOS are nearly equivalent. However, as turbines are upscaled, TCC increase more rapidly than BOS costs, underscoring the need to focus on TCC to achieve cost reductions. While upscaling increases AEP, it also raises Initial Capital Costs, including both TCC and BOS. This trade-off suggests that the additional energy generated from larger turbines does not sufficiently offset the higher Initial Capital Costs, negatively impacting the LCoE and highlighting the need for cost reductions.

\pagebreak

The authors conclude that upscaling without design modifications fails to reduce costs and may increase them. The mass and cost of the tower rise due to higher loads and increased tower-top component mass. To address these challenges, load alleviation and mass reduction strategies are needed. For upscaling to be a viable cost-reduction strategy, it should proceed incrementally, incorporating advanced technologies at each stage to mitigate negative impacts and ensure economic viability

\section{Optimization Studies Review} \label{sec:review}

This section provides a review of key studies that apply optimization techniques to the design of OWT structures, with a particular emphasis on tower optimization. Structural optimization of wind turbine towers has gained significant attention due to its critical impact on both structural performance and cost-effectiveness, particularly as turbines scale up in size. As highlighted in \autoref{sec:economics}, towers are among the most costly and heaviest components of wind turbines, making them a central focus for optimization.

Several optimization approaches have been explored in the literature, ranging from traditional techniques such as gradient-based optimization and genetic algorithms to more advanced methodologies like topology optimization and AI-driven models. These methods aim to reduce material usage, lower manufacturing costs, and address the challenges associated with scaling larger turbines, as highlighted in \autoref{sec:upscaling_tower}.

\subsection{Goals}
The aim of this literature review is to identify the most effective strategies for optimizing OWT towers and to provide a comprehensive analysis of current research trends. By synthesizing the methodologies, tools, and outcomes from key studies, this review offers valuable insights that support the scalability and efficiency of wind turbine systems. These findings are crucial for researchers in both academia and industry, offering a foundation for advancing existing techniques or adopting new approaches in the optimization and design of wind turbine towers.

\subsection{Tower Components}
The primary focus of this review is the optimization of OWT towers. However, given the limited research available in this emerging area, insights from onshore wind turbine towers will also be considered due to their structural similarities. These parallels allow for valuable extrapolation of findings from onshore to offshore environments.

Moreover, due to the limited number of studies specifically focusing on tower optimization through topology optimization and considering the potential of this method to reduce mass and costs, this review also includes research on offshore foundations, such as jacket and tripod structures. These foundations share important structural, methodological, and environmental characteristics with turbine towers, offering valuable insights into optimization techniques in offshore applications.

\subsection{Review Parameters}
This literature review explores key aspects of wind turbine tower design, including turbine types, optimization methods, software tools, load conditions, design variables, design constraints, analysis techniques, and results. Understanding turbine types clarifies the most commonly used models and their capacities, while optimization methods reveal effective strategies for improving performance and reducing costs. The review of software tools highlights the best platforms for simulating complex design scenarios. Load conditions are examined to uncover critical forces impacting tower design, while the analysis of design variables and constraints identifies crucial parameters that can be adjusted and restricted to enhance structural efficiency and durability. Finally, the analysis techniques provide essential methods for evaluating and comparing original and optimized designs, while the results highlight the key outcomes of the studies.

\subsection{Methodology}

The selection process for this review of wind turbine tower design optimization was guided by specific inclusion criteria. Each study had to employ at least one optimization technique and be directly related to the engineering design of wind turbine towers or their foundations with a focus on TO. Studies unrelated to tower design, such as those centered on energy management, grid layout, power system optimization, or wind farm configuration, were excluded to ensure a focused review.

Relevant studies were identified through targeted keyword searches in Google Scholar, emphasizing topics closely related to wind turbine tower design and optimization. The selection of studies was based on their direct relevance to the review topic, as assessed by two independent raters (the co-authors of this review). Each rater independently evaluated the studies for inclusion based on the title, abstract, and keywords, ensuring objectivity. Any disagreements between the raters regarding the relevance of specific studies were resolved through discussion among all authors, thereby maintaining consistency and rigor in the selection process.

Of the 32 papers reviewed, 15 focused on the design optimization of onshore turbine towers, 10 on offshore turbine towers, and 7 on the application of TO methods for offshore foundations. These studies can be categorized by component type. A total of 25 papers focused on turbine towers, with 10 on offshore towers and 15 on onshore towers, while 7 focused on foundations. A total of 17 studies focused on offshore components, with 10 on offshore towers and 7 on offshore foundations, while 15 focused on onshore turbines.

\subsection{Review Results} 
A detailed analysis of 32 optimization studies is provided, categorized into onshore towers, offshore towers, and offshore foundations. The studies on onshore and offshore towers are summarized in \autoref{table:lr_towers_on} and \autoref{table:lr_towers_off}, respectively, while \autoref{table:optimization_offshore_foundations} focuses on the topology optimization of offshore foundations. A detailed analysis of each key parameter is provided, combining qualitative insights and quantitative data to offer a complete understanding of the application of optimization techniques in wind turbine tower design.

\subsubsection{Turbine} \label{sec:review_turbine}
The turbines were analyzed by capacity, model, and tower design type. Related to turbine capacities, \autoref{table:turbine_capacity_studies} shows that the 5 MW turbine is the most frequently studied, appearing in 44\% (14 out of 32) of the reviewed studies. The analysis reveals that onshore studies generally focus on smaller turbines, with capacities ranging from 100 kW to 5 MW, with a significant portion of studies centered around 2 MW models. In contrast, offshore studies predominantly investigate larger turbines, ranging from 5 MW to 25 MW, with the majority concentrating on the 5 MW model. This trend aligns with the observation discussed in \autoref{subsec:ontooff}, where, as turbine capacities increase, there is a noticeable transition from onshore to offshore applications. This shift is motivated by the higher wind potential in offshore environments and the corresponding demand for increased power generation capacity.

\begin{table}[h!]
\scriptsize
\centering
\caption{Number of studies by turbine capacity (MW).}
\begin{tabular}{@{}p{1.7cm}p{0.6cm}p{0.6cm}p{1cm}p{3.2cm}@{}}
\toprule
\textbf{Turbine Capacity (MW)} & \textcolor{specialred}{\textbf{Onshore Towers}} & \textcolor{specialblue}{\textbf{Offshore Towers}} & \textcolor{specialcyan}{\textbf{Offshore Foundations}} & \textbf{Total \break Studies} \\ 
\midrule
0.1  & 1 & 0 & 0 & \phantom{1}1 \textcolor{specialred}{\rule{2mm}{2mm}} \\ \addlinespace[0.1cm]
0.4  & 2 & 0 & 0 & \phantom{1}2 \textcolor{specialred}{\rule{4mm}{2mm}} \\ \addlinespace[0.1cm]
1.5  & 2 & 0 & 0 & \phantom{1}2 \textcolor{specialred}{\rule{4mm}{2mm}} \\ \addlinespace[0.1cm]
2    & 5 & 1 & 0 & \phantom{1}6 \textcolor{specialred}{\rule{10mm}{2mm}}\textcolor{specialblue}{\rule{2mm}{2mm}}  \\ \addlinespace[0.1cm]
3    & 1 & 0 & 0 & \phantom{1}1 \textcolor{specialred}{\rule{2mm}{2mm}}\\ \addlinespace[0.1cm]
5    & 4 & 3 & 7 & 14 \textcolor{specialred}{\rule{8mm}{2mm}}\textcolor{specialblue}{\rule{6mm}{2mm}}\textcolor{specialcyan}{\rule{14mm}{2mm}}\\ \addlinespace[0.1cm]
6    & 0 & 1 & 0 & \phantom{1}1 \textcolor{specialblue}{\rule{2mm}{2mm}}\\ \addlinespace[0.1cm]
10   & 0 & 1 & 0 & \phantom{1}1 \textcolor{specialblue}{\rule{2mm}{2mm}}\\ \addlinespace[0.1cm]
15   & 0 & 1 & 0 & \phantom{1}1 \textcolor{specialblue}{\rule{2mm}{2mm}}\\ \addlinespace[0.1cm]
25   & 0  & 1 & 0 & \phantom{1}1 \textcolor{specialblue}{\rule{2mm}{2mm}} \\ \addlinespace[0.1cm]
Unlisted & 0  & 2 & 0 & \phantom{1}2 \textcolor{specialblue}{\rule{4mm}{2mm}}\\
\bottomrule
\end{tabular}
\label{table:turbine_capacity_studies}
\end{table}

Regarding turbine models, as shown in \autoref{table:turbine_model_studies}, the NREL 5 MW turbine is the most commonly used, featured in 31\% (10 out of 32) of the studies specifying a turbine model. This dominance is likely due to its comprehensive documentation and broad acceptance as a reference standard. Although larger turbines, such as the 10 MW, 15 MW and 22 MW models discussed in \autoref{sec:wt_reference}, are gaining attention due to their potential for increased energy capture, they remain underexplored due to their recent development. Further research is needed to address scaling up challenges, particularly offshore where structural complexity significantly increases.

\begin{table}[h!]
\scriptsize
\centering
\caption{Number of studies by turbine model.}
\begin{tabular}{@{}p{1.1cm}p{0.6cm}p{0.6cm}p{1cm}p{3.8cm}@{}}
\toprule
\textbf{Turbine Model} & \textcolor{specialred}{\textbf{Onshore Towers}} & \textcolor{specialblue}{\textbf{Offshore Towers}} & \textcolor{specialcyan}{\textbf{Offshore Foundations}} & \textbf{Total \break Studies} \\ 
\midrule
WindPACT 1.5 MW  & 1 & 0 & 0 & \phantom{1}1 \textcolor{specialred}{\rule{2mm}{2mm}} \\ \addlinespace[0.1cm]
Gamesa 2.0 MW  & 1 & 0 & 0 & \phantom{1}1 \textcolor{specialred}{\rule{2mm}{2mm}} \\ \addlinespace[0.1cm]
Vestas V90 3 MW   & 1 & 0 & 0 & \phantom{1}1 \textcolor{specialred}{\rule{2mm}{2mm}} \\ \addlinespace[0.1cm]
IEA Wind 15 MW    & 0 & 1 & 0 & \phantom{1}1 \textcolor{specialblue}{\rule{2mm}{2mm}}  \\ \addlinespace[0.1cm]
NOWITECH 10 MW   & 0 & 1 & 0 & \phantom{1}1 \textcolor{specialblue}{\rule{2mm}{2mm}}\\ \addlinespace[0.1cm]
NREL \break 5 MW & 2 & 2 & 6 & 10 \textcolor{specialred}{\rule{4mm}{2mm}}\textcolor{specialblue}{\rule{4mm}{2mm}}\textcolor{specialcyan}{\rule{12mm}{2mm}}\\ \addlinespace[0.1cm]
Unlisted & 10  & 6 & 1 & 17 \textcolor{specialred}{\rule{20mm}{2mm}}\textcolor{specialblue}{\rule{12mm}{2mm}}\textcolor{specialcyan}{\rule{2mm}{2mm}}\\ \addlinespace[0.1cm]
\bottomrule
\end{tabular}
\label{table:turbine_model_studies}
\end{table}

In terms of tower design type, among the 25 studies focused on tower structures, tubular steel towers are the most studied, accounting for 60\% (15 out of 25) of these studies, as shown in \autoref{table:tower_type_studies}. Lattice towers and hybrid towers are also investigated, representing 16\% (4 out of 25) and 12\% (3 out of 25) of the studies, respectively. The use of composite or alternative materials, such as bamboo, remains limited, but as noted in \autoref{sec:tower_design_types}, they are being explored in remote areas to reduce costs and enhance sustainability.

\begin{table}[h!]
\scriptsize
\centering
\caption{Number of studies by tower design type.}
\begin{tabular}{@{}p{1.9cm}p{0.6cm}p{0.6cm}p{3.4cm}@{}}
\toprule
\textbf{Tower \break Design}       & \textcolor{specialred}{\textbf{Onshore Towers}} & \textcolor{specialblue}{\textbf{Offshore Towers}} & \textbf{Total  \break Studies} \\ 
\midrule
Concrete             & 1 & 0 & \phantom{1}1 \textcolor{specialred}{\rule{2mm}{2mm}} \\ 
\addlinespace[0.1cm]
Composite/Bamboo     & 1 & 1 & \phantom{1}2 \textcolor{specialred}{\rule{2mm}{2mm}}\textcolor{specialblue}{\rule{2mm}{2mm}} \\ 
\addlinespace[0.1cm]
Hybrid               & 3 & 0 & \phantom{1}3 \textcolor{specialred}{\rule{6mm}{2mm}} \\ 
\addlinespace[0.1cm]
Lattice              & 3 & 1 & \phantom{1}4 \textcolor{specialred}{\rule{6mm}{2mm}}\textcolor{specialblue}{\rule{2mm}{2mm}} \\ 
\addlinespace[0.1cm]
Tubular Steel        & 7 & 8 & 15 \textcolor{specialred}{\rule{14mm}{2mm}}\textcolor{specialblue}{\rule{16mm}{2mm}} \\ 
\bottomrule
\end{tabular}
\label{table:tower_type_studies}
\end{table}

\subsubsection{Software}
The software used in the reviewed optimization studies is categorized into three main groups: Finite Element Analysis (FEA), AHSE, and TO. A summary of these classifications is shown in \autoref{table:software_studies}. In the FEA category, Ansys is the most frequently used software, appearing in approximately 44\% (14 out of 32) of the studies. For AHSE, Bladed and OpenFAST are predominantly employed in offshore projects, used in about 22\% (7 out of 32) and 13\% (4 out of 32) of the studies, respectively. In the TO category, no single software dominates, OptiStruct is the most used, featured in approximately 9\% (3 out of 32) of the studies, followed by FENDAC, Ansys, FE-Design, and HyperWorks, each used in just one study.

\begin{table}[h!]
\scriptsize
\centering
\caption{Number of studies by software type: (a) FEA; (b) AHSE; (c) TO.}
\begin{tabular}{@{}p{1cm}p{0.6cm}p{0.6cm}p{1cm}p{3.2cm}@{}}
\toprule
\textbf{Software}        & \textcolor{specialred}{\textbf{Onshore Towers}} & \textcolor{specialblue}{\textbf{Offshore Towers}} & \textcolor{specialcyan}{\textbf{Offshore Foundations}} & \textbf{Total \break Studies} \\ 
\midrule
\multicolumn{5}{l}{\textbf{(a) FEA}} \\
\midrule
OpenSees                  & 1 & 0 & 0 & \phantom{1}1 \textcolor{specialred}{\rule{2mm}{2mm}} \\ \addlinespace[0.1cm]
SAP2000                   & 1 & 0 & 0 & \phantom{1}1 \textcolor{specialred}{\rule{2mm}{2mm}} \\ \addlinespace[0.1cm]
FEDEM           & 0 & 1 & 0 & \phantom{1}1 \textcolor{specialblue}{\rule{2mm}{2mm}} \\ \addlinespace[0.1cm]
Nastran                    & 0 & 0 & 1 & \phantom{1}1 \textcolor{specialcyan}{\rule{2mm}{2mm}} \\ \addlinespace[0.1cm]
Sesam                      & 0 & 0 & 1 & \phantom{1}1 \textcolor{specialcyan}{\rule{2mm}{2mm}} \\ \addlinespace[0.1cm]
ZEUS NL                    & 3 & 0 & 0 & \phantom{1}3 \textcolor{specialred}{\rule{6mm}{2mm}} \\ \addlinespace[0.1cm]
Ansys                      & 7 & 5 & 2 & 14 \textcolor{specialred}{\rule{14mm}{2mm}}\textcolor{specialblue}{\rule{10mm}{2mm}}\textcolor{specialcyan}{\rule{4mm}{2mm}} \\ 
\midrule
\multicolumn{5}{l}{\textbf{(b) AHSE}} \\
\midrule
OpenFAST                   & 1 & 3 & 0 & \phantom{1}4 \textcolor{specialred}{\rule{2mm}{2mm}}\textcolor{specialblue}{\rule{8mm}{2mm}} \\ \addlinespace[0.1cm]
Bladed                     & 1 & 0 & 6 & \phantom{1}7 \textcolor{specialred}{\rule{2mm}{2mm}}\textcolor{specialcyan}{\rule{14mm}{2mm}} \\ \addlinespace[0.1cm]
\midrule
\multicolumn{5}{l}{\textbf{(c) TO}} \\
\midrule
FENDAC                     & 1 & 0 & 0 & \phantom{1}1 \textcolor{specialred}{\rule{2mm}{2mm}} \\ \addlinespace[0.1cm]
Ansys                      & 0 & 0 & 1 & \phantom{1}1 \textcolor{specialcyan}{\rule{2mm}{2mm}} \\ \addlinespace[0.1cm]
FE-Design                  & 0 & 0 & 1 & \phantom{1}1 \textcolor{specialcyan}{\rule{2mm}{2mm}} \\ \addlinespace[0.1cm]
HyperWorks                 & 0 & 0 & 1 & \phantom{1}1 \textcolor{specialcyan}{\rule{2mm}{2mm}} \\ \addlinespace[0.1cm]
Optistruct                 & 0 & 0 & 3 & \phantom{1}3 \textcolor{specialcyan}{\rule{6mm}{2mm}} \\ 
\bottomrule
\end{tabular}
\label{table:software_studies}
\end{table}

\subsubsection{Loads}

As highlighted in \autoref{sec:towerloads}, different load types are essential for evaluating the performance and reliability of wind turbine towers. Among the reviewed studies, gravitational loads are included in 100\% (32 out of 32) of cases, wind loads in 94\% (30 out of 32), aerodynamic loads in 81\% (26 out of 32), and hydrodynamic loads in 82\% (14 out of 17) for offshore applications, as summarized in \autoref{table:turbine_load_studies}. In contrast, earthquake loads are considered in only 4 studies, reflecting their relative rarity compared to other load types.

\begin{table}[h!]
\scriptsize
\centering
\caption{Number of studies by load type.}
\begin{tabular}{@{}p{1.2cm}p{0.6cm}p{0.6cm}p{1cm}p{3.5cm}@{}}
\toprule
\textbf{Load} & \textcolor{specialred}{\textbf{Onshore Towers}} & \textcolor{specialblue}{\textbf{Offshore Towers}} & \textcolor{specialcyan}{\textbf{Offshore Foundations}} & \textbf{Total \break Studies} \\ 
\midrule
Earthquake  & 4 & 0 & 0 & \phantom{3}4 \textcolor{specialred}{\rule{4mm}{2mm}} \\ \addlinespace[0.1cm]
Hydrodynamic  & 0 & 7 & 7 & 14 \textcolor{specialblue}{\rule{7mm}{2mm}}\textcolor{specialcyan}{\rule{7mm}{2mm}} \\ \addlinespace[0.1cm]
Aerodynamic   & 11 & 8 & 7 & 26 \textcolor{specialred}{\rule{11mm}{2mm}}\textcolor{specialblue}{\rule{8mm}{2mm}}\textcolor{specialcyan}{\rule{7mm}{2mm}} \\ \addlinespace[0.1cm]
Wind    & 15 & 8 & 7 & 30 \textcolor{specialred}{\rule{15mm}{2mm}}\textcolor{specialblue}{\rule{8mm}{2mm}}\textcolor{specialcyan}{\rule{7mm}{2mm}}  \\ \addlinespace[0.1cm]
Gravitational   & 15 & 10 & 7 & 32 \textcolor{specialred}{\rule{15mm}{2mm}}\textcolor{specialblue}{\rule{9mm}{2mm}}\textcolor{specialcyan}{\rule{7mm}{2mm}}\\ 
\bottomrule
\end{tabular}
\label{table:turbine_load_studies}
\end{table}

Regarding DLCs, the reviewed studies examine various turbine conditions, each represented by different DLCs. These include power production scenarios (DLC 1.2, 1.3, and 1.6), power production with fault occurrence (DLC 2.1), start-up conditions (DLC 3.2), and states where the turbine is parked, either standing still or idling (DLC 6.1, 6.2, and 6.3). These DLCs are primarily applied in offshore studies. Full descriptions and the parameters defining each DLC are available in \cite{Faraggiana_Sirigu_Ghigo_Bracco_Mattiazzo_2022}. 

Among these, DLC 1.2 and 1.3 are the most commonly used, as shown in \autoref{table:turbine_dlc_studies}. Specifically, DLC 1.2, used in approximately 19\% (6 out of 32) of studies, evaluates fatigue loads during power production, while DLC 1.3, applied in 16\% (5 out of 32) of studies, assesses the ultimate design load during turbine operation. Following these, DLC 6.1 and 6.2, each applied in 12.5\% (4 out of 32) of studies, focus on calculating ultimate design loads when the turbine is in a non-operational state.

\begin{table}[h!]
\scriptsize
\centering
\caption{Number of studies by DLC.}
\begin{tabular}{@{}p{0.3cm}p{0.6cm}p{0.6cm}p{1cm}p{1.5cm}@{}}
\toprule
\textbf{DLC} & \textcolor{specialred}{\textbf{Onshore Towers}} & \textcolor{specialblue}{\textbf{Offshore Towers}} & \textcolor{specialcyan}{\textbf{Offshore Foundations}} & \textbf{Total \break Studies} \\ 
\midrule
1.2  & 0 & 2 & 4 & 6 \textcolor{specialblue}{\rule{4mm}{2mm}}\textcolor{specialcyan}{\rule{8mm}{2mm}} \\ \addlinespace[0.1cm]
1.3 & 1 & 0 & 4 & 5 \textcolor{specialred}{\rule{2mm}{2mm}}\textcolor{specialcyan}{\rule{8mm}{2mm}} \\ \addlinespace[0.1cm]
1.6   & 0 & 1 & 0 & 1 \textcolor{specialblue}{\rule{2mm}{2mm}} \\ \addlinespace[0.1cm]
2.1  & 0 & 1 & 0 & 1 \textcolor{specialblue}{\rule{2mm}{2mm}} \\ \addlinespace[0.1cm]
3.2   & 1 & 0 & 0 & 1 \textcolor{specialred}{\rule{2mm}{2mm}} \\ \addlinespace[0.1cm]
6.1  & 1 & 3 & 0 & 4 \textcolor{specialred}{\rule{3mm}{2mm}}\textcolor{specialblue}{\rule{6mm}{2mm}}\\ \addlinespace[0.1cm]
6.2 & 0 & 0 & 4 & 4 \textcolor{specialcyan}{\rule{8mm}{2mm}}\\ \addlinespace[0.1cm]
6.3  & 0 & 1 & 0 & 1 \textcolor{specialblue}{\rule{2mm}{2mm}} \\  
\bottomrule
\end{tabular}
\label{table:turbine_dlc_studies}
\end{table}

\subsubsection{Optimization Method}
This review identifies several optimization methods employed in the design of wind turbine towers, including Gradient-Based Optimization (GBO), Iterative Optimization (IO), MDO, SQP, Differential Evolution (DE), Kriging Model (KM), PSO, Tabu Search (TS), Reliability-Based Design Optimization (RBDO), and GA. Of these, GA is the most commonly used, appearing in 32\% (8 out of 25) of the studies focused on tower design optimization, as shown in \autoref{table:tower_method_studies}.

\begin{table}[h!]
\scriptsize
\centering
\caption{Number of studies by optimization method.}
\begin{tabular}{@{}p{1.2cm}p{0.6cm}p{0.6cm}p{1.9cm}@{}}
\toprule
\textbf{Optimization Method}       & \textcolor{specialred}{\textbf{Onshore Towers}} & \textcolor{specialblue}{\textbf{Offshore Towers}} & \textbf{Total  \break Studies} \\ 
\midrule
GBO             & 0 & 1 & 1 \textcolor{specialblue}{\rule{2mm}{2mm}} \\ 
\addlinespace[0.1cm]
IO     & 0 & 1 & 1 \textcolor{specialblue}{\rule{2mm}{2mm}}\\ 
\addlinespace[0.1cm]
MDO              & 0 & 1 & 1 \textcolor{specialblue}{\rule{2mm}{2mm}} \\ 
\addlinespace[0.1cm]
SQP              & 0 & 1 & 1 \textcolor{specialblue}{\rule{2mm}{2mm}} \\ 
\addlinespace[0.1cm]
DE        & 2 & 0 & 2 \textcolor{specialred}{\rule{4mm}{2mm}} \\ \addlinespace[0.1cm]
KM             & 1 & 1 & 2 \textcolor{specialred}{\rule{2mm}{2mm}}\textcolor{specialblue}{\rule{2mm}{2mm}} \\ 
\addlinespace[0.1cm]
TO             & 1 & 1 & 2 \textcolor{specialred}{\rule{2mm}{2mm}}\textcolor{specialblue}{\rule{2mm}{2mm}} \\ 
\addlinespace[0.1cm]
PSO     & 3 & 0 & 3 \textcolor{specialred}{\rule{6mm}{2mm}}\\ 
\addlinespace[0.1cm]
TS              & 3 & 0 & 3 \textcolor{specialred}{\rule{6mm}{2mm}} \\ 
\addlinespace[0.1cm]
RBDO              & 1 & 3 & 4 \textcolor{specialred}{\rule{2mm}{2mm}}\textcolor{specialblue}{\rule{6mm}{2mm}} \\ 
\addlinespace[0.1cm]
GA        & 6 & 2 & 8 \textcolor{specialred}{\rule{12mm}{2mm}}\textcolor{specialblue}{\rule{4mm}{2mm}} \\ 
\bottomrule
\end{tabular}
\label{table:tower_method_studies}
\end{table}

These optimization methods aim to minimize key variables. As shown in \autoref{table:tower_obj_func_studies}, mass reduction is the most frequently targeted variable, appearing in 52\% (13 out of 25) of tower design studies, followed by cost minimization in 28\% (7 out of 25) of the studies. This trend highlights the primary goal of enhancing cost efficiency, either through direct cost minimization or by focusing on mass reduction. Prioritizing mass reduction is particularly effective as it leads to lower material costs, reduced transportation expenses, and decreased operational costs.

\begin{table}[h!]
\scriptsize
\centering
\caption{Number of studies by types of objective function variables targeted
for minimization.}
\begin{tabular}{@{}p{1.2cm}p{0.6cm}p{0.6cm}p{3cm}@{}}
\toprule
\textbf{Objective Function}       & \textcolor{specialred}{\textbf{Onshore Towers}} & \textcolor{specialblue}{\textbf{Offshore Towers}} & \textbf{Total  \break Studies} \\ 
\midrule
Displacement             & 1 & 0 & \phantom{1}1 \textcolor{specialred}{\rule{2mm}{2mm}} \\ 
\addlinespace[0.1cm]
Reliability     & 1 & 0 & \phantom{1}1 \textcolor{specialred}{\rule{2mm}{2mm}}\\ 
\addlinespace[0.1cm]
Volume              & 0 & 1 & \phantom{1}1 \textcolor{specialblue}{\rule{2mm}{2mm}} \\ 
\addlinespace[0.1cm]
Compliance              & 1 & 1 & \phantom{1}2 \textcolor{specialred}{\rule{2mm}{2mm}}\textcolor{specialblue}{\rule{2mm}{2mm}} \\ 
\addlinespace[0.1cm]
Cost        & 6 & 1 & \phantom{1}7 \textcolor{specialred}{\rule{12mm}{2mm}}\textcolor{specialblue}{\rule{2mm}{2mm}} \\ \addlinespace[0.1cm]
Mass             & 7 & 6 & 13 \textcolor{specialred}{\rule{14mm}{2mm}}\textcolor{specialblue}{\rule{12mm}{2mm}} \\ 
\bottomrule
\end{tabular}
\label{table:tower_obj_func_studies}
\end{table}

\subsubsection{Design Variables}
Given the variety of wind turbine tower designs, comparing them using a single set of variables can be challenging. However, most studies focus on key geometric parameters such as tower height,  top and bottom dimensions, and wall thickness or sectional thickness across different sections of the tower. Focusing on the most commonly used design, the tubular steel tower, as outlined in \autoref{sec:review_turbine}, the critical design variables across these 15 studies (7 onshore and 8 offshore) are detailed in \autoref{table:tower_design_var_studies}.

\begin{table}[h!]
\scriptsize
\centering
\caption{Number of studies by design variables types using the tubular steel design.}
\begin{tabular}{@{}p{1.7cm}p{0.6cm}p{0.6cm}p{3cm}@{}}
\toprule
\textbf{Design \break Variable}       & \textcolor{specialred}{\textbf{Onshore Towers}} & \textcolor{specialblue}{\textbf{Offshore Towers}} & \textbf{Total  \break Studies} \\ 
\midrule
Section Thickness             & 2 & 6 & \phantom{1}8 \textcolor{specialred}{\rule{4mm}{2mm}}\textcolor{specialblue}{\rule{12mm}{2mm}} \\ 
\addlinespace[0.1cm]
Top/bottom Tower Thickness     & 7 & 1 & \phantom{1}8 \textcolor{specialred}{\rule{14mm}{2mm}}\textcolor{specialblue}{\rule{2mm}{2mm}} \\ 
\addlinespace[0.1cm]
Top/bottom Tower Diameter               & 6 & 7 & 13 \textcolor{specialred}{\rule{12mm}{2mm}}\textcolor{specialblue}{\rule{14mm}{2mm}} \\ 
\bottomrule
\end{tabular}
\label{table:tower_design_var_studies}
\end{table}

The most frequent variable is the tower's top and bottom diameter, appearing in 87\% (13 out of 15) of these studies. Additionally, the top and bottom thickness, along with the section thickness are important variables, each analyzed in 53\% (8 out of 15) of the studies.

\subsubsection{Design Constraints}
The reviewed studies incorporate various design constraints that are vital to ensuring the safety and reliability of wind turbine structures under operational conditions. As shown in \autoref{table:tower_design_constraint_studies}, geometry limits are the most commonly applied, appearing in approximately 81\% (26 out of 32) of the studies. These constraints are crucial for defining the physical dimensions and shapes of turbines, ensuring they can be manufactured and meet regulatory standards.

\begin{table}[h!]
\scriptsize
\centering
\caption{Number of studies by design constraints type.}
\begin{tabular}{@{}p{1.9cm}p{0.6cm}p{0.6cm}p{1cm}p{3cm}@{}}
\toprule
\textbf{Design \break Constraint}        & \textcolor{specialred}{\textbf{Onshore Towers}} & \textcolor{specialblue}{\textbf{Offshore Towers}} & \textcolor{specialcyan}{\textbf{Offshore Foundations}} & \textbf{Total  \break Studies} \\ 
\midrule
Symmetry \break Planes                 & 1 & 0 & 5 & \phantom{1}6 \textcolor{specialred}{\rule{1mm}{2mm}}\textcolor{specialcyan}{\rule{5mm}{2mm}} \\ 
\addlinespace[0.1cm]
Maximum Original Mass/Volume    & 1 & 1 & 5 & \phantom{1}7 \textcolor{specialred}{\rule{1mm}{2mm}}\textcolor{specialblue}{\rule{1mm}{2mm}}\textcolor{specialcyan}{\rule{5mm}{2mm}} \\ 
\addlinespace[0.1cm]
Minimum Fatigue Life            & 7 & 3 & 1 & 11 \textcolor{specialred}{\rule{7mm}{2mm}}\textcolor{specialblue}{\rule{3mm}{2mm}}\textcolor{specialcyan}{\rule{1mm}{2mm}} \\ 
\addlinespace[0.1cm]
Minimum Buckling Load Multiplier & 9 & 7 & 1 & 17 \textcolor{specialred}{\rule{9mm}{2mm}}\textcolor{specialblue}{\rule{7mm}{2mm}}\textcolor{specialcyan}{\rule{1mm}{2mm}} \\ 
\addlinespace[0.1cm]
Maximum Deflection              & 9 & 5 & 3 & 17 \textcolor{specialred}{\rule{9mm}{2mm}}\textcolor{specialblue}{\rule{5mm}{2mm}}\textcolor{specialcyan}{\rule{3mm}{2mm}} \\ 
\addlinespace[0.1cm]
Frequency \break Limits                & 11 & 7 & 2 & 20 \textcolor{specialred}{\rule{11mm}{2mm}}\textcolor{specialblue}{\rule{7mm}{2mm}}\textcolor{specialcyan}{\rule{2mm}{2mm}} \\ 
\addlinespace[0.1cm]
Maximum \break Stress                  & 11 & 8 & 3 & 22 \textcolor{specialred}{\rule{11mm}{2mm}}\textcolor{specialblue}{\rule{8mm}{2mm}}\textcolor{specialcyan}{\rule{3mm}{2mm}} \\ 
\addlinespace[0.1cm]
Geometry \break Limits                 & 14 & 9 & 3 & 26 \textcolor{specialred}{\rule{14mm}{2mm}}\textcolor{specialblue}{\rule{9mm}{2mm}}\textcolor{specialcyan}{\rule{3mm}{2mm}} \\ 
\bottomrule
\end{tabular}
\label{table:tower_design_constraint_studies}
\end{table}

Maximum stress constraints follow closely, utilized in approximately 69\% (22 out of 32) of the studies to prevent material failure by keeping structural stresses within the material's strength limits. 

Frequency limits are used in approximately 63\% (20 out of 32) of the studies, ensuring that the natural frequencies of the turbine structures do not coincide with operational frequencies, thereby preventing resonance and excessive vibrations. 

Maximum deflection and minimum buckling load multipliers, each employed in approximately 53\% (17 out of 32) of the studies, help prevent excessive deformation and structural collapse under compressive forces. 

Minimum fatigue life constraints, applied in approximately 34\% (11 out of 32) of the studies, ensure that turbines can endure repeated loading cycles throughout their lifetimes. 

Lastly, symmetry planes and maximum original mass or volume constraints, which are less frequently used, are typically applied in topology optimization studies, especially in offshore foundation designs, to balance structural efficiency with material minimization.

\subsubsection{Analysis}
The distribution of studies by the types of analyses conducted for both original and optimized geometries is summarized in \autoref{table:analysis_type_studies}. For original geometries, modal analysis is the most commonly employed method, appearing in 41\% (13 out of 32) of the studies. Stress and deflection analyses follow closely, each representing 22\% (7 out of 32). Buckling analysis is less common, occurring in 13\% (4 out of 32) of studies, while fatigue analysis is the least studied, present in only 6\% (2 out of 32) of the studies.

\begin{table}[h!]
\scriptsize
\centering
\caption{Number of studies by analysis type: (a) Original Geometry; (b) Optimized Geometry.}
\begin{tabular}{@{}p{0.8cm}p{0.6cm}p{0.6cm}p{1cm}p{4cm}@{}}
\toprule
\textbf{Analysis}        & \textcolor{specialred}{\textbf{Onshore Towers}} & \textcolor{specialblue}{\textbf{Offshore Towers}} & \textcolor{specialcyan}{\textbf{Offshore Foundations}} & \textbf{Total \break Studies} \\ 
\midrule
\multicolumn{4}{l}{\textbf{(a) Original Geometry}}&\\
\midrule
Stress Analysis               & 0 & 1 & 6 & \phantom{1}7 \textcolor{specialblue}{\rule{2mm}{2mm}}\textcolor{specialcyan}{\rule{12mm}{2mm}} \\ \addlinespace[0.1cm]
Deflection Analysis           & 0 & 1 & 6 & \phantom{1}7 \textcolor{specialblue}{\rule{2mm}{2mm}}\textcolor{specialcyan}{\rule{12mm}{2mm}} \\ \addlinespace[0.1cm]
Modal Analysis                & 5 & 2 & 6 & 13 \textcolor{specialred}{\rule{10mm}{2mm}}\textcolor{specialblue}{\rule{4mm}{2mm}}\textcolor{specialcyan}{\rule{12mm}{2mm}} \\ \addlinespace[0.1cm]
Buckling Analysis             & 1 & 1 & 2 & \phantom{1}4 \textcolor{specialred}{\rule{2mm}{2mm}}\textcolor{specialblue}{\rule{2mm}{2mm}}\textcolor{specialcyan}{\rule{4mm}{2mm}} \\ \addlinespace[0.1cm]
Fatigue Analysis              & 0 & 0 & 2 & \phantom{1}2 \textcolor{specialcyan}{\rule{4mm}{2mm}} \\ 
\midrule
\multicolumn{5}{l}{\textbf{(b) Optimized Geometry}} \\
\midrule
Stress Analysis               & 4 & 6 & 6 & 16 \textcolor{specialred}{\rule{8mm}{2mm}}\textcolor{specialblue}{\rule{12mm}{2mm}}\textcolor{specialcyan}{\rule{12mm}{2mm}} \\ \addlinespace[0.1cm]
Deflection Analysis           & 4 & 3 & 4 & 11 \textcolor{specialred}{\rule{8mm}{2mm}}\textcolor{specialblue}{\rule{6mm}{2mm}}\textcolor{specialcyan}{\rule{8mm}{2mm}} \\ \addlinespace[0.1cm]
Modal Analysis                & 8 & 5 & 6 & 19 \textcolor{specialred}{\rule{16mm}{2mm}}\textcolor{specialblue}{\rule{10mm}{2mm}}\textcolor{specialcyan}{\rule{12mm}{2mm}} \\ \addlinespace[0.1cm]
Buckling Analysis             & 5 & 5 & 2 & 12 \textcolor{specialred}{\rule{10mm}{2mm}}\textcolor{specialblue}{\rule{10mm}{2mm}}\textcolor{specialcyan}{\rule{4mm}{2mm}} \\ \addlinespace[0.1cm]
Fatigue Analysis              & 4 & 3 & 3 & 10 \textcolor{specialred}{\rule{8mm}{2mm}}\textcolor{specialblue}{\rule{6mm}{2mm}}\textcolor{specialcyan}{\rule{6mm}{2mm}} \\ 
\bottomrule
\end{tabular}
\label{table:analysis_type_studies}
\end{table}

For optimized geometries, modal analysis remains predominant, applied in 60\% (19 out of 32) of the studies. Stress analysis follows closely at 50\% (16 out of 32). While buckling, deflection, and fatigue analyses are less frequent, their relevance increases in optimized studies, appearing in 38\% (12 out of 32), 34\% (11 out of 32), and 31\% (10 out of 32), respectively.

\begin{figure}[ht!]
	\centering
    \includegraphics[width=0.485\textwidth]{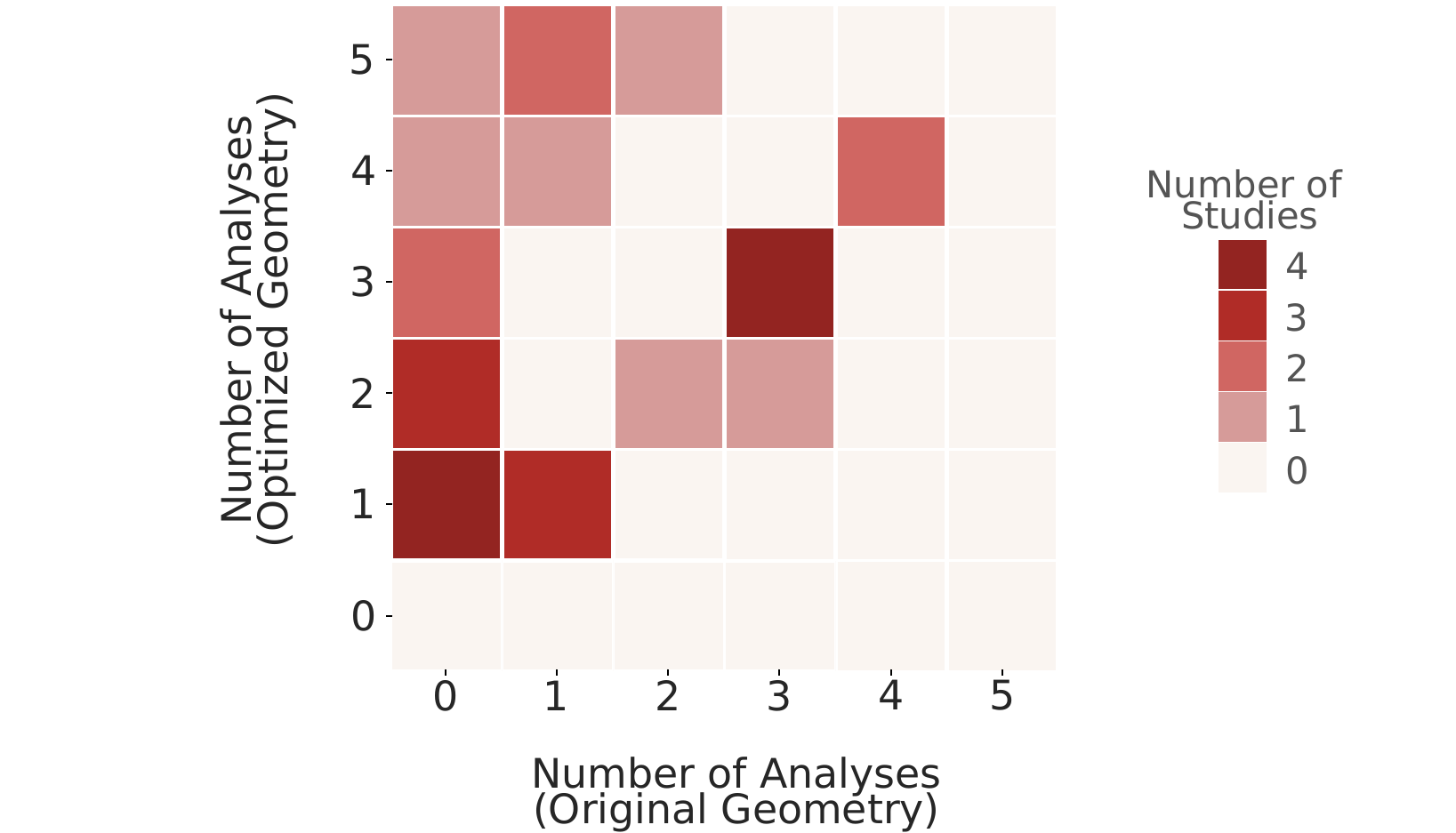}
    \caption{Number of studies based on different combinations of the number of analyses performed on original and optimized geometries.}
    \label{fig:offshore_tower_original_vs_optimized_results}
\end{figure}

A comparison between original and optimized geometries highlights a clear trend: the number of studies for each analysis type increases when transitioning from original to optimized geometry. \autoref{fig:offshore_tower_original_vs_optimized_results}, provides a detailed examination of the number of analysis types conducted on original versus optimized geometries (ranging from 0 to 5). Most studies cluster in three key regions: the top-left, the left side, and along the secondary diagonal.

The top-left region shows that more studies conduct multiple analyses on optimized geometries compared to the original ones. This trend suggests that the original geometry is often considered sufficiently tested, and after optimization, further analyses are necessary to ensure that the refined geometry meets the required design criteria.

The left side of the figure shows a darkening gradient from top to bottom, indicating an increase in study count as the number of analyses on optimized geometries decreases. Although comprehensive analyzes are essential to validate optimized designs across various conditions, this pattern suggests that the complexity and resource intensity of certain analyzes may lead to a reduction in the types conducted, with studies often prioritizing critical aspects such as modal or stress characteristics in optimized geometries.

The secondary diagonal region represents studies in which the same number of analyzes was performed for both original and optimized geometries. This reflects efforts to directly compare the performance of the two geometries.

These findings emphasize the need for further studies that conduct a broader range of analyzes, incorporating more comprehensive evaluations for both original and optimized geometries. A shift towards the top-right region of the figure would indicate a greater density of studies, highlighting the potential for more thorough and multifaceted optimization research.

\subsubsection{Results}
The results of these optimization studies demonstrate significant advancements in the design process, ensuring that structures are robust enough to meet imposed design constraints while optimizing key variables.  

The optimization methods used in these studies effectively reduce both the mass and the cost of tower and foundation structures. For onshore towers, mass reductions range from 6.3\% to 35.5\%, while offshore towers exhibit reductions between 0.8\% and 54.8\%. Offshore foundations show a similar trend, with mass reductions ranging from 7.4\% to 38.2\%. 

These reductions highlight the effectiveness of the optimization process in achieving lighter and more cost-efficient structures without compromising structural integrity.

\section{Wind Turbine Offshore Tower Design: AI-Driven Future Directions} \label{sec:futureprespectives}

While traditional optimization has enhanced OWT tower design and reduced costs, evolving design requirements now call for more adaptive, dynamic solutions where AI can play a transformative role. Despite its expanding presence in engineering, AI’s potential in OWT tower optimization remains less prevalent. AI-driven models offer the potential to move beyond conventional methods, delivering precise, efficient designs that dynamically adapt to complex requirements and real-time data inputs \cite{10.1007/978-3-030-77977-1_36, Alam2024From, 10.1115/1.4064245, Edwards_Man_Ahmed_2024}. Bridging this gap, DT technology seamlessly integrates with AI to enhance and actualize these dynamic solutions in OWT tower design \cite{LENG2021119, LO2021101297, s22145396, alexopoulos2020digital}.

This section explores seven future directions, highlighting how AI can address current challenges, drive innovation, and support a more sustainable and efficient approach to OWT tower design.

\subsection{Datasets for AI in Tower Design} 
\subsubsection{Current Limitations of Datasets for AI in Tower Design} 
AI models for tower design, such as those presented in the following sections, rely on large and detailed datasets. Various datasets support wind turbine research, such as the Fraunhofer dataset \cite{Mostafavi2024} with monitoring data from a 750 W turbine and NREL resources \cite{nrel_wind_data_tools}, including the Wind Resource Database (WRDB) and Metocean Data for offshore conditions. The U.S. Wind Turbine Database (USWTDB) \cite{hoen2018uswtdb} also provides geospatial data on over 70,000 turbines. A broader survey of wind resources, capacities, and open-source datasets is available in \cite{en13184702, https://doi.org/10.1002/we.2766}. However, despite their value for turbine studies, these datasets often lack critical engineering details such as load values, structural dimensions, and material properties, which are essential for tower design. Proprietary datasets with such details are often inaccessible, forcing researchers to rely on costly simulations or incomplete data for AI-driven tower design.

\subsubsection{Current AI Dataset Developments in Engineering Design} 
Several 2D/3D shape datasets offer millions of CAD models, such as TraceParts \cite{traceparts} and Thangs \cite{thangs}, supporting 3D printing, object recognition, and classification. Specialized datasets in engineering design include OpenSketch \cite{10.1145/3355089.3356533} for design sketches, SHIP-D \cite{bagazinski2023shipdshiphulldataset} for ship hulls, Autosurf and Aircraft Lift and Drag \cite{autosurf, Edwards2021DesignFA} for aircraft, and airfoil datasets \cite{doi:10.2514/6.2019-2351, 10.1145/3447548.3467414}. Metamaterial datasets include Truss metamaterials \cite{Zheng2023}, focusing on truss lattices, and METASET \cite{10.1115/1.4048629} for 3D metamaterial unit cells. Multi-modal datasets, integrating data types like images and simulation results, are increasingly common, with examples like DrivAerNet++ \cite{elrefaie2024drivaernetlargescalemultimodalcar} for aerodynamic data on 3D car models and BIKED++ \cite{regenwetter2024bikedmultimodaldataset14} for bicycles \cite{10.1115/1.4063954}. In structural analysis, there are few available datasets, such as simJEB \cite{simjeb} and Mechanical MNIST \cite{MNIST}. SimuStruct \cite{ribeiro2023simustruct} is another large-scale dataset for structural analysis that provides 2D structural parts, specifically plates with holes, along with stress, strain, and displacement data. Despite these efforts, there is an inherent trade-off between the dataset size, its real-world applicability, and its generalizability and current engineering datasets do not match the scale and breadth. 

\subsubsection{Future of Datasets for AI in Tower Design} 
The future of AI in tower design will depend on hybrid datasets combining synthetic and real-world data to address complex multiphysics challenges. Synthetic datasets, generated through physics-based simulations, are becoming more feasible with advancements in computational power, parallelized simulations, GPU acceleration, and the potential of quantum computing \cite{klusch2024quantumartificialintelligencebrief}. These improvements support efficient modeling of phenomena such as large-scale wind loads, fluid-structure interactions, and material degradation. Real-time sensor data, essential for assessing actual damage, provides high-fidelity insights into tower performance and enables structural adjustments in response to evolving wind, ocean, and environmental stresses. Key datasets will include high-resolution wind-ocean interaction data, geometric studies, material fatigue records, and operational data capturing load variations and maintenance activities.

Future datasets for wind tower design should prioritize realism, ensuring they accurately represent actual operational environments, and diversity in design and performance dimensions to enhance AI applications. They should also incorporate active-learning frameworks to dynamically refine data, employ multifidelity modeling for computational efficiency, and extend documentation and metadata standards to enhance usability and reproducibility. Additionally, datasets must support high granularity temporal resolutions to capture dynamic environmental changes, detail advanced process, material, and geometric properties under diverse operational stresses, and integrate predictive maintenance data to improve decision-making processes in tower design.
Multi-modal datasets are also essential for integrating geometry, environmental, structural, and operational data within a unified framework, enhancing predictive accuracy and optimizing. Collaborative open-access initiatives for creating, sharing, and maintaining these resources and research partnerships are critical for generating accessible and scalable datasets for advanced AI models and DT applications in tower design.

\subsection{Surrogate Models in AI for Tower Design}
\label{subsec:ai_speedup_simu}
\subsubsection{Current Limitations of Structural Analysis Methods for Tower Design}
Currently, structural analysis for tower design primarily relies on FEM simulations to calculate structural responses to different loads, as detailed in \autoref{sec:towerloads}. While accurate, FEM is computationally intensive and slow, especially for OWT towers with varying boundary conditions and large geometries. The iterative nature of the design cycle necessitates repeated simulations and adjustments to parameters, making the process time-consuming and demanding on computational resources. This time and resource demand limits thorough exploration of the design space, often resulting in suboptimal outcomes.

\subsubsection{Current AI Developments in Surrogate Models for Structural Analysis}\label{subsec:ai_speedup_simu_ai}
AI, particularly DL, has proven effective in accelerating FEM simulations as surrogate models \cite{kudela_recent_2022}. DNNs enable faster stress predictions and lower computational costs through ROMs \cite{doi:10.1177/1464420721992445, mlp_ribeiro, koeppe_model_2016, schilders_model_2008, group_faster_2020}. Other models, such as CNNs \cite{liang_deep_2018, nie_stress_2019},  conditional GANs \cite{jiang_stressgan_2021}, and spectral shape encoding \cite{mlpcmu}, have enhanced stress analysis however face challenges in generalizing across varying geometries. PINNs improve FEM approximations by embedding governing equations into networks \cite{HAGHIGHAT2021113741}, while GNNs transform FEM meshes into graphs for dynamic predictions, though generalization issues remain \cite{maurizi_predicting_2022, ribeiro2023simustruct}. Operator learning approaches, such as DeepONet \cite{Lu2021} and Fourier Neural Operators (FNOs) \cite{li2021fourier},  offer better efficiency and generalization for partial differential equations. Recently, generative AI models, such as diffusion models and transformers, have shown the ability to predict stress fields in complex geometries \cite{JADHAV2023116343, difusionmarkus}. Additionally, LLMs have automated code generation for FEM, optimizing the simulation process and stress field generation \cite{NI2024102131}.

\subsubsection{Future of Surrogate Models in Structural Analysis for Tower Design}
The future of surrogate models in tower design is likely to focus on refining AI techniques to enhance computational efficiency, accuracy, and generalization across diverse design scenarios. Methods that embed physical laws into the learning (e.g., PINNs and operator learning models) could improve structural analysis by allowing ML models to capture design and performance spaces more efficiently, including variability in material properties and deviations between designed and manufactured geometries. This integration may reduce the reliance on costly, time-consuming simulations by improving both accuracy and convergence speed. However, a key challenge remains in the lack of a unified representation of designs, which results in the need for different types of AI models to handle various data representations such as CAD sequences, point clouds, meshes, voxels, and images. To address this, approaches GNNs are emerging as potent solutions. GNNs can effectively handle mesh representations, enhancing the ability of surrogate models to generalize and provide real-time predictions of structural responses under varying conditions. Generative AI models, such as diffusion models and transformers, are likely to help optimize structural configurations by efficiently exploring design spaces and leveraging existing data to predict performance, potentially reducing design iteration time. Additionally, LLMs could further optimize the code generation process for FEM simulations, improving the integration of physical models and enhancing the accuracy of stress and strain field predictions. These innovations may accelerate physics simulations and design iterations, improve predictive accuracy, and optimize OWT tower designs.

\subsection{AI-Driven Multiphysics Simulations for Tower Design}
\subsubsection{Current Limitations in Multiphysics Simulations}
Multiphysics simulations are critical for accurately capturing the complex interactions in OWT tower design, integrating structural, fluid, and electromagnetic fields. Tools like OpenFAST \cite{Jonkman_2013} provide open-source platforms for multi-fidelity and coupled dynamic simulations. Fluid-structure interaction (FSI), a key component of multiphysics methods, models the mutual influences between fluid and structural forces. While advancements in computational resources have made FSI simulations more accessible, they remain computationally demanding, particularly for floating OWT designs \cite{YAN2016155, doi:10.1080/17445302.2019.1565295, WIEGARD2021102970}. Furthermore, these simulations require advanced algorithms to ensure stability and accuracy across domains, highlighting the need for specialized expertise to handle the integration of diverse physics models \cite{multiphysics_review}.

\subsubsection{Current AI Developments in Multiphysics Simulations}
Advances in AI have significantly improved both the accuracy and speed of physics simulations, with similar progress in multiphysics applications. As discussed in \autoref{subsec:ai_speedup_simu_ai}, DL approaches have been effectively applied to accelerate FEM simulations, with comparable advancements observed in CFD using methods like PINNs \cite{Cai2021}, GNNs \cite{elrefaie2024drivaernetlargescalemultimodalcar}, and diffusion models \cite{Li2024}. An overview of recent ML applications in CFD is provided in \cite{wang2024recentadvancesmachinelearning}. In FSI, where FEM and CFD are coupled, accelerating individual components, such as solving structural dynamics or fluid flow, can significantly enhance overall simulation efficiency. For FSI, various DNN models have been deployed, including ANNs \cite{app10020705}, CNNs \cite{WANG2023367}, combinations of CNNs and Long Short-Term Memory (LSTM) networks \cite{10.1063/5.0183290}, hybrid models integrating CNNs, LSTMs, and PINNs \cite{FAN2024112584}, GNNs \cite{10.1063/5.0182801, NEURIPS2022_59593615}, and DNN-based ROMs \cite{101063_50096432_Han}. Tools like NVIDIA’s Modulus \cite{10.1007/978-3-030-77977-1_36} serves as an open-source platform for developing physics-informed ML models across domains such as CFD, structural analysis, and electromagnetics. While these approaches enhance efficiency, further advances are needed to improve accuracy and generalization in multiphysics scenarios.

\subsubsection{Future of AI in Multiphysics Simulations for Tower Design}
Advancements in AI-driven multiphysics simulations are likely to address challenges in coupled domains, such as FSI, critical for floating OWTs. Physics-Guided, PhysicsInformed, and Physics-Encoded learning methods could play a pivotal role in embedding governing equations across interacting physical fields, enabling more accurate and computationally efficient solutions for coupled simulations. However, the methods may need to work well with higher geometric and physics-complexity and identify ways to capture implicit and explicit requirements. Data-driven GNNs, Transformers, and Generative AI models like diffusion models could also provide ways to enhance scalability by representing interactions between physical phenomena, such as structural mechanics and fluid dynamics. These models hold promise for parameter estimation and exploring design variations in multiphysics contexts. Hybrid AI-numerical approaches may help tackle stability and convergence in non-linear coupled simulations, potentially reducing computational demands and enabling real-time predictive capabilities for multiphysics OWT design.

\subsection{AI-Enhanced Digital Twins for Tower Design}
\subsubsection{Current Limitations of Digital Twins for Tower Design} 
As discussed in \autoref{sec:dt_of_benefits}, Digital Twins (DTs) enable continuous monitoring, predictive maintenance, and enhanced turbine performance, improving reliability, optimizing energy capture, minimizing structural loads through precise wind field mapping, and facilitating real-time fault diagnosis in OWTs. However, as highlighted in \autoref{sec:dt_of_challenges}, current DT models face limitations such as the lack of high-resolution multiphysics simulations and challenges in integrating SCADA and sensor data, which reduces the accuracy of real-time insights. Signal processing, noise management, and fault detection further compromise system reliability. Additionally, limited access to comprehensive public datasets restricts the development and validation of DT models, affecting their effectiveness in predictive maintenance and operational analysis.

\subsubsection{Current AI Developments in Digital Twins for Tower Design}
AI models integrated into DT systems, as outlined in \autoref{sec:dt_of_ai}, advance wind turbine operations by employing techniques such as regression, ANN, PINN, TCN, and DCLSTMNN for SCADA-based monitoring, forecasting, and anomaly detection. These approaches enhance predictive accuracy and facilitate the early identification of structural anomalies, enabling proactive maintenance strategies. Despite these advancements, further research is essential to fully integrate AI with DTs specifically for tower design, with a focus on refining predictive precision, optimizing structural performance, and ensuring adaptability to complex and dynamic environmental conditions.

\subsubsection{Future of AI in Digital Twins for Tower Design}
The future of AI-integrated DTs in OWT tower design is likely to focus on developing adaptive, data-driven frameworks that address the limitations of current real-time simulations and operational monitoring. Future DTs could incorporate high-resolution multiphysics simulations to improve FSI modeling and address environmental variability through techniques like mesh refinement and parallel processing. Enhanced data fusion methods may improve the integration of SCADA and sensor data, enabling more accurate real-time operational insights. AI algorithms could ensure data quality by reducing noise and minimizing errors in noisy or incomplete datasets, supporting reliable performance. A key advancement may involve the application of principles from Dynamic Data Driven Application Systems (DDDAS) \cite{blasch2022handbook} into OWT DTs, enabling bidirectional data flows between sensors and DTs. In this setup, the DT can continuously refines its model based on sensor inputs, while the model itself could adaptively optimize sensor configurations, enhancing system performance. This dynamic interaction may allow for real-time adjustments to the tower design based on operational conditions, improving fault detection and enabling proactive predictive maintenance. Additionally, open-access data initiatives could expand the availability of high-quality datasets, improving predictive accuracy and extending the operational lifespan of OWT structures.

\subsection{Generative AI for Tower Design Optimization} \label{sec:generative_ai_tower_design}
\subsubsection{Current Limitations of Tower Design Optimization}  
Current tower design optimization focuses on achieving essential performance objectives, such as maximizing structural resilience and minimizing mass, often relying on computationally intensive simulations. Design methodologies are detailed in \autoref{sec:tower_desing_process}, with specific optimization strategies for wind turbines in \autoref{sec:opt_wt} and commonly used optimization algorithms in \autoref{sec:opt_alg_wt}. However, these methods face significant limitations, including high computational costs and slow iterative processes, which restrict the number of feasible design iterations. Furthermore, traditional approaches often explore a narrow design space, limiting the discovery of innovative configurations and relying on simplifying assumptions that reduce solution diversity and fidelity to complex real-world conditions.

\subsubsection{Current Generative AI Developments for Engineering Design}
Generative AI provides effective solutions by reducing computational demands through efficient simulations and expanding design exploration. Despite conventional data constraints, it can generate diverse configurations, uncovering innovative solutions beyond traditional methods. Deep generative models, such as NNs, GANs, VAEs, and reinforcement learning frameworks, synthesize new designs and have demonstrated potential in structural optimization, materials discovery, and shape generation \cite{10.1115/1.4053859}. Recently, diffusion models have emerged as transformative tools in engineering design, iteratively refining initial structures into optimized geometries \cite{HERRON2024103707}. For instance, \citet{Mazé_Ahmed_2023} showed that diffusion models outperform GANs in optimization tasks, while other studies applied them to multi-objective ship hull design \cite{jmse11122215} and the generation of editable 3D CAD models \cite{alam2024gencadimageconditionedcomputeraideddesign}. Furthermore, \citet{NEURIPS2023_a2fe4bb5} highlight that combining diffusion models with optimization techniques enhances constrained design generation and efficiency. Generative models could potentially be used to augment existing optimization approaches in OWT design by amortizing the cost of optimization and enabling efficient design space exploration.

\subsubsection{Future of Generative AI in Tower Design Optimization}
Generative AI has the potential to significantly influence tower design optimization by addressing key design constraints, including fatigue resistance, material strength, stress distribution, and buckling stability. Advanced GANs and VAEs, utilizing extensive datasets, are likely to explore diverse design solutions tailored to the unique demands of tower structures and may enable more accurate modeling of complex structural behaviors, such as load distribution and dynamic responses. Diffusion models, recognized for their iterative refinement capabilities, could play a crucial role in transforming initial designs into optimized geometries by leveraging probabilistic approaches to explore complex design spaces. To fully realize these capabilities, generative AI must overcome computational cost barriers to enable rapid design iterations and effectively manage intricate structural complexities. Advancements in training methodologies and model interpretability will likely be pivotal in ensuring that generated designs achieve superior performance, resilience, and adaptability to the operational demands of OWTs.

\subsection{AI-Driven Material and Manufacturing Innovations for Tower Design}
\subsubsection{Current Limitations in Materials and Manufacturing for Tower Design}
OWT tower design faces critical material and manufacturing challenges that constrain structural integrity and scalability. Defects introduced during manufacturing or transportation can act as stress concentrators, weakening critical structural components and increasing susceptibility to buckling \cite{WAGNER2025112577} and fatigue \cite{YETER2022110271}. These challenges are magnified by the limitations of current manufacturing processes, which lack the precision and flexibility needed to produce the complex geometries generated by generative AI, as detailed in \autoref{sec:generative_ai_tower_design}. Moreover, the development of advanced designs is hindered by the insufficient availability of lighter materials \cite{Li_Lu_2014}, sustainable alternatives \cite{MOROZOVSKA2024142578}, and corrosion-resistant methods essential for offshore environments \cite{JUHL2024107998}. Together, these deficiencies in materials and manufacturing processes limit the scalability of lighter, more resilient designs required to advance OWT structures, as discussed in \autoref{sec:upscaling}.

\subsubsection{Current AI Developments in Material Discovery and Manufacturing}
AI and ML are revolutionizing material discovery by predicting material properties with precision, designing novel materials, and uncovering complex material behaviors that surpass the capabilities of traditional approaches \cite{D0MH01451F}. In manufacturing, AI enhances defect detection during inspections, reducing the risk of defective structures entering the market and mitigating premature failures \cite{Zheng2021}. These advancements also extend to 3D printing, where AI-driven insights improve precision and quality control, enabling the production of resilient, high-performance components that meet stricter manufacturing standards. For a detailed review of ML applications in 3D printing, see \cite{Goh2021}.

\subsubsection{Future of AI for Material Discovery and Manufacturing in Tower Design}
AI integration with 3D printing is likely be essential for creating complex tower geometries that traditional methods may not be able to achieve, enabling customized and high-precision structures. In materials research, AI will drive the discovery of high-strength and lightweight materials to reduce tower weight while preserving structural integrity, which is critical for scaling OWTs. As AI-generated designs may feature more open or exposed structures, advanced models will likely be critical for identifying materials with enhanced corrosion resistance to withstand the harsh marine environments of offshore installations. Additionally, AI-driven insights could promote the development of sustainable materials to minimize environmental impact. By combining optimized geometries with materials that provide durability, corrosion resistance, low weight, and sustainability, AI may enable the creation of long-lasting and low-maintenance structures for the future of tower design.

\subsection{AI-Tailored Tower Design in Wind Farms}
\subsubsection{Current Limitations of Tower Design in Wind Farms}
An alternative to isolated tower design is to consider the wind farm context, as discussed in \autoref{sec:wind_farm}. As noted in \autoref{sec:wf_opt}, efforts to optimize both turbine layouts and tower designs face significant computational challenges, particularly when simulating wake effects across large farms. These challenges arise from the complex interactions between turbines, where each wake influences downstream performance. The complexity is further amplified by the increasing number of design variables, such as tower height, base diameter, taper ratio, and wall thickness. Fully optimizing these factors at scale becomes computationally prohibitive, particularly when high-fidelity methods like CFD are employed. Consequently, designers often rely on simplified models with a reduced number of design parameters, such as tower height and base and top diameters. While computationally efficient, these simplifications often lead to suboptimal designs and higher operational costs, underscoring the need for advanced, integrated optimization approaches.

\subsubsection{Current AI Developments in Tower Design in Wind Farms}
ML models, such as ANNs, PINNs, GANs, and GNNs, have demonstrated significant potential in wind farm design by accelerating wind flow simulations and optimizing turbine layouts, as discussed in \autoref{sec:wf_ai}. These AI-driven approaches effectively capture complex relationships between input parameters and performance metrics, substantially reducing the computational costs associated with traditional CFD methods and enabling more efficient layout optimization. However, their application to refining turbine tower designs within wind farms, particularly with layout optimization, remains limited and requires further investigation.

\subsubsection{Future of AI in Tower Design in Wind Farms}
AI is poised to revolutionize wind farm tower design by optimizing key parameters, such as height, base diameter, wall thickness, and taper ratio, while addressing the complexities of wake effects. Future advancements are likely to integrate high-fidelity CFD simulations with ML techniques for precise and efficient wake modeling, potentially overcoming current computational constraints and the limitations of isolated tower design approaches. AI-driven DTs could play a central role by dynamically adapting tower-specific configurations to localized environmental conditions and wake interactions using real-time sensor data and predictive algorithms. These feedback loops may not only refine operational strategies but also guide the iterative improvement of future tower designs to meet site-specific and long-term performance requirements. By combining dynamic modeling with adaptive frameworks, AI may transform isolated methodologies into resilient, data-driven systems that reduce wake-induced losses, increase energy capture, and enhance material efficiency for sustainable wind farm operations.

\section{Conclusions}
\textbf{\textit{Overview of This Paper:}} 
This paper presents an in-depth review of recent advancements and challenges and AI-driven future directions in design optimizing of OWT structures, with a primary focus on towers. 

\textbf{\textit{Comprehensive Background on OWT Tower Design:}}
This review outlined key aspects of tower design optimization, covering design types, design methodologies, analysis techniques, monitoring components, DT technologies, software tools, standards, reference turbines, economic factors, and optimization strategies,  presenting a thorough foundation for current research.

\textbf{\textit{The Tower’s Role in OWT Upscaling:}} 
As energy demands rise, larger turbines are needed to boost power generation and lower the LCoE, which represents the average cost of electricity over a project's lifespan. However, scaling up turbine sizes introduces engineering challenges, especially in designing supporting structures like towers. Towers must withstand the increased loads of larger turbines while ensuring structural integrity, cost efficiency, and transportability, making them crucial to offshore wind project success.

\textbf{\textit{Tower Optimization Studies Review:}} 
This review analyzes 32 studies on tower optimization, with a focus on tubular steel towers using the NREL 5 MW model. Ansys (FEA) is widely used, addressing complex load cases including gravitational, wind, aerodynamic, and hydrodynamic forces, often using DLC 1.2. Mass minimization is a common objective, with design variables such as tower diameter and section thickness, and constraints like mechanical stress, natural frequencies, deflection, and buckling load factors. Modal analyses are typically conducted on initial designs, while optimized designs undergo more extensive evaluations, such as stress and modal analysis. Results consistently show reductions in mass and cost, highlighting the potential of advanced algorithms for OWT tower designs, though further advancements are needed.

\textbf{\textit{AI-Driven Future Directions in Tower Design:}}
AI-driven approaches offer solutions to current challenges in tower design, with future directions focusing on dataset generation, development of surrogate models, multiphysics simulations for analyzing complex interactions, DTs for predictive maintenance, generative AI for innovative designs, advancements in materials and manufacturing, and optimization of tower design within wind farms. These strategies aim to enhance design efficiency, lower computational costs, and enable the creation of resilient and scalable towers essential for advancing OWT.

\textbf{\textit{Final Remarks on OWT Tower Design:}} 
This paper emphasizes the vital role of tower optimization in advancing offshore wind energy. By combining current methodologies with AI-driven strategies, it addresses challenges like complex loads and material efficiency, while also identifying opportunities to improve scalability, resilience, and cost-effectiveness. Through its comprehensive approach to the state-of-the-art and future directions, this article serves as a valuable resource for guiding further advancements. These insights support the development of larger, more sustainable turbines, accelerating the transition to renewable energy.

\section*{CRediT authorship contribution statement}

\textbf{João Alves Ribeiro}: Conceptualization, Methodology, Formal Analysis, Investigation, Writing -- Original Draft, Visualization. \textbf{Bruno Alves Ribeiro}: Conceptualization, Methodology, Investigation, Validation, Writing -- Review \& Editing. \textbf{Francisco Pimenta}: Methodology, Investigation, Validation, Writing -- Review \& Editing. \textbf{Sérgio M. O. Tavares}: Conceptualization, Methodology, Validation, Investigation, Writing -- Review \& Editing, Supervision. \textbf{Jie Zhang}: Conceptualization, Methodology, Validation, Writing -- Review \& Editing, Supervision. \textbf{Faez Ahmed}: Conceptualization, Methodology, Validation, Investigation, Writing -- Review \& Editing, Supervision.

\section*{Declaration of competing interest}
The authors declare that they have no known competing financial interests or personal relationships that could have appeared to influence the work reported in this paper.

\section*{Acknowledgements}
\textbf{João Alves Ribeiro} acknowledges funding from the Luso-American Development Foundation (FLAD) and the doctoral grant SFRH/BD/151362/2021 (DOI: \href{https://doi.org/10.54499/SFRH/BD/151364/2021}{10.54499/SFRH/BD/151364/2021} (accessed on November 18, 2024)), financed by the Portuguese Foundation for Science and Technology (FCT), Ministério da Ciência, Tecnologia e Ensino Superior (MCTES), Portugal, with funds from the State Budget (OE), European Social Fund (ESF), and PorNorte under the MIT Portugal Program.

\textbf{Bruno Alves Ribeiro} acknowledges financial support from FCT through the doctoral grant 2021/08659/BD.

\textbf{Francisco Pimenta} acknowledges the financial support for project 2022.08120.PTDC, M4WIND (DOI: \href{https://doi.org/10.54499/2022.08120.PTDC}{10.54499/2022.08120.PTDC} (accessed on November 18, 2024)), funded by national funds through FCT/MCTES (PIDDAC), and for UID/ECI/04708/2020-CONSTRUCT-Instituto de I\&D em Estruturas e Construções, also funded by national funds through FCT/MCTES (PIDDAC).

\appendix
\onecolumn
\begin{landscape}
\section{Optimization Studies: Onshore Wind Turbine Towers} \label{A:a}

\scriptsize	
\centering
\begin{longtable}{@{}p{0.4cm}p{1.8cm}p{1.8cm}p{3cm}p{2.3cm}p{3cm}p{3cm}p{3cm}p{2.1cm}@{}}
\caption{Literature review of optimization studies on onshore wind turbine towers.}
\label{table:lr_towers_on}\\  
\toprule
\textbf{Study}  & \textbf{Turbine} & \textbf{Software} & \textbf{Loads} & \textbf{Method}  & \textbf{Design Variables} &  \textbf{Design Constraints}  & \textbf{Analysis} & \textbf{Results} \\ \midrule

\cite{Gencturk_Attar_Tort_2015} &
\begin{minipage}[t]{1.8cm}
\begin{itemize}[label={-}, leftmargin=*, topsep=0pt, partopsep=0pt, parsep=0pt, itemsep=0pt]
    \item Onshore: Lattice tower.
    \item 100-400 kW.
\end{itemize}
\end{minipage} &
\begin{minipage}[t]{1.8cm}
\begin{itemize}[label={-}, leftmargin=*, topsep=0pt, partopsep=0pt, parsep=0pt, itemsep=0pt]
    \item PLS-TOWER (modeling).
    \item ZEUS NL (FEA).
\end{itemize}
\end{minipage} &
\begin{minipage}[t]{3cm}
\begin{itemize}[label={-}, leftmargin=*, topsep=0pt, partopsep=0pt, parsep=0pt, itemsep=0pt]
    \item Types: earthquake, wind, and gravitational.
\end{itemize}
\end{minipage} &
\begin{minipage}[t]{2.3cm}
\begin{itemize}[label={-}, leftmargin=*, topsep=0pt, partopsep=0pt, parsep=0pt, itemsep=0pt]
    \item TS. 
    \item Objective function: minimize cost.
\end{itemize}
\end{minipage} &
\begin{minipage}[t]{3cm}
\begin{itemize}[label={-}, leftmargin=*, topsep=0pt, partopsep=0pt, parsep=0pt, itemsep=0pt]
    \item Section sizes (3 vars.).
    \item Section combinations.
\end{itemize}
\end{minipage} &
\begin{minipage}[t]{3cm}
\begin{itemize}[label={-}, leftmargin=*, topsep=0pt, partopsep=0pt, parsep=0pt, itemsep=0pt]
    \item Geometry limits.
\end{itemize}
\end{minipage} &
\begin{minipage}[t]{3cm}
\begin{itemize}[label={-}, leftmargin=*, topsep=0pt, partopsep=0pt, parsep=0pt, itemsep=0pt]
    \item Original: modal analysis. 
    \item Optimized: modal analysis.
\end{itemize}
\end{minipage} &
\begin{minipage}[t]{2.1cm}
\begin{itemize}[label={-}, leftmargin=*, topsep=0pt, partopsep=0pt, parsep=0pt, itemsep=0pt]
    \item 14.6-35.5\% mass reduction.
    \item 20\% cost reduction.
\end{itemize}
\end{minipage} 
\\ 
\addlinespace[0.2cm]

\cite{Gencturk_Attar_2012} &
\begin{minipage}[t]{1.8cm}
\begin{itemize}[label={-}, leftmargin=*, topsep=0pt, partopsep=0pt, parsep=0pt, itemsep=0pt]
    \item Onshore: Lattice tower.
    \item 100 kW.
\end{itemize}
\end{minipage} &
\begin{minipage}[t]{1.8cm}
\begin{itemize}[label={-}, leftmargin=*, topsep=0pt, partopsep=0pt, parsep=0pt, itemsep=0pt]
    \item PLS-TOWER (modeling).
    \item ZEUS NL (FEA).
\end{itemize}
\end{minipage} &
\begin{minipage}[t]{3cm}
\begin{itemize}[label={-}, leftmargin=*, topsep=0pt, partopsep=0pt, parsep=0pt, itemsep=0pt]
    \item Types: earthquake, wind, and gravitational.
\end{itemize}
\end{minipage} &
\begin{minipage}[t]{2.3cm}
\begin{itemize}[label={-}, leftmargin=*, topsep=0pt, partopsep=0pt, parsep=0pt, itemsep=0pt]
    \item TS. 
    \item Objective function: minimize cost.
\end{itemize}
\end{minipage} &
\begin{minipage}[t]{3cm}
\begin{itemize}[label={-}, leftmargin=*, topsep=0pt, partopsep=0pt, parsep=0pt, itemsep=0pt]
    \item Section sizes (3 vars.).
    \item Section combinations.
\end{itemize}
\end{minipage} &
\begin{minipage}[t]{3cm}
\begin{itemize}[label={-}, leftmargin=*, topsep=0pt, partopsep=0pt, parsep=0pt, itemsep=0pt]
    \item Geometry limits.
\end{itemize}
\end{minipage} &
\begin{minipage}[t]{3cm}
\begin{itemize}[label={-}, leftmargin=*, topsep=0pt, partopsep=0pt, parsep=0pt, itemsep=0pt]
    \item Optimized: modal analysis.
\end{itemize}
\end{minipage} &
\begin{minipage}[t]{2.1cm}
\begin{itemize}[label={-}, leftmargin=*, topsep=0pt, partopsep=0pt, parsep=0pt, itemsep=0pt]
    \item 22.5\% mass reduction.
    \item Key design constraint: fatigue.
\end{itemize}
\end{minipage} 
\\ 
\addlinespace[0.2cm]

\cite{Gencturk_Attar_Tort_2014} &
\begin{minipage}[t]{1.8cm}
\begin{itemize}[label={-}, leftmargin=*, topsep=0pt, partopsep=0pt, parsep=0pt, itemsep=0pt]
    \item Onshore: Lattice tower.
    \item 100-400 kW.
\end{itemize}
\end{minipage} &
\begin{minipage}[t]{1.8cm}
\begin{itemize}[label={-}, leftmargin=*, topsep=0pt, partopsep=0pt, parsep=0pt, itemsep=0pt]
    \item PLS-TOWER (modeling).
    \item ZEUS NL (FEA).
\end{itemize}
\end{minipage} &
\begin{minipage}[t]{3cm}
\begin{itemize}[label={-}, leftmargin=*, topsep=0pt, partopsep=0pt, parsep=0pt, itemsep=0pt]
    \item Types: earthquake, wind, and gravitational.
\end{itemize}
\end{minipage} &
\begin{minipage}[t]{2.3cm}
\begin{itemize}[label={-}, leftmargin=*, topsep=0pt, partopsep=0pt, parsep=0pt, itemsep=0pt]
    \item TS. 
    \item Objective function: minimize cost.
\end{itemize}
\end{minipage} &
\begin{minipage}[t]{3cm}
\begin{itemize}[label={-}, leftmargin=*, topsep=0pt, partopsep=0pt, parsep=0pt, itemsep=0pt]
    \item Section sizes (3 vars.).
    \item Section combinations (8 vars.).
\end{itemize}
\end{minipage} &
\begin{minipage}[t]{3cm}
\begin{itemize}[label={-}, leftmargin=*, topsep=0pt, partopsep=0pt, parsep=0pt, itemsep=0pt]
    \item Geometry limits.
\end{itemize}
\end{minipage} & &
\begin{minipage}[t]{2.1cm}
\begin{itemize}[label={-}, leftmargin=*, topsep=0pt, partopsep=0pt, parsep=0pt, itemsep=0pt]
    \item 30\% mass reduction.
\end{itemize}
\end{minipage} 
\\ 
\addlinespace[0.2cm]

\cite{de_Barros_Ribeiro_2021} &
\begin{minipage}[t]{1.8cm}
\begin{itemize}[label={-}, leftmargin=*, topsep=0pt, partopsep=0pt, parsep=0pt, itemsep=0pt]
    \item Onshore: Prestressed concrete tower.
    \item 5 MW.
\end{itemize}
\end{minipage} &
\begin{minipage}[t]{1.8cm}
\begin{itemize}[label={-}, leftmargin=*, topsep=0pt, partopsep=0pt, parsep=0pt, itemsep=0pt]
    \item Ansys (FEA).
\end{itemize}
\end{minipage} &
\begin{minipage}[t]{3cm}
\begin{itemize}[label={-}, leftmargin=*, topsep=0pt, partopsep=0pt, parsep=0pt, itemsep=0pt]
    \item Types: aerodynamic, wind, and gravitational. 
    \item Values: 2 cases.
\end{itemize}
\end{minipage} &
\begin{minipage}[t]{2.3cm}
\begin{itemize}[label={-}, leftmargin=*, topsep=0pt, partopsep=0pt, parsep=0pt, itemsep=0pt]
    \item GA. 
    \item Objective function: minimize mass.
\end{itemize}
\end{minipage} &
\begin{minipage}[t]{3cm}
\begin{itemize}[label={-}, leftmargin=*, topsep=0pt, partopsep=0pt, parsep=0pt, itemsep=0pt]
    \item Top/bottom tower thickness (2 vars.). 
    \item Top/bottom tower diameter (2 vars.).
    \item Steel cables number and radius (2 vars.).
\end{itemize}
\end{minipage} &
\begin{minipage}[t]{3cm}
\begin{itemize}[label={-}, leftmargin=*, topsep=0pt, partopsep=0pt, parsep=0pt, itemsep=0pt]
    \item Geometry limits. 
    \item Maximum stress. 
    \item Minimum buckling load multiplier. 
    \item Minimum fatigue life. 
    \item Frequency limits.
\end{itemize}
\end{minipage} &
\begin{minipage}[t]{3cm}
\begin{itemize}[label={-}, leftmargin=*, topsep=0pt, partopsep=0pt, parsep=0pt, itemsep=0pt]
    \item Optimized: stress, deflection, modal, buckling, and fatigue analysis.
\end{itemize}
\end{minipage} 
& 
\begin{minipage}[t]{2.1cm}
\begin{itemize}[label={-}, leftmargin=*, topsep=0pt, partopsep=0pt, parsep=0pt, itemsep=0pt]
    \item Reduction of stress concentrations. 
    \item Cost reduction in construction.
\end{itemize}
\end{minipage} 
\\ 
\addlinespace[0.2cm]

\cite{Chen_Yang_Ma_Li_2016} &
\begin{minipage}[t]{1.8cm}
\begin{itemize}[label={-}, leftmargin=*, topsep=0pt, partopsep=0pt, parsep=0pt, itemsep=0pt]
    \item Onshore: Lattice/tubular steel tower.
    \item 1.5 MW.
\end{itemize}
\end{minipage} &
\begin{minipage}[t]{1.8cm}
\begin{itemize}[label={-}, leftmargin=*, topsep=0pt, partopsep=0pt, parsep=0pt, itemsep=0pt]
    \item Ansys (FEA). 
    \item MATLAB (O).
\end{itemize}
\end{minipage} &
\begin{minipage}[t]{3cm}
\begin{itemize}[label={-}, leftmargin=*, topsep=0pt, partopsep=0pt, parsep=0pt, itemsep=0pt]
    \item Types: aerodynamic, wind, and gravitational.
    \item Values: 17 cases.
\end{itemize}
\end{minipage} &
\begin{minipage}[t]{2.3cm}
\begin{itemize}[label={-}, leftmargin=*, topsep=0pt, partopsep=0pt, parsep=0pt, itemsep=0pt]
    \item PSO. \newline
    \item Objective function: minimize lattice mass.
\end{itemize}
\end{minipage} &
\begin{minipage}[t]{3cm}
\begin{itemize}[label={-}, leftmargin=*, topsep=0pt, partopsep=0pt, parsep=0pt, itemsep=0pt]
    \item Bottom width (1 var.). 
    \item Component width (15 vars.). 
    \item Component thickness (12 vars.).
\end{itemize}
\end{minipage} &
\begin{minipage}[t]{3cm}
\begin{itemize}[label={-}, leftmargin=*, topsep=0pt, partopsep=0pt, parsep=0pt, itemsep=0pt]
    \item Geometry limits. 
    \item Maximum deflection. 
    \item Maximum stress. 
    \item Frequency limits.
\end{itemize}
\end{minipage} &
&
\begin{minipage}[t]{2.1cm}
\begin{itemize}[label={-}, leftmargin=*, topsep=0pt, partopsep=0pt, parsep=0pt, itemsep=0pt]
    \item Fewer joints and simpler configuration. 
    \item Reduced total weight.
\end{itemize}
\end{minipage} 
\\ 
\addlinespace[0.2cm]

\cite{Cheng_Zhao_Qi_Zhou_2024} &
\begin{minipage}[t]{1.8cm}
\begin{itemize}[label={-}, leftmargin=*, topsep=0pt, partopsep=0pt, parsep=0pt, itemsep=0pt]
    \item Onshore: Tubular steel/concrete hybrid tower. 
    \item 5 MW.
\end{itemize}
\end{minipage} &
\begin{minipage}[t]{1.8cm}
\begin{itemize}[label={-}, leftmargin=*, topsep=0pt, partopsep=0pt, parsep=0pt, itemsep=0pt]
    \item OpenSees (FEA).
\end{itemize}
\end{minipage} &
\begin{minipage}[t]{3cm}
\begin{itemize}[label={-}, leftmargin=*, topsep=0pt, partopsep=0pt, parsep=0pt, itemsep=0pt]
    \item Types: aerodynamic, wind, and gravitational.
\end{itemize}
\end{minipage} &
\begin{minipage}[t]{2.3cm}
\begin{itemize}[label={-}, leftmargin=*, topsep=0pt, partopsep=0pt, parsep=0pt, itemsep=0pt]
    \item GA and DE.
    \item Objective function: minimize cost.
\end{itemize}
\end{minipage} &
\begin{minipage}[t]{3cm}
\begin{itemize}[label={-}, leftmargin=*, topsep=0pt, partopsep=0pt, parsep=0pt, itemsep=0pt]
    \item Top/bottom concrete tower diameter (2 vars.). 
    \item Top/bottom concrete tower thickness (2 vars.). \newline
    \item Top/bottom steel tower diameter (2 vars.). 
    \item Top/bottom steel tower thickness (2 vars.). 
    \item Concrete tower height (1 var.).
\end{itemize}
\end{minipage} &
\begin{minipage}[t]{3cm}
\begin{itemize}[label={-}, leftmargin=*, topsep=0pt, partopsep=0pt, parsep=0pt, itemsep=0pt]
    \item Geometry limits. 
    \item Maximum deflection. 
    \item Maximum stress. 
    \item Minimum buckling load multiplier. 
    \item Minimum fatigue life. 
    \item Frequency limits.
\end{itemize}
\end{minipage} &

&
\begin{minipage}[t]{2.1cm}
\begin{itemize}[label={-}, leftmargin=*, topsep=0pt, partopsep=0pt, parsep=0pt, itemsep=0pt]
    \item 11\% cost reduction in construction. 
    \item Lowest cost: 65\% tower in prestressed concrete.
\end{itemize}
\end{minipage} 
\\ 
\addlinespace[0.2cm]

\cite{Li_Chen_Xu_Ge_2021} &
\begin{minipage}[t]{1.8cm}
\begin{itemize}[label={-}, leftmargin=*, topsep=0pt, partopsep=0pt, parsep=0pt, itemsep=0pt]
    \item Onshore: Tubular steel/concrete hybrid tower.     \item 2 MW.
\end{itemize}
\end{minipage} &
&
\begin{minipage}[t]{3cm}
\begin{itemize}[label={-}, leftmargin=*, topsep=0pt, partopsep=0pt, parsep=0pt, itemsep=0pt]
    \item Types: earthquake, aerodynamic, wind, and gravitational.
\end{itemize}
\end{minipage} &
\begin{minipage}[t]{2.3cm}
\begin{itemize}[label={-}, leftmargin=*, topsep=0pt, partopsep=0pt, parsep=0pt, itemsep=0pt]
    \item Parallel Updated PSO.
    \item Objective function: minimize LCOE.
\end{itemize}
\end{minipage} &
\begin{minipage}[t]{3cm}
\begin{itemize}[label={-}, leftmargin=*, topsep=0pt, partopsep=0pt, parsep=0pt, itemsep=0pt]
    \item Top/bottom concrete tower diameter (2 vars.). 
    \item Top/bottom concrete tower thickness (2 vars.). 
    \item Prestressed reinforcement area (1 var.). 
    \item Top/bottom steel tower diameter (2 vars.). 
    \item Steel section thickness (1 var.). 
    \item Steel segments (1 var.). 
    \item Concrete/steel section length (2 vars.).
\end{itemize}
\end{minipage} &
\begin{minipage}[t]{3cm}
\begin{itemize}[label={-}, leftmargin=*, topsep=0pt, partopsep=0pt, parsep=0pt, itemsep=0pt]
    \item Geometry limits. 
    \item Maximum deflection.
    \item Maximum stress.
    \item Minimum buckling load multiplier. 
    \item Minimum fatigue life.
    \item Frequency limits.
\end{itemize}
\end{minipage} &
\begin{minipage}[t]{3cm}
\begin{itemize}[label={-}, leftmargin=*, topsep=0pt, partopsep=0pt, parsep=0pt, itemsep=0pt]
    \item Original: modal analysis.
    \item Optimized: modal analysis.
\end{itemize}
\end{minipage} &
\begin{minipage}[t]{2.1cm}
\begin{itemize}[label={-}, leftmargin=*, topsep=0pt, partopsep=0pt, parsep=0pt, itemsep=0pt]
    \item 82\% steel reduction, 56\% prestressed concrete increase. 
    \item 60-110\% optimization efficiency improvement. 
    \item Steel segment should be 30\% of tower height.
\end{itemize}
\end{minipage} 
\\ 
\addlinespace[0.2cm]

\cite{Dong_Zheng_Li_He_2022} &
\begin{minipage}[t]{1.8cm}
\begin{itemize}[label={-}, leftmargin=*, topsep=0pt, partopsep=0pt, parsep=0pt, itemsep=0pt]
    \item Onshore: Bionic bamboo tower. 
    \item 2 MW.
\end{itemize}
\end{minipage} &
\begin{minipage}[t]{1.8cm}
\begin{itemize}[label={-}, leftmargin=*, topsep=0pt, partopsep=0pt, parsep=0pt, itemsep=0pt]
    \item Bladed (AHSE).
    \item Ansys (FEA).
\end{itemize}
\end{minipage} &
\begin{minipage}[t]{3cm}
\begin{itemize}[label={-}, leftmargin=*, topsep=0pt, partopsep=0pt, parsep=0pt, itemsep=0pt]
    \item Types: aerodynamic, wind, and gravitational.
    \item Ultimate design load case: DLC 1.3, 3.2, and 6.1 (IEC61400-3).
\end{itemize}
\end{minipage} &
\begin{minipage}[t]{2.3cm}
\begin{itemize}[label={-}, leftmargin=*, topsep=0pt, partopsep=0pt, parsep=0pt, itemsep=0pt]
    \item KM and Nondominated Sorting GA-II. 
    \item Objective function: minimize mass and displacement.
\end{itemize}
\end{minipage} &
\begin{minipage}[t]{3cm}
\begin{itemize}[label={-}, leftmargin=*, topsep=0pt, partopsep=0pt, parsep=0pt, itemsep=0pt]
    \item Section thickness (1 var.). 
    \item Section radius (3 vars.).
\end{itemize}
\end{minipage} &
\begin{minipage}[t]{3cm}
\begin{itemize}[label={-}, leftmargin=*, topsep=0pt, partopsep=0pt, parsep=0pt, itemsep=0pt]
    \item Geometry limits.
    \item Maximum stress. 
    \item Frequency limits.
\end{itemize}
\end{minipage} &
\begin{minipage}[t]{3cm}
\begin{itemize}[label={-}, leftmargin=*, topsep=0pt, partopsep=0pt, parsep=0pt, itemsep=0pt]
    \item Original: buckling analysis. 
    \item Optimized: buckling analysis.
\end{itemize}
\end{minipage} &
\begin{minipage}[t]{2.1cm}
\begin{itemize}[label={-}, leftmargin=*, topsep=0pt, partopsep=0pt, parsep=0pt, itemsep=0pt]
    \item Increase stability.
    \item Reduced displacement. 
    \item Reduced total weight.
\end{itemize}
\end{minipage} 
\\ 
\addlinespace[0.2cm]

\cite{Al-Sanad_Parol_Wang_Kolios_2023} &
\begin{minipage}[t]{1.8cm}
\begin{itemize}[label={-}, leftmargin=*, topsep=0pt, partopsep=0pt, parsep=0pt, itemsep=0pt]
    \item Onshore: Tubular steel tower. 
    \item 2 MW.
\end{itemize}
\end{minipage} &
\begin{minipage}[t]{1.8cm}
\begin{itemize}[label={-}, leftmargin=*, topsep=0pt, partopsep=0pt, parsep=0pt, itemsep=0pt]
    \item OpenFAST (AHSE). 
    \item Ansys (FEA).
\end{itemize}
\end{minipage} &
\begin{minipage}[t]{3cm}
\begin{itemize}[label={-}, leftmargin=*, topsep=0pt, partopsep=0pt, parsep=0pt, itemsep=0pt]
    \item Types: aerodynamic, wind, and gravitational. 
    \item Values: 1 case.
\end{itemize}
\end{minipage} &
\begin{minipage}[t]{2.3cm}
\begin{itemize}[label={-}, leftmargin=*, topsep=0pt, partopsep=0pt, parsep=0pt, itemsep=0pt]
    \item GA. \newline
    \item Objective function: minimize mass.
\end{itemize}
\end{minipage} &
\begin{minipage}[t]{3cm}
\begin{itemize}[label={-}, leftmargin=*, topsep=0pt, partopsep=0pt, parsep=0pt, itemsep=0pt]
    \item Section thickness (6 vars.).
    \item Top/bottom tower diameter (2 vars.).
\end{itemize}
\end{minipage} &
\begin{minipage}[t]{3cm}
\begin{itemize}[label={-}, leftmargin=*, topsep=0pt, partopsep=0pt, parsep=0pt, itemsep=0pt]
    \item Geometry limits. 
    \item Maximum deflection. 
    \item Maximum stress. 
    \item Minimum buckling load multiplier. 
    \item Frequency limits.
\end{itemize}
\end{minipage} & &
\begin{minipage}[t]{2.1cm}
\begin{itemize}[label={-}, leftmargin=*, topsep=0pt, partopsep=0pt, parsep=0pt, itemsep=0pt]
    \item Partial safety factors are calibrated on the basis of reliability. 
    \item 2.9\% mass reduction.
\end{itemize}
\end{minipage} 
\\ 
\addlinespace[0.2cm]

\cite{Furlanetto_Gomes_de_Almeida_2020} &
\begin{minipage}[t]{1.8cm}
\begin{itemize}[label={-}, leftmargin=*, topsep=0pt, partopsep=0pt, parsep=0pt, itemsep=0pt]
    \item Onshore: Tubular steel tower. 
    \item WindPACT 1.5 MW.
\end{itemize}
\end{minipage} &
\begin{minipage}[t]{1.8cm}
\begin{itemize}[label={-}, leftmargin=*, topsep=0pt, partopsep=0pt, parsep=0pt, itemsep=0pt]
    \item MATLAB (O).
\end{itemize}
\end{minipage} &
\begin{minipage}[t]{3cm}
\begin{itemize}[label={-}, leftmargin=*, topsep=0pt, partopsep=0pt, parsep=0pt, itemsep=0pt]
    \item Types: aerodynamic, wind, and gravitational.
\end{itemize}
\end{minipage} &
\begin{minipage}[t]{2.3cm}
\begin{itemize}[label={-}, leftmargin=*, topsep=0pt, partopsep=0pt, parsep=0pt, itemsep=0pt]
    \item Quantum PSO. 
    \item Objective function: minimize mass.
\end{itemize}
\end{minipage} &
\begin{minipage}[t]{3cm}
\begin{itemize}[label={-}, leftmargin=*, topsep=0pt, partopsep=0pt, parsep=0pt, itemsep=0pt]
    \item Section thickness (4 vars.). 
    \item Top/bottom tower diameter (2 vars.).
\end{itemize}
\end{minipage} &
\begin{minipage}[t]{3cm}
\begin{itemize}[label={-}, leftmargin=*, topsep=0pt, partopsep=0pt, parsep=0pt, itemsep=0pt]
    \item Geometry limits. 
    \item Maximum deflection. 
    \item Maximum stress. 
    \item Minimum buckling load multiplier.
    \item Minimum fatigue life.
    \item Frequency limits.
\end{itemize}
\end{minipage} &
\begin{minipage}[t]{3cm}
\begin{itemize}[label={-}, leftmargin=*, topsep=0pt, partopsep=0pt, parsep=0pt, itemsep=0pt]
    \item Optimized: modal analysis. 
\end{itemize}
\end{minipage} 
&
\begin{minipage}[t]{2.1cm}
\begin{itemize}[label={-}, leftmargin=*, topsep=0pt, partopsep=0pt, parsep=0pt, itemsep=0pt]
    \item 12\% mass reduction.
\end{itemize}
\end{minipage} 
\\
\addlinespace[0.2cm]

\cite{Wang_Kolios_Luengo_Liu_2016} &
\begin{minipage}[t]{1.8cm}
\begin{itemize}[label={-}, leftmargin=*, topsep=0pt, partopsep=0pt, parsep=0pt, itemsep=0pt]
    \item Onshore: Tubular steel tower. 
    \item NREL 5 MW.
\end{itemize}
\end{minipage} &
\begin{minipage}[t]{1.8cm}
\begin{itemize}[label={-}, leftmargin=*, topsep=0pt, partopsep=0pt, parsep=0pt, itemsep=0pt]
    \item Ansys (FEA).
\end{itemize}
\end{minipage} &
\begin{minipage}[t]{3cm}
\begin{itemize}[label={-}, leftmargin=*, topsep=0pt, partopsep=0pt, parsep=0pt, itemsep=0pt]
    \item Types: aerodynamic, wind, and gravitational. 
    \item Fatigue and ultimate load cases. 
    \item Values: 1 case.
\end{itemize}
\end{minipage} &
\begin{minipage}[t]{2.3cm}
\begin{itemize}[label={-}, leftmargin=*, topsep=0pt, partopsep=0pt, parsep=0pt, itemsep=0pt]
    \item GA. \newline
    \item Objective function: minimize mass.
\end{itemize}
\end{minipage} &
\begin{minipage}[t]{3cm}
\begin{itemize}[label={-}, leftmargin=*, topsep=0pt, partopsep=0pt, parsep=0pt, itemsep=0pt]
    \item Section thickness (16 vars.). 
    \item Top/bottom tower diameter (2 vars.).
\end{itemize}
\end{minipage} &
\begin{minipage}[t]{3cm}
\begin{itemize}[label={-}, leftmargin=*, topsep=0pt, partopsep=0pt, parsep=0pt, itemsep=0pt]
    \item Geometry limits. 
    \item Maximum deflection. 
    \item Maximum stress. 
    \item Minimum buckling load multiplier. 
    \item Minimum fatigue life. 
    \item Frequency limits.
\end{itemize}
\end{minipage} &
\begin{minipage}[t]{3cm}
\begin{itemize}[label={-}, leftmargin=*, topsep=0pt, partopsep=0pt, parsep=0pt, itemsep=0pt]
    \item Original: modal analysis. 
    \item Optimized: stress, deflection, modal, buckling, and fatigue analysis.
\end{itemize}
\end{minipage} &
\begin{minipage}[t]{2.1cm}
\begin{itemize}[label={-}, leftmargin=*, topsep=0pt, partopsep=0pt, parsep=0pt, itemsep=0pt]
    \item 6.3\% mass reduction. 
    \item Key design constraint: fatigue.
\end{itemize}
\end{minipage} 
\\ 
\addlinespace[0.2cm]

\cite{doi:10.1080/14786451.2021.1953495} &
\begin{minipage}[t]{1.8cm}
\begin{itemize}[label={-}, leftmargin=*, topsep=0pt, partopsep=0pt, parsep=0pt, itemsep=0pt]
    \item Onshore: Tubular steel tower. 
    \item 2 MW.
\end{itemize}
\end{minipage} &
\begin{minipage}[t]{1.8cm}
\begin{itemize}[label={-}, leftmargin=*, topsep=0pt, partopsep=0pt, parsep=0pt, itemsep=0pt]
    \item Ansys (FEA).
\end{itemize}
\end{minipage} &
\begin{minipage}[t]{3cm}
\begin{itemize}[label={-}, leftmargin=*, topsep=0pt, partopsep=0pt, parsep=0pt, itemsep=0pt]
    \item Types: aerodynamic, wind, and gravitational.
    \item Fatigue and ultimate load cases. 
    \item Values: 1 case.
\end{itemize}
\end{minipage} &
\begin{minipage}[t]{2.3cm}
\begin{itemize}[label={-}, leftmargin=*, topsep=0pt, partopsep=0pt, parsep=0pt, itemsep=0pt]
    \item GA. \newline
    \item Objective function: minimize mass.
\end{itemize}
\end{minipage} &
\begin{minipage}[t]{3cm}
\begin{itemize}[label={-}, leftmargin=*, topsep=0pt, partopsep=0pt, parsep=0pt, itemsep=0pt]
    \item Section thickness (6 vars.).
    \item Top/bottom tower diameter (2 vars.).
\end{itemize}
\end{minipage} &
\begin{minipage}[t]{3cm}
\begin{itemize}[label={-}, leftmargin=*, topsep=0pt, partopsep=0pt, parsep=0pt, itemsep=0pt]
    \item Geometry limits. 
    \item Maximum deflection. 
    \item Maximum stress. 
    \item Minimum buckling load multiplier. 
    \item Minimum fatigue life. 
    \item Frequency limits.
\end{itemize}
\end{minipage} &
\begin{minipage}[t]{3cm}
\begin{itemize}[label={-}, leftmargin=*, topsep=0pt, partopsep=0pt, parsep=0pt, itemsep=0pt]
    \item Original: modal analysis. 
    \item Optimized: stress, deflection, modal, buckling, and fatigue analysis.
\end{itemize}
\end{minipage} &
\begin{minipage}[t]{2.1cm}
\begin{itemize}[label={-}, leftmargin=*, topsep=0pt, partopsep=0pt, parsep=0pt, itemsep=0pt]
    \item 13.5\% mass reduction. 
    \item Key design constraint: fatigue.
\end{itemize}
\end{minipage} 
\\ 
\addlinespace[0.2cm]

\cite{Al-Sanad_Wang_Parol_Kolios_2021} &
\begin{minipage}[t]{1.8cm}
\begin{itemize}[label={-}, leftmargin=*, topsep=0pt, partopsep=0pt, parsep=0pt, itemsep=0pt]
    \item Onshore: Tubular steel tower. 
    \item Gamesa 2.0 MW.
\end{itemize}
\end{minipage} &
\begin{minipage}[t]{1.8cm}
\begin{itemize}[label={-}, leftmargin=*, topsep=0pt, partopsep=0pt, parsep=0pt, itemsep=0pt]
    \item Ansys (FEA). 
    \item MATLAB (O).
\end{itemize}
\end{minipage} &
\begin{minipage}[t]{3cm}
\begin{itemize}[label={-}, leftmargin=*, topsep=0pt, partopsep=0pt, parsep=0pt, itemsep=0pt]
    \item Types: aerodynamic, wind, and gravitational.
\end{itemize}
\end{minipage} &
\begin{minipage}[t]{2.3cm}
\begin{itemize}[label={-}, leftmargin=*, topsep=0pt, partopsep=0pt, parsep=0pt, itemsep=0pt]
    \item RBDO. 
    \item Objective function: minimize reliability and target index.
\end{itemize}
\end{minipage} &
\begin{minipage}[t]{3cm}
\begin{itemize}[label={-}, leftmargin=*, topsep=0pt, partopsep=0pt, parsep=0pt, itemsep=0pt]
    \item Top/bottom thickness (2 vars.).
    \item Top/bottom diameter (2 vars.).
\end{itemize}
\end{minipage} &
\begin{minipage}[t]{3cm}
\begin{itemize}[label={-}, leftmargin=*, topsep=0pt, partopsep=0pt, parsep=0pt, itemsep=0pt]
    \item Geometry limits. 
    \item Maximum deflection. 
    \item Maximum stress. 
    \item Minimum fatigue life. 
    \item Minimum buckling load multiplier. 
    \item Frequency limits.
\end{itemize}
\end{minipage} &
\begin{minipage}[t]{3cm}
\begin{itemize}[label={-}, leftmargin=*, topsep=0pt, partopsep=0pt, parsep=0pt, itemsep=0pt]
    \item Original: modal analysis. 
    \item Optimized: stress, modal, buckling, and fatigue analysis.
\end{itemize}
\end{minipage} &
\begin{minipage}[t]{2.1cm}
\begin{itemize}[label={-}, leftmargin=*, topsep=0pt, partopsep=0pt, parsep=0pt, itemsep=0pt]
    \item 15.1\% mass reduction.
\end{itemize}
\end{minipage} 
\\ 
\addlinespace[0.2cm]
\cite{Warshawsky2015PracticalAO} &
\begin{minipage}[t]{1.8cm}
\begin{itemize}[label={-}, leftmargin=*, topsep=0pt, partopsep=0pt, parsep=0pt, itemsep=0pt]
    \item Onshore: Tubular steel tower. 
    \item NREL 5 MW.
\end{itemize}
\end{minipage} &
\begin{minipage}[t]{1.8cm}
\begin{itemize}[label={-}, leftmargin=*, topsep=0pt, partopsep=0pt, parsep=0pt, itemsep=0pt]
    \item FENDAC (TO).
\end{itemize}
\end{minipage} &
\begin{minipage}[t]{3cm}
\begin{itemize}[label={-}, leftmargin=*, topsep=0pt, partopsep=0pt, parsep=0pt, itemsep=0pt]
    \item Types: aerodynamic, wind, and gravitational. 
    \item Values: 1 case.
\end{itemize}
\end{minipage} &
\begin{minipage}[t]{2.3cm}
\begin{itemize}[label={-}, leftmargin=*, topsep=0pt, partopsep=0pt, parsep=0pt, itemsep=0pt]
    \item TO: SIMP. 
    \item Objective function: minimize compliance.
\end{itemize}
\end{minipage} &
\begin{minipage}[t]{3cm}
\begin{itemize}[label={-}, leftmargin=*, topsep=0pt, partopsep=0pt, parsep=0pt, itemsep=0pt]
    \item Section thickness (1 var.).
    \item Top, bottom and taper diameter (3 vars.).
\end{itemize}
\end{minipage} &
\begin{minipage}[t]{3cm}
\begin{itemize}[label={-}, leftmargin=*, topsep=0pt, partopsep=0pt, parsep=0pt, itemsep=0pt]
    \item Design: 4 symmetry planes. 
    \item Maximum deflection.
    \item Response: 25\%, 37.5\%, 50\%, and 75\% original volume.
\end{itemize}
\end{minipage} &
\begin{minipage}[t]{3cm}
\begin{itemize}[label={-}, leftmargin=*, topsep=0pt, partopsep=0pt, parsep=0pt, itemsep=0pt]
    \item Optimized: deflection analysis.
\end{itemize}
\end{minipage} &
\begin{minipage}[t]{2.1cm}
\begin{itemize}[label={-}, leftmargin=*, topsep=0pt, partopsep=0pt, parsep=0pt, itemsep=0pt]
    \item 35\% mass reduction.
\end{itemize}
\end{minipage} 
\\ 
\addlinespace[0.2cm]

\cite{Lagaros_Karlaftis_2016} &
\begin{minipage}[t]{1.8cm}
\begin{itemize}[label={-}, leftmargin=*, topsep=0pt, partopsep=0pt, parsep=0pt, itemsep=0pt]
    \item Onshore: Tubular steel tower. 
    \item Vestas V90-3 MW.
\end{itemize}
\end{minipage} &
\begin{minipage}[t]{1.8cm}
\begin{itemize}[label={-}, leftmargin=*, topsep=0pt, partopsep=0pt, parsep=0pt, itemsep=0pt]
    \item SAP2000 (FEA). 
    \item Optimization Computing Platform (O).
\end{itemize}
\end{minipage} &
\begin{minipage}[t]{3cm}
\begin{itemize}[label={-}, leftmargin=*, topsep=0pt, partopsep=0pt, parsep=0pt, itemsep=0pt]
    \item Types: wind and gravitational.
\end{itemize}
\end{minipage} &
\begin{minipage}[t]{2.3cm}
\begin{itemize}[label={-}, leftmargin=*, topsep=0pt, partopsep=0pt, parsep=0pt, itemsep=0pt]
    \item DE. 
    \item Objective function: minimize construction cost.
\end{itemize}
\end{minipage} &
\begin{minipage}[t]{3cm}
\begin{itemize}[label={-}, leftmargin=*, topsep=0pt, partopsep=0pt, parsep=0pt, itemsep=0pt]
    \item Section thickness. 
    \item Section diameter.
\end{itemize}
\end{minipage} &
\begin{minipage}[t]{3cm}
\begin{itemize}[label={-}, leftmargin=*, topsep=0pt, partopsep=0pt, parsep=0pt, itemsep=0pt]
    \item Geometry limits. 
    \item Maximum deflection. 
    \item Maximum stress. 
    \item Minimum buckling load multiplier. 
    \item Frequency limits.
\end{itemize}
\end{minipage} & &
\begin{minipage}[t]{2.3cm}
\begin{itemize}[label={-}, leftmargin=*, topsep=0pt, partopsep=0pt, parsep=0pt, itemsep=0pt]
    \item Structural optimization is vital for cost-efficient design.
\end{itemize}
\end{minipage} 
\\ 
\bottomrule
\end{longtable}
\end{landscape}

\begin{landscape}
\section{Optimization Studies: Offshore Wind Turbine Towers} \label{A:b}

\scriptsize	
\centering
\begin{longtable}{@{}p{0.4cm}p{1.8cm}p{1.8cm}p{3cm}p{2.3cm}p{3cm}p{3cm}p{3cm}p{2.1cm}@{}}
\caption{Literature review of optimization studies on OWT towers.}
\label{table:lr_towers_off}\\  
\toprule
\textbf{Study}  & \textbf{Turbine} & \textbf{Software} & \textbf{Loads} & \textbf{Method}  & \textbf{Design Variables} &  \textbf{Design Constraints}  & 
 \textbf{Analysis} & \textbf{Results} \\ \midrule

\cite{Zwick_Muskulus_Moe_2012} 
& 
\begin{minipage}[t]{1.8cm}
\begin{itemize}[label={-}, leftmargin=*, topsep=0pt, partopsep=0pt, parsep=0pt, itemsep=0pt]
    \item Offshore: Lattice tower.
    \item 10 MW NOWITECH.
\end{itemize}
\end{minipage} &
\begin{minipage}[t]{1.8cm}
\begin{itemize}[label={-}, leftmargin=*, topsep=0pt, partopsep=0pt, parsep=0pt, itemsep=0pt]
    \item FEDEM (FEA).
\end{itemize}
\end{minipage} &
\begin{minipage}[t]{3cm}
\begin{itemize}[label={-}, leftmargin=*, topsep=0pt, partopsep=0pt, parsep=0pt, itemsep=0pt]
    \item Types: hydrodynamic, wind, and gravitational.
\end{itemize}
\end{minipage} &
\begin{minipage}[t]{3cm}
\begin{itemize}[label={-}, leftmargin=*, topsep=0pt, partopsep=0pt, parsep=0pt, itemsep=0pt]
    \item IO.
\end{itemize}
\end{minipage} &
\begin{minipage}[t]{3cm}
\begin{itemize}[label={-}, leftmargin=*, topsep=0pt, partopsep=0pt, parsep=0pt, itemsep=0pt]
    \item Leg number (1 var.).
    \item Section number (1 var.).
    \item Leg/brace member diameter and thickness (2 vars.).
    \item Constant brace angle or constant section height (1 var.).
\end{itemize}
\end{minipage} &
\begin{minipage}[t]{3cm}
\begin{itemize}[label={-}, leftmargin=*, topsep=0pt, partopsep=0pt, parsep=0pt, itemsep=0pt]
    \item Geometry limits.
    \item Maximum stress.
    \item Minimum fatigue life.
\end{itemize}
\end{minipage} &
\begin{minipage}[t]{3cm}
\begin{itemize}[label={-}, leftmargin=*, topsep=0pt, partopsep=0pt, parsep=0pt, itemsep=0pt]
    \item Optimized: stress and fatigue analysis.
\end{itemize}
\end{minipage} &
\begin{minipage}[t]{2.1cm}
\begin{itemize}[label={-}, leftmargin=*, topsep=0pt, partopsep=0pt, parsep=0pt, itemsep=0pt]
    \item Process efficiency.
    \item Reduced total weight.
\end{itemize}
\end{minipage} 
\\ 
\addlinespace[0.2cm]

\cite{10.1063/1.4984259} &
\begin{minipage}[t]{1.8cm}
\begin{itemize}[label={-}, leftmargin=*, topsep=0pt, partopsep=0pt, parsep=0pt, itemsep=0pt]
    \item Offshore: Composite floating tower.
    \item 6 MW.
\end{itemize}
\end{minipage} &
\begin{minipage}[t]{1.8cm}
\begin{itemize}[label={-}, leftmargin=*, topsep=0pt, partopsep=0pt, parsep=0pt, itemsep=0pt]
    \item OpenFAST (AHSE).
    \item Ansys (FEA).
    \item MATLAB (O).
\end{itemize}
\end{minipage} &
\begin{minipage}[t]{3cm}
\begin{itemize}[label={-}, leftmargin=*, topsep=0pt, partopsep=0pt, parsep=0pt, itemsep=0pt]
    \item Types: aerodynamic, hydrodynamic, wind, and gravitational.
    \item Ultimate design load case: DLC 1.6, 6.1, 6.3, and Survival (IEC61400-3).
\end{itemize}
\end{minipage} &
\begin{minipage}[t]{2.3cm}
\begin{itemize}[label={-}, leftmargin=*, topsep=0pt, partopsep=0pt, parsep=0pt, itemsep=0pt]
    \item GA.
    \item Objective function: minimize mass.
\end{itemize}
\end{minipage} &
\begin{minipage}[t]{3cm}
\begin{itemize}[label={-}, leftmargin=*, topsep=0pt, partopsep=0pt, parsep=0pt, itemsep=0pt]
    \item Section number.
    \item Tower section elevation.
    \item Section diameter.
    \item Material layers thickness.
\end{itemize}
\end{minipage} &
\begin{minipage}[t]{3cm}
\begin{itemize}[label={-}, leftmargin=*, topsep=0pt, partopsep=0pt, parsep=0pt, itemsep=0pt]
    \item Geometry limits.
    \item Maximum stress.
    \item Minimum buckling load multiplier.
    \item Frequency limits.
\end{itemize}
\end{minipage} &
\begin{minipage}[t]{3cm}
\begin{itemize}[label={-}, leftmargin=*, topsep=0pt, partopsep=0pt, parsep=0pt, itemsep=0pt]
    \item Original: stress, modal, and buckling analysis.
    \item Optimized: stress, modal, and buckling analysis.
\end{itemize}
\end{minipage} &
\begin{minipage}[t]{2.1cm}
\begin{itemize}[label={-}, leftmargin=*, topsep=0pt, partopsep=0pt, parsep=0pt, itemsep=0pt]
    \item 54.8\% mass reduction.
\end{itemize}
\end{minipage} 
\\ 
\addlinespace[0.2cm]

\cite{Mendoza_Griffith_Qin_Loth_Johnson_2022} &
\begin{minipage}[t]{1.8cm}
\begin{itemize}[label={-}, leftmargin=*, topsep=0pt, partopsep=0pt, parsep=0pt, itemsep=0pt]
    \item Offshore: Tubular steel tower.
    \item 25 MW.
\end{itemize}
\end{minipage} &
\begin{minipage}[t]{1.8cm}
\begin{itemize}[label={-}, leftmargin=*, topsep=0pt, partopsep=0pt, parsep=0pt, itemsep=0pt]
    \item BModes (frequency).
    \item Ansys (FEA).
    \item MATLAB (O).
\end{itemize}
\end{minipage} &
\begin{minipage}[t]{3cm}
\begin{itemize}[label={-}, leftmargin=*, topsep=0pt, partopsep=0pt, parsep=0pt, itemsep=0pt]
    \item Types: aerodynamic and gravitational.
\end{itemize}
\end{minipage} &
\begin{minipage}[t]{2.3cm}
\begin{itemize}[label={-}, leftmargin=*, topsep=0pt, partopsep=0pt, parsep=0pt, itemsep=0pt]
    \item SQP.
    \item Objective function: minimize mass.
\end{itemize}
\end{minipage} &
\begin{minipage}[t]{3cm}
\begin{itemize}[label={-}, leftmargin=*, topsep=0pt, partopsep=0pt, parsep=0pt, itemsep=0pt]
    \item Top/taper thickness (2 vars.).
    \item Top/taper diameter (2 vars.).
\end{itemize}
\end{minipage} &
\begin{minipage}[t]{3cm}
\begin{itemize}[label={-}, leftmargin=*, topsep=0pt, partopsep=0pt, parsep=0pt, itemsep=0pt]
    \item Geometry limits.
    \item Maximum stress.
    \item Minimum buckling load multiplier.
\end{itemize}
\end{minipage} &
\begin{minipage}[t]{3cm}
\begin{itemize}[label={-}, leftmargin=*, topsep=0pt, partopsep=0pt, parsep=0pt, itemsep=0pt]
    \item Optimized: stress, modal, and buckling analysis.
\end{itemize}
\end{minipage} &
\begin{minipage}[t]{2.1cm}
\begin{itemize}[label={-}, leftmargin=*, topsep=0pt, partopsep=0pt, parsep=0pt, itemsep=0pt]
    \item 0.8-14\% mass reduction.
\end{itemize}
\end{minipage} 
\\ 
\addlinespace[0.2cm]

\cite{10.1115/1.4000401} &
\begin{minipage}[t]{1.8cm}
\begin{itemize}[label={-}, leftmargin=*, topsep=0pt, partopsep=0pt, parsep=0pt, itemsep=0pt]
    \item Offshore: Tubular steel tower.
\end{itemize}
\end{minipage} &
\begin{minipage}[t]{1.8cm}
\begin{itemize}[label={-}, leftmargin=*, topsep=0pt, partopsep=0pt, parsep=0pt, itemsep=0pt]
    \item IMSL (O).
\end{itemize}
\end{minipage} &
\begin{minipage}[t]{3cm}
\begin{itemize}[label={-}, leftmargin=*, topsep=0pt, partopsep=0pt, parsep=0pt, itemsep=0pt]
    \item Types: hydrodynamic and gravitational.
\end{itemize}
\end{minipage} &
\begin{minipage}[t]{2.3cm}
\begin{itemize}[label={-}, leftmargin=*, topsep=0pt, partopsep=0pt, parsep=0pt, itemsep=0pt]
    \item RBDO.
    \item Objective function: minimize mass.
\end{itemize}
\end{minipage} &
\begin{minipage}[t]{3cm}
\begin{itemize}[label={-}, leftmargin=*, topsep=0pt, partopsep=0pt, parsep=0pt, itemsep=0pt]
    \item Section thickness (6 vars.).
    \item Top/bottom tower thickness (2 vars.).
    \item Top/bottom tower diameter (2 vars.).
\end{itemize}
\end{minipage} &
\begin{minipage}[t]{3cm}
\begin{itemize}[label={-}, leftmargin=*, topsep=0pt, partopsep=0pt, parsep=0pt, itemsep=0pt]
    \item Geometry limits.
    \item Maximum stress.
    \item Minimum buckling load multiplier.
    \item Frequency limits.
\end{itemize}
\end{minipage} &
\begin{minipage}[t]{3cm}
\begin{itemize}[label={-}, leftmargin=*, topsep=0pt, partopsep=0pt, parsep=0pt, itemsep=0pt]
    \item Optimized: stress, modal, and buckling analysis.
\end{itemize}
\end{minipage} &
\begin{minipage}[t]{2.1cm}
\begin{itemize}[label={-}, leftmargin=*, topsep=0pt, partopsep=0pt, parsep=0pt, itemsep=0pt]
    \item Maintain safety while being economically efficient.
\end{itemize}
\end{minipage} 
\\ 
\addlinespace[0.2cm]

\cite{Meng_Yang_Jesus_Zhu_2023} &
\begin{minipage}[t]{1.8cm}
\begin{itemize}[label={-}, leftmargin=*, topsep=0pt, partopsep=0pt, parsep=0pt, itemsep=0pt]
    \item Offshore: Tubular steel tower.
    \item 5 MW.
\end{itemize}
\end{minipage} &
\begin{minipage}[t]{1.8cm}
\begin{itemize}[label={-}, leftmargin=*, topsep=0pt, partopsep=0pt, parsep=0pt, itemsep=0pt]
    \item Ansys (FEA).
\end{itemize}
\end{minipage} &
\begin{minipage}[t]{3cm}
\begin{itemize}[label={-}, leftmargin=*, topsep=0pt, partopsep=0pt, parsep=0pt, itemsep=0pt]
    \item Types: aerodynamic, hydrodynamic, wind, and gravitational.
\end{itemize}
\end{minipage} &
\begin{minipage}[t]{2.3cm}
\begin{itemize}[label={-}, leftmargin=*, topsep=0pt, partopsep=0pt, parsep=0pt, itemsep=0pt]
    \item KM and RBDO.
    \item Objective function: minimize mass.
\end{itemize}
\end{minipage} &
\begin{minipage}[t]{3cm}
\begin{itemize}[label={-}, leftmargin=*, topsep=0pt, partopsep=0pt, parsep=0pt, itemsep=0pt]
    \item Top/bottom tower thickness (2 vars.).
    \item Top/bottom tower diameter (2 vars.).
    \item Monopile thickness (1 var.).
\end{itemize}
\end{minipage} &
\begin{minipage}[t]{3cm}
\begin{itemize}[label={-}, leftmargin=*, topsep=0pt, partopsep=0pt, parsep=0pt, itemsep=0pt]
    \item Geometry limits.
    \item Maximum deflection.
    \item Maximum stress.
    \item Minimum buckling load multiplier.
    \item Frequency limits.
\end{itemize}
\end{minipage} &
\begin{minipage}[t]{3cm}
\begin{itemize}[label={-}, leftmargin=*, topsep=0pt, partopsep=0pt, parsep=0pt, itemsep=0pt]
    \item Optimized: stress, deflection, modal, and buckling analysis.
\end{itemize}
\end{minipage} &
\begin{minipage}[t]{2.1cm}
\begin{itemize}[label={-}, leftmargin=*, topsep=0pt, partopsep=0pt, parsep=0pt, itemsep=0pt]
    \item 6.2\% mass reduction.
\end{itemize}
\end{minipage} 
\\ 
\addlinespace[0.2cm]

\cite{Gentils_Wang_Kolios_2017} &
\begin{minipage}[t]{1.8cm}
\begin{itemize}[label={-}, leftmargin=*, topsep=0pt, partopsep=0pt, parsep=0pt, itemsep=0pt]
    \item Offshore: Tubular steel tower.
    \item NREL 5 MW.
\end{itemize}
\end{minipage} &
\begin{minipage}[t]{1.8cm}
\begin{itemize}[label={-}, leftmargin=*, topsep=0pt, partopsep=0pt, parsep=0pt, itemsep=0pt]
    \item Ansys (FEA).
\end{itemize}
\end{minipage} &
\begin{minipage}[t]{3cm}
\begin{itemize}[label={-}, leftmargin=*, topsep=0pt, partopsep=0pt, parsep=0pt, itemsep=0pt]
    \item Types: aerodynamic, hydrodynamic, wind, and gravitational.
    \item Ultimate design load case: DLC 2.1 and 6.1b (IEC61400-3).
    \item Fatigue design load case: DLC 1.2 (IEC61400-3).
\end{itemize}
\end{minipage} &
\begin{minipage}[t]{2.3cm}
\begin{itemize}[label={-}, leftmargin=*, topsep=0pt, partopsep=0pt, parsep=0pt, itemsep=0pt]
    \item GA.
    \item Objective function: minimize mass.
\end{itemize}
\end{minipage} &
\begin{minipage}[t]{3cm}
\begin{itemize}[label={-}, leftmargin=*, topsep=0pt, partopsep=0pt, parsep=0pt, itemsep=0pt]
    \item Section thickness (25 vars.).
    \item Top/bottom tower thickness (2 vars.).
    \item Top/bottom tower diameter (2 vars.).
    \item Transient piece thickness (1 var.).
    \item Top/bottom monopile diameter (2 vars.).
\end{itemize}
\end{minipage} &
\begin{minipage}[t]{3cm}
\begin{itemize}[label={-}, leftmargin=*, topsep=0pt, partopsep=0pt, parsep=0pt, itemsep=0pt]
    \item Geometry limits.
    \item Maximum deflection.
    \item Maximum stress.
    \item Minimum buckling load multiplier.
    \item Minimum fatigue life.
    \item Frequency limits.
\end{itemize}
\end{minipage} &
\begin{minipage}[t]{3cm}
\begin{itemize}[label={-}, leftmargin=*, topsep=0pt, partopsep=0pt, parsep=0pt, itemsep=0pt]
    \item Original: deflection and modal analysis.
    \item Optimized: stress, deflection, modal, buckling, and fatigue analysis.
\end{itemize}
\end{minipage} &
\begin{minipage}[t]{2.1cm}
\begin{itemize}[label={-}, leftmargin=*, topsep=0pt, partopsep=0pt, parsep=0pt, itemsep=0pt]
    \item 19.8\% mass reduction.
    \item Key design constraint: fatigue and natural frequency.
\end{itemize}
\end{minipage} 
\\ 
\addlinespace[0.2cm]

\cite{Kamel_Dammak_El_Hami_Ben_Jdidia_Hammami_Haddar_2022} &
\begin{minipage}[t]{1.8cm}
\begin{itemize}[label={-}, leftmargin=*, topsep=0pt, partopsep=0pt, parsep=0pt, itemsep=0pt]
    \item Offshore: Tubular steel tower.
    \item 2 MW.
\end{itemize}
\end{minipage} &
\begin{minipage}[t]{1.8cm}
\begin{itemize}[label={-}, leftmargin=*, topsep=0pt, partopsep=0pt, parsep=0pt, itemsep=0pt]
    \item Ansys (FEA).
    \item MATLAB (O).
\end{itemize}
\end{minipage} &
\begin{minipage}[t]{3cm}
\begin{itemize}[label={-}, leftmargin=*, topsep=0pt, partopsep=0pt, parsep=0pt, itemsep=0pt]
    \item Types: aerodynamic, hydrodynamic, wind, and gravitational.
    \item Values: 1 case.
\end{itemize}
\end{minipage} &
\begin{minipage}[t]{2.3cm}
\begin{itemize}[label={-}, leftmargin=*, topsep=0pt, partopsep=0pt, parsep=0pt, itemsep=0pt]
    \item RBDO.
    \item Objective function: minimize volume.
\end{itemize}
\end{minipage} &
\begin{minipage}[t]{3cm}
\begin{itemize}[label={-}, leftmargin=*, topsep=0pt, partopsep=0pt, parsep=0pt, itemsep=0pt]
    \item Top tower thickness (1 var.).
    \item Top/bottom tower diameter (2 vars.).
\end{itemize}
\end{minipage} &
\begin{minipage}[t]{3cm}
\begin{itemize}[label={-}, leftmargin=*, topsep=0pt, partopsep=0pt, parsep=0pt, itemsep=0pt]
    \item Geometry limits.
    \item Maximum deflection.
    \item Maximum stress.
    \item Minimum buckling load multiplier.
    \item Frequency limits.
\end{itemize}
\end{minipage} &
\begin{minipage}[t]{3cm}
\begin{itemize}[label={-}, leftmargin=*, topsep=0pt, partopsep=0pt, parsep=0pt, itemsep=0pt]
    \item Optimized: stress and deflection analysis.
\end{itemize}
\end{minipage} &
\begin{minipage}[t]{2.1cm}
\begin{itemize}[label={-}, leftmargin=*, topsep=0pt, partopsep=0pt, parsep=0pt, itemsep=0pt]
    \item 50\% mass reduction.
    \item Improved feasibility and constraint adherence.
    \item Better performance and calculation time.
\end{itemize}
\end{minipage} 
\\ 
\addlinespace[0.2cm]

\cite{Pollini_Pegalajar-Jurado_Bredmose_2023} &
\begin{minipage}[t]{1.8cm}
\begin{itemize}[label={-}, leftmargin=*, topsep=0pt, partopsep=0pt, parsep=0pt, itemsep=0pt]
    \item Offshore: Tubular steel tower.
    \item TetraSpar floater.
    \item IEA Wind 15 MW.
\end{itemize}
\end{minipage} &
\begin{minipage}[t]{1.8cm}
\begin{itemize}[label={-}, leftmargin=*, topsep=0pt, partopsep=0pt, parsep=0pt, itemsep=0pt]
    \item OpenFAST (AHSE).
    \item MATLAB (O).
\end{itemize}
\end{minipage} &
\begin{minipage}[t]{3cm}
\begin{itemize}[label={-}, leftmargin=*, topsep=0pt, partopsep=0pt, parsep=0pt, itemsep=0pt]
    \item Types: aerodynamic, hydrodynamic, wind, and gravitational.
\end{itemize}
\end{minipage} &
\begin{minipage}[t]{2.3cm}
\begin{itemize}[label={-}, leftmargin=*, topsep=0pt, partopsep=0pt, parsep=0pt, itemsep=0pt]
    \item GBO.
    \item Objective function: minimize mass.
\end{itemize}
\end{minipage} &
\begin{minipage}[t]{3cm}
\begin{itemize}[label={-}, leftmargin=*, topsep=0pt, partopsep=0pt, parsep=0pt, itemsep=0pt]
    \item Top/bottom tower thickness (2 vars.).
    \item Top/bottom tower diameter (2 vars.).
    \item Mooring line length and anchor radius (2 vars.).
    \item Fairlead connection and keel depth (2 vars.).
    \item Brace diameter and length (4 vars.).
    \item Central column diameter and height (2 vars.).
\end{itemize}
\end{minipage} &
\begin{minipage}[t]{3cm}
\begin{itemize}[label={-}, leftmargin=*, topsep=0pt, partopsep=0pt, parsep=0pt, itemsep=0pt]
    \item Geometry limits.
    \item Maximum deflection.
    \item Frequency limits.
\end{itemize}
\end{minipage} & &
\begin{minipage}[t]{2.1cm}
\begin{itemize}[label={-}, leftmargin=*, topsep=0pt, partopsep=0pt, parsep=0pt, itemsep=0pt]
    \item 11\% cost reduction.
    \item Fatigue constraint at the tower base.
\end{itemize}
\end{minipage} 
\\ 
\addlinespace[0.2cm]

\cite{Zakhama_Abdalla_Gürdal_Smaoui_2010} &
\begin{minipage}[t]{1.8cm}
\begin{itemize}[label={-}, leftmargin=*, topsep=0pt, partopsep=0pt, parsep=0pt, itemsep=0pt]
    \item Offshore: Tubular steel tower.
\end{itemize}
\end{minipage} &
\begin{minipage}[t]{1.8cm}
\end{minipage} &
\begin{minipage}[t]{3cm}
\begin{itemize}[label={-}, leftmargin=*, topsep=0pt, partopsep=0pt, parsep=0pt, itemsep=0pt]
    \item Types: aerodynamic, wind, and gravitational.
    \item Values: 1 case.
\end{itemize}
\end{minipage} &
\begin{minipage}[t]{2.3cm}
\begin{itemize}[label={-}, leftmargin=*, topsep=0pt, partopsep=0pt, parsep=0pt, itemsep=0pt]
    \item TO: SIMP.
    \item Objective function: minimize compliance.
\end{itemize}
\end{minipage} &
\begin{minipage}[t]{3cm}
\end{minipage} &
\begin{minipage}[t]{3cm}
\begin{itemize}[label={-}, leftmargin=*, topsep=0pt, partopsep=0pt, parsep=0pt, itemsep=0pt]
    \item Response: Maximum original volume.
    \item Explicit: Clear solid/void separation.
\end{itemize}
\end{minipage} &
\begin{minipage}[t]{3cm}
\end{minipage} &
\begin{minipage}[t]{2.1cm}
\begin{itemize}[label={-}, leftmargin=*, topsep=0pt, partopsep=0pt, parsep=0pt, itemsep=0pt]
    \item 37-40\% stiffer with wind loads.
    \item Accounts for wind pressure.
\end{itemize}
\end{minipage} 
\\ 
\addlinespace[0.2cm]

\cite{Ashuri_Zaaijer_Martins_van_Bussel_van_Kuik_2014} &
\begin{minipage}[t]{1.8cm}
\begin{itemize}[label={-}, leftmargin=*, topsep=0pt, partopsep=0pt, parsep=0pt, itemsep=0pt]
    \item Offshore: Tubular steel tower.
    \item NREL 5 MW.
\end{itemize}
\end{minipage} &
\begin{minipage}[t]{1.8cm}
\begin{itemize}[label={-}, leftmargin=*, topsep=0pt, partopsep=0pt, parsep=0pt, itemsep=0pt]
    \item OpenFAST (AHSE).
\end{itemize}
\end{minipage} &
\begin{minipage}[t]{3cm}
\begin{itemize}[label={-}, leftmargin=*, topsep=0pt, partopsep=0pt, parsep=0pt, itemsep=0pt]
    \item Types: aerodynamic, wind, and gravitational.
    \item Ultimate design load case: DLC 6.1 (IEC61400-3).
    \item Fatigue design load case: DLC 1.2 (IEC61400-3).
\end{itemize}
\end{minipage} &
\begin{minipage}[t]{2.3cm}
\begin{itemize}[label={-}, leftmargin=*, topsep=0pt, partopsep=0pt, parsep=0pt, itemsep=0pt]
    \item MDO.
    \item Objective function: minimize LCOE.
\end{itemize}
\end{minipage} &
\begin{minipage}[t]{3cm}
\begin{itemize}[label={-}, leftmargin=*, topsep=0pt, partopsep=0pt, parsep=0pt, itemsep=0pt]
    \item Top/bottom tower thickness (2 vars.).
    \item Top/bottom tower diameter (2 vars.).
    \item Tower height (1 var.).
    \item Blade length (1 var.).
    \item Rated rotational speed (1 var.).
    \item Station twist (2 vars.).
    \item Station chord (4 vars.).
    \item Station skin/web/span thickness (10 vars.).
\end{itemize}
\end{minipage} &
\begin{minipage}[t]{3cm}
\begin{itemize}[label={-}, leftmargin=*, topsep=0pt, partopsep=0pt, parsep=0pt, itemsep=0pt]
    \item Geometry limits.
    \item Maximum deflection.
    \item Maximum stress.
    \item Minimum fatigue life.
    \item Minimum buckling load multiplier.
    \item Frequency limits.
\end{itemize}
\end{minipage} &
\begin{minipage}[t]{3cm}
\begin{itemize}[label={-}, leftmargin=*, topsep=0pt, partopsep=0pt, parsep=0pt, itemsep=0pt]
    \item Optimized: fatigue analysis.
\end{itemize}
\end{minipage} &
\begin{minipage}[t]{2.1cm}
\begin{itemize}[label={-}, leftmargin=*, topsep=0pt, partopsep=0pt, parsep=0pt, itemsep=0pt]
    \item 2.3\% levelized cost of energy reduction.
\end{itemize}
\end{minipage} 
\\ 
\bottomrule
\end{longtable}
\end{landscape}

\begin{landscape}
\section{Topology Optimization Studies: Offshore Wind Turbine Foundations} \label{A:c}

\scriptsize	
\centering
\begin{longtable}{@{}p{0.4cm}p{1.8cm}p{1.8cm}p{3cm}p{2.3cm}p{3cm}p{3cm}p{3cm}p{2.1cm}@{}}
\caption{Literature review of topology optimization studies on foundations of OWTs.}
\label{table:optimization_offshore_foundations}\\  
\toprule
\textbf{Study}  & \textbf{Turbine} & \textbf{Software} & \textbf{Loads} & \textbf{Method}  & \textbf{Design Variables} &  \textbf{Design Constraints}  & 
 \textbf{Analysis} & \textbf{Results} \\ \midrule

\cite{Marjan_Huang_2023} 
& 
\begin{minipage}[t]{1.8cm}
\begin{itemize}[label={-}, leftmargin=*, topsep=0pt, partopsep=0pt, parsep=0pt, itemsep=0pt]
    \item OC4 Jacket foundation.
    \item NREL 5 MW.
\end{itemize}
\end{minipage} & 
\begin{minipage}[t]{1.8cm}
\begin{itemize}[label={-}, leftmargin=*, topsep=0pt, partopsep=0pt, parsep=0pt, itemsep=0pt]
    \item Sesam (modeling, FEA).
    \item Bladed (AHSE).
    \item Ansys (TO).
\end{itemize}
\end{minipage} &
\begin{minipage}[t]{3cm}
\begin{itemize}[label={-}, leftmargin=*, topsep=0pt, partopsep=0pt, parsep=0pt, itemsep=0pt]
    \item Types: aerodynamic, hydrodynamic, wind, and gravitational.
    \item Loads in the same direction.
    \item Values: 1 case.
\end{itemize}
\end{minipage} &
\begin{minipage}[t]{2.3cm}
\begin{itemize}[label={-}, leftmargin=*, topsep=0pt, partopsep=0pt, parsep=0pt, itemsep=0pt]
    \item TO: SIMP.
    \item Objective function: minimize compliance.
\end{itemize}
\end{minipage} &
\begin{minipage}[t]{3cm}
\begin{itemize}[label={-}, leftmargin=*, topsep=0pt, partopsep=0pt, parsep=0pt, itemsep=0pt]
    \item Section thickness (7 vars.).
    \item Section diameter (7 vars.).
\end{itemize}
\end{minipage} &
\begin{minipage}[t]{3cm}
\begin{itemize}[label={-}, leftmargin=*, topsep=0pt, partopsep=0pt, parsep=0pt, itemsep=0pt]
    \item Geometry limits.
    \item Maximum deflection.
    \item Maximum stress.
    \item Response: 35\% original mass.
    \item Design: 2 symmetry planes.
\end{itemize}
\end{minipage} &
\begin{minipage}[t]{3cm}
\begin{itemize}[label={-}, leftmargin=*, topsep=0pt, partopsep=0pt, parsep=0pt, itemsep=0pt]
    \item Original: deflection and modal analysis.
    \item Optimized: modal and fatigue analysis.
    \item Optimized geometry is reconstructed.
\end{itemize}
\end{minipage} &
\begin{minipage}[t]{2.1cm}
\begin{itemize}[label={-}, leftmargin=*, topsep=0pt, partopsep=0pt, parsep=0pt, itemsep=0pt]
    \item 35.2\% mass reduction.
    \item 37.2\% improvement in fatigue life.
\end{itemize}
\end{minipage} 
\\ 
\addlinespace[0.2cm]

\cite{Tian_Sun_Liu_Deng_Wang_Li_Li_2022} &
\begin{minipage}[t]{1.8cm}
\begin{itemize}[label={-}, leftmargin=*, topsep=0pt, partopsep=0pt, parsep=0pt, itemsep=0pt]
    \item OC4 Jacket foundation.
    \item NREL 5 MW.
\end{itemize}
\end{minipage} &
\begin{minipage}[t]{1.8cm}
\begin{itemize}[label={-}, leftmargin=*, topsep=0pt, partopsep=0pt, parsep=0pt, itemsep=0pt]
    \item Ansys (FEA).
    \item HyperWorks (TO).
\end{itemize}
\end{minipage} &
\begin{minipage}[t]{3cm}
\begin{itemize}[label={-}, leftmargin=*, topsep=0pt, partopsep=0pt, parsep=0pt, itemsep=0pt]
    \item Types: aerodynamic, hydrodynamic, wind, and gravitational.
    \item Loads in the same direction.
    \item Ultimate design load case: DLC 1.3 (IEC61400-3).
    \item Values: 1 case.
\end{itemize}
\end{minipage} &
\begin{minipage}[t]{2.3cm}
\begin{itemize}[label={-}, leftmargin=*, topsep=0pt, partopsep=0pt, parsep=0pt, itemsep=0pt]
    \item TO: SIMP.
    \item Objective function: minimize compliance.
    \item Size and shape optimization.
\end{itemize}
\end{minipage} &
\begin{minipage}[t]{3cm}
\begin{itemize}[label={-}, leftmargin=*, topsep=0pt, partopsep=0pt, parsep=0pt, itemsep=0pt]
    \item Joint point height (15 vars.).
    \item Tube thickness (8 vars.).
    \item Tube diameter (2 vars.).
\end{itemize}
\end{minipage} &
\begin{minipage}[t]{3cm}
\begin{itemize}[label={-}, leftmargin=*, topsep=0pt, partopsep=0pt, parsep=0pt, itemsep=0pt]
    \item Geometry limits.
    \item Maximum deflection.
    \item Maximum stress.
    \item Minimum buckling load multiplier.
    \item Frequency limits.
\end{itemize}
\end{minipage} &
\begin{minipage}[t]{3cm}
\begin{itemize}[label={-}, leftmargin=*, topsep=0pt, partopsep=0pt, parsep=0pt, itemsep=0pt]
    \item Original: stress, deflection, modal, and buckling analysis.
    \item Optimized: stress, deflection, modal, and buckling analysis.
    \item Optimized geometry is reconstructed.
\end{itemize}
\end{minipage} &
\begin{minipage}[t]{2.1cm}
\begin{itemize}[label={-}, leftmargin=*, topsep=0pt, partopsep=0pt, parsep=0pt, itemsep=0pt]
    \item 38.2\% mass reduction.
    \item 9.6\% reduction in maximum stress.
\end{itemize}
\end{minipage} 
\\ 
\addlinespace[0.2cm]

\cite{Lee_González_Lee_Kim_Park_Han_2016} &
\begin{minipage}[t]{1.8cm}
\begin{itemize}[label={-}, leftmargin=*, topsep=0pt, partopsep=0pt, parsep=0pt, itemsep=0pt]
    \item Jacket foundation.
    \item NREL 5 MW.
\end{itemize}
\end{minipage} &
\begin{minipage}[t]{1.8cm}
\begin{itemize}[label={-}, leftmargin=*, topsep=0pt, partopsep=0pt, parsep=0pt, itemsep=0pt]
    \item Bladed (AHSE).
    \item Nastran (FEA).
    \item FE-Design (TO).
\end{itemize}
\end{minipage} &
\begin{minipage}[t]{3cm}
\begin{itemize}[label={-}, leftmargin=*, topsep=0pt, partopsep=0pt, parsep=0pt, itemsep=0pt]
    \item Types: aerodynamic, hydrodynamic, wind, and gravitational.
    \item Loads in different directions.
    \item Ultimate design load case: DLC 6.2b (IEC61400-3).
    \item Fatigue design load case: DLC 1.2 (IEC61400-3).
    \item Values: 1 case.
\end{itemize}
\end{minipage} &
\begin{minipage}[t]{2.3cm}
\begin{itemize}[label={-}, leftmargin=*, topsep=0pt, partopsep=0pt, parsep=0pt, itemsep=0pt]
    \item TO: SIMP.
    \item Objective function: minimize compliance.
\end{itemize}
\end{minipage} &
\begin{minipage}[t]{3cm}
\end{minipage} &
\begin{minipage}[t]{3cm}
\begin{itemize}[label={-}, leftmargin=*, topsep=0pt, partopsep=0pt, parsep=0pt, itemsep=0pt]
    \item Design: 2 symmetry planes.
    \item Response: 10\% original volume.
\end{itemize}
\end{minipage} &
\begin{minipage}[t]{3cm}
\begin{itemize}[label={-}, leftmargin=*, topsep=0pt, partopsep=0pt, parsep=0pt, itemsep=0pt]
    \item Original: stress, deflection, and fatigue analysis.
    \item Optimized: stress and fatigue analysis.
\end{itemize}
\end{minipage} &
\begin{minipage}[t]{2.1cm}
\begin{itemize}[label={-}, leftmargin=*, topsep=0pt, partopsep=0pt, parsep=0pt, itemsep=0pt]
    \item 7.4\% mass reduction.
    \item 14\% reduction in maximum stress.
    \item Improved fatigue life 2-4 times.
\end{itemize}
\end{minipage} 
\\ 
\addlinespace[0.2cm]

\cite{Zhang_Long_Zhang_Lu_Bai_Jia_2022} &
\begin{minipage}[t]{1.8cm}
\begin{itemize}[label={-}, leftmargin=*, topsep=0pt, partopsep=0pt, parsep=0pt, itemsep=0pt]
    \item OC4 Jacket foundation.
    \item NREL 5 MW.
\end{itemize}
\end{minipage} &
\begin{minipage}[t]{1.8cm}
\begin{itemize}[label={-}, leftmargin=*, topsep=0pt, partopsep=0pt, parsep=0pt, itemsep=0pt]
    \item Bladed (AHSE).
    \item Ansys (FEA).
    \item Optistruct (TO).
\end{itemize}
\end{minipage} &
\begin{minipage}[t]{3cm}
\begin{itemize}[label={-}, leftmargin=*, topsep=0pt, partopsep=0pt, parsep=0pt, itemsep=0pt]
    \item Types: aerodynamic, hydrodynamic, wind, and gravitational.
    \item Wind and wave misaligned.
    \item Ultimate design load case: DLC 1.3 and 6.2 (IEC61400-3).
    \item Fatigue design load case: DLC 1.2 (IEC61400-3).
    \item Values: 14 cases.
\end{itemize}
\end{minipage} &
\begin{minipage}[t]{2.3cm}
\begin{itemize}[label={-}, leftmargin=*, topsep=0pt, partopsep=0pt, parsep=0pt, itemsep=0pt]
    \item TO: SIMP.
    \item Objective function: minimize compliance.
\end{itemize}
\end{minipage} &
\begin{minipage}[t]{3cm}
\begin{itemize}[label={-}, leftmargin=*, topsep=0pt, partopsep=0pt, parsep=0pt, itemsep=0pt]
    \item Tube thickness (6 vars.).
    \item Tube diameter (6 vars.).
\end{itemize}
\end{minipage} &
\begin{minipage}[t]{3cm}
\begin{itemize}[label={-}, leftmargin=*, topsep=0pt, partopsep=0pt, parsep=0pt, itemsep=0pt]
    \item Design: 2 symmetry planes.
    \item Response: 20\% original volume.
\end{itemize}
\end{minipage} &
\begin{minipage}[t]{3cm}
\begin{itemize}[label={-}, leftmargin=*, topsep=0pt, partopsep=0pt, parsep=0pt, itemsep=0pt]
    \item Original: stress, deflection, and modal analysis.
    \item Optimized: stress, deflection, and modal analysis.
    \item Optimized geometry is reconstructed.
\end{itemize}
\end{minipage} &
\begin{minipage}[t]{2.1cm}
\begin{itemize}[label={-}, leftmargin=*, topsep=0pt, partopsep=0pt, parsep=0pt, itemsep=0pt]
    \item 38.2\% mass reduction.
    \item 9.6\% reduction in maximum stress.
\end{itemize}
\end{minipage} 
\\
\addlinespace[0.2cm]

\cite{Lan_Long_Saeed_Geng_Chen_Zhang_Tao_Liu_2024} &
\begin{minipage}[t]{1.8cm}
\begin{itemize}[label={-}, leftmargin=*, topsep=0pt, partopsep=0pt, parsep=0pt, itemsep=0pt]
    \item OC4 Jacket foundation.
    \item 5 MW.
\end{itemize}
\end{minipage} &
\begin{minipage}[t]{1.8cm}
\begin{itemize}[label={-}, leftmargin=*, topsep=0pt, partopsep=0pt, parsep=0pt, itemsep=0pt]
    \item Bladed (AHSE).
    \item Ansys (FEA).
    \item Optistruct (TO).
\end{itemize}
\end{minipage} &
\begin{minipage}[t]{3cm}
\begin{itemize}[label={-}, leftmargin=*, topsep=0pt, partopsep=0pt, parsep=0pt, itemsep=0pt]
    \item Types: aerodynamic, hydrodynamic, wind, and gravitational.
    \item Ultimate design load case: DLC 1.3 and 6.2 (IEC61400-3).
    \item Fatigue design load case: DLC 1.2 (IEC61400-3).
    \item Values: 14 cases.
\end{itemize}
\end{minipage} &
\begin{minipage}[t]{2.3cm}
\begin{itemize}[label={-}, leftmargin=*, topsep=0pt, partopsep=0pt, parsep=0pt, itemsep=0pt]
    \item TO: SIMP.
    \item Objective function: minimize compliance with fail-safe concept.
\end{itemize}
\end{minipage} &
\begin{minipage}[t]{3cm}
\begin{itemize}[label={-}, leftmargin=*, topsep=0pt, partopsep=0pt, parsep=0pt, itemsep=0pt]
    \item Tube thickness.
    \item Tube diameter.
\end{itemize}
\end{minipage} &
\begin{minipage}[t]{3cm}
\begin{itemize}[label={-}, leftmargin=*, topsep=0pt, partopsep=0pt, parsep=0pt, itemsep=0pt]
    \item Design: 2 symmetry planes.
    \item Response: 20\% original volume.
\end{itemize}
\end{minipage} &
\begin{minipage}[t]{3cm}
\begin{itemize}[label={-}, leftmargin=*, topsep=0pt, partopsep=0pt, parsep=0pt, itemsep=0pt]
    \item Original: stress, deflection, and modal analysis.
    \item Optimized: stress, deflection, and modal analysis.
    \item Optimized geometry is reconstructed.
\end{itemize}
\end{minipage} &
\begin{minipage}[t]{2.1cm}
\begin{itemize}[label={-}, leftmargin=*, topsep=0pt, partopsep=0pt, parsep=0pt, itemsep=0pt]
    \item Reduced maximum deflection.
    \item Reduced maximum stress.
\end{itemize}
\end{minipage} 
\\ 
\addlinespace[0.2cm]

\cite{Tian_Liu_Deng_Xie_Wang_2024} &
\begin{minipage}[t]{1.8cm}
\begin{itemize}[label={-}, leftmargin=*, topsep=0pt, partopsep=0pt, parsep=0pt, itemsep=0pt]
    \item OC4 Jacket foundation.
    \item NREL 5 MW.
\end{itemize}
\end{minipage} &
\begin{minipage}[t]{1.8cm}
\begin{itemize}[label={-}, leftmargin=*, topsep=0pt, partopsep=0pt, parsep=0pt, itemsep=0pt]
    \item Bladed (AHSE).
\end{itemize}
\end{minipage} &
\begin{minipage}[t]{3cm}
\begin{itemize}[label={-}, leftmargin=*, topsep=0pt, partopsep=0pt, parsep=0pt, itemsep=0pt]
    \item Types: aerodynamic, hydrodynamic, wind, and gravitational.
\end{itemize}
\end{minipage} &
\begin{minipage}[t]{2.3cm}
\begin{itemize}[label={-}, leftmargin=*, topsep=0pt, partopsep=0pt, parsep=0pt, itemsep=0pt]
    \item TO: SIMP.
    \item Objective function: minimize fatigue damage.
\end{itemize}
\end{minipage} &
\begin{minipage}[t]{3cm}
\begin{itemize}[label={-}, leftmargin=*, topsep=0pt, partopsep=0pt, parsep=0pt, itemsep=0pt]
    \item Tube thickness (4 vars.).
    \item Tube diameter (5 vars.).
    \item Connection heights (5 vars.).
    \item Intermediate node heights (6 vars.).
\end{itemize}
\end{minipage} &
\begin{minipage}[t]{3cm}
\begin{itemize}[label={-}, leftmargin=*, topsep=0pt, partopsep=0pt, parsep=0pt, itemsep=0pt]
    \item Geometry limits.
    \item Maximum deflection.
    \item Maximum stress.
    \item Minimum fatigue life.
    \item Frequency limits.
\end{itemize}
\end{minipage} &
\begin{minipage}[t]{3cm}
\begin{itemize}[label={-}, leftmargin=*, topsep=0pt, partopsep=0pt, parsep=0pt, itemsep=0pt]
    \item Original: stress, deflection, modal, and buckling analysis.
    \item Optimized: stress, deflection, modal, and buckling analysis.
    \item Optimized geometry is reconstructed.
\end{itemize}
\end{minipage} &
\begin{minipage}[t]{2.1cm}
\begin{itemize}[label={-}, leftmargin=*, topsep=0pt, partopsep=0pt, parsep=0pt, itemsep=0pt]
    \item 14.6\% mass reduction.
    \item Reduced maximum stress.
    \item More uniform stress distribution.
    \item Increased critical buckling load.
\end{itemize}
\end{minipage} 
\\ 
\addlinespace[0.2cm]

\cite{Lu_Long_Zhang_Zhang_Tao_2023} &
\begin{minipage}[t]{1.8cm}
\begin{itemize}[label={-}, leftmargin=*, topsep=0pt, partopsep=0pt, parsep=0pt, itemsep=0pt]
    \item Tripod foundation.
    \item NREL 5 MW.
\end{itemize}
\end{minipage} &
\begin{minipage}[t]{1.8cm}
\begin{itemize}[label={-}, leftmargin=*, topsep=0pt, partopsep=0pt, parsep=0pt, itemsep=0pt]
    \item Bladed (AHSE).
    \item Nastran (FEA).
    \item Optistruct (TO).
    \item Designlife (fatigue).
\end{itemize}
\end{minipage} &
\begin{minipage}[t]{3cm}
\begin{itemize}[label={-}, leftmargin=*, topsep=0pt, partopsep=0pt, parsep=0pt, itemsep=0pt]
    \item Types: aerodynamic, hydrodynamic, wind, and gravitational.
    \item Ultimate design load case: DLC 1.3 and 6.2 (IEC61400-3).
    \item Fatigue design load case: DLC 1.2 (IEC61400-3).
    \item Values: 14 cases.
\end{itemize}
\end{minipage} &
\begin{minipage}[t]{2.3cm}
\begin{itemize}[label={-}, leftmargin=*, topsep=0pt, partopsep=0pt, parsep=0pt, itemsep=0pt]
    \item TO: SIMP.
    \item Objective function: minimize compliance.
\end{itemize}
\end{minipage} &
\begin{minipage}[t]{3cm}
\begin{itemize}[label={-}, leftmargin=*, topsep=0pt, partopsep=0pt, parsep=0pt, itemsep=0pt]
    \item Tube thickness.
    \item Tube diameter.
\end{itemize}
\end{minipage} &
\begin{minipage}[t]{3cm}
\begin{itemize}[label={-}, leftmargin=*, topsep=0pt, partopsep=0pt, parsep=0pt, itemsep=0pt]
    \item Design: 3 symmetry planes.
    \item Response: 30\% original volume.
\end{itemize}
\end{minipage} &
\begin{minipage}[t]{3cm}
\begin{itemize}[label={-}, leftmargin=*, topsep=0pt, partopsep=0pt, parsep=0pt, itemsep=0pt]
    \item Original: stress, modal, and fatigue analysis.
    \item Optimized: stress, modal, and fatigue analysis.
    \item Optimized geometry is reconstructed.
\end{itemize}
\end{minipage} &
\begin{minipage}[t]{2.1cm}
\begin{itemize}[label={-}, leftmargin=*, topsep=0pt, partopsep=0pt, parsep=0pt, itemsep=0pt]
    \item 16.3\% mass reduction.
    \item Improved fatigue life 2-3 times.
\end{itemize}
\end{minipage} 
\\ 
\bottomrule
\end{longtable}
\end{landscape}
\twocolumn


\bibliographystyle{elsarticle-num-names} 

\bibliography{references}





\end{document}